\documentclass[tighten, times, twocolumn]{aastex631}  
\usepackage{multirow}

\usepackage{soul}
\newcommand{\add}[1]{\textbf{\textcolor{blue}{#1}}}

\shorttitle{FAST GPPS survey: VIII. 116 binary pulsars}
\shortauthors{P. F. Wang et al.}

\begin{document}

\title{The FAST Galactic Plane Pulsar Snapshot survey:
  VIII. 116 binary pulsars}

\author[0000-0002-6437-0487]{P.~F. Wang}
\affiliation{National Astronomical Observatories, Chinese Academy of Sciences, Jia-20 Datun Road, ChaoYang District, Beijing 100012, China}
\affiliation{School of Astronomy and Space Science, University of Chinese Academy of Sciences, Beijing 100049, China}
\affiliation{Key Laboratory of Radio Astronomy and Technology,  Chinese Academy of Sciences, Beijing 100101, China }

\author[0000-0002-9274-3092]{J.~L. Han} 
\affiliation{National Astronomical Observatories, Chinese Academy of Sciences, Jia-20 Datun Road, ChaoYang District, Beijing 100012, China}
\affiliation{School of Astronomy and Space Science, University of Chinese Academy of Sciences, Beijing 100049, China}
\affiliation{Key Laboratory of Radio Astronomy and Technology,  Chinese Academy of Sciences, Beijing 100101, China }

\author[0009-0009-6590-1540]{Z.~L. Yang}
\affiliation{National Astronomical Observatories, Chinese Academy of Sciences, Jia-20 Datun Road, ChaoYang District, Beijing 100012, China}
\affiliation{School of Astronomy and Space Science, University of Chinese Academy of Sciences, Beijing 100049, China}

\author[0000-0002-4704-5340]{T. Wang}
\affiliation{National Astronomical Observatories, Chinese Academy of Sciences, Jia-20 Datun Road, ChaoYang District, Beijing 100012, China}

\author[0009-0004-3433-2027]{C. Wang}
\affiliation{National Astronomical Observatories, Chinese Academy of Sciences, Jia-20 Datun Road, ChaoYang District, Beijing 100012, China}
\affiliation{School of Astronomy and Space Science, University of Chinese Academy of Sciences, Beijing 100049, China}
\affiliation{Key Laboratory of Radio Astronomy and Technology,  Chinese Academy of Sciences, Beijing 100101, China }

\author[0009-0003-2212-4792]{W.~Q. Su}
\affiliation{National Astronomical Observatories, Chinese Academy of Sciences, Jia-20 Datun Road, ChaoYang District, Beijing 100012, China}
\affiliation{School of Astronomy and Space Science, University of Chinese Academy of Sciences, Beijing 100049, China}

\author[0000-0003-1778-5580]{J. Xu}
\affiliation{National Astronomical Observatories, Chinese Academy of Sciences, Jia-20 Datun Road, ChaoYang District, Beijing 100012, China}
\affiliation{Key Laboratory of Radio Astronomy and Technology,  Chinese Academy of Sciences, Beijing 100101, China }

\author[0000-0002-6423-6106]{D.~J. Zhou}
\affiliation{National Astronomical Observatories, Chinese Academy of Sciences, Jia-20 Datun Road, ChaoYang District, Beijing 100012, China}

\author[0009-0008-1612-9948]{Yi Yan}
\affiliation{National Astronomical Observatories, Chinese Academy of Sciences, Jia-20 Datun Road, ChaoYang District, Beijing 100012, China}

\author[0000-0002-1056-5895]{W.~C. Jing}
\affiliation{National Astronomical Observatories, Chinese Academy of Sciences, Jia-20 Datun Road, ChaoYang District, Beijing 100012, China}
\affiliation{School of Astronomy and Space Science, University of Chinese Academy of Sciences, Beijing 100049, China}

\author[0000-0002-5915-5539]{N.~N. Cai}
\affiliation{National Astronomical Observatories, Chinese Academy of Sciences, Jia-20 Datun Road, ChaoYang District, Beijing 100012, China}

\author{J.~P. Yuan}
\affiliation{Xinjiang Astronomical Observatory, Chinese Academy of Sciences, 150 Science 1-street, Urumqi 830011, China}

\author{R.~X. Xu}
\affil{Department of Astronomy, Peking University, Beijing 100871, China}
\affil{Kavli Institute for Astronomy and Astrophysics, Peking University, Beijing 100871, China}

\author{H.~G. Wang}         
\affil{Department of Astronomy, School of Physics and Materials Science, Guangzhou University, Guangzhou 510006, China}

\author{X.~P. You}
\affil{School of Physical Science and Technology, Southwest University, Chongqing 400715, China}

\email{Corresponding to: hjl@nao.cas.cn; pfwang@nao.cas.cn}


\begin{abstract}
Finding pulsars in binaries are important for measurements of the masses of neutron stars, for tests of gravity theories, and for studies of star evolution. We are carrying out the Galactic Plane Pulsar Snapshot survey (GPPS) by using the the Five-hundred-meter Aperture Spherical radio Telescope (FAST). Here we present the Keplerian parameters for 116 newly discovered pulsars in the FAST GPPS survey, and obtain  timing solutions for 29 pulsars. Companions of these pulsars are He white dwarfs, CO/ONe white dwarfs, neutron stars, main sequence stars and ultra light objects or even planets. Our observations uncover eclipses of 8 binary systems. The optical counterpart for the companion of PSR J1908+1036 is identified. The Post-Keplerian parameter $\dot{\omega}$ for the double neutron star systems PSR J0528+3529 and J1844-0128 have been measured, with which the total masses of the binary systems are determined.  
\end{abstract}


\section{Introduction}           
\label{sect:intro}

Pulsars are rapidly rotating neutron stars with spin periods ranging from about 1.39 millisecond to several tens of seconds  \citep[][see updated catalogue 2.4.0\footnote{https://www.atnf.csiro.au/people/pulsar/psrcat/}]{mhth05}. They are formed through supernova explosion of massive stars (8-25$M_{\sun}$) at the end of stellar evolution or accretion induced collapse of massive white dwarfs (WDs). Pulsars serve as important laboratories for studying fundamental physics due to their super high density, extremely strong magnetic field and gravitional field. They have been employed to detect low-frequency gravitational waves through pulsar timing arrays \citep[e.g.][]{ltm+15, aab+21, rsc+21, xcg+23}, test the theory of gravity \citep[e.g.][]{fwe+12}, constrain the equation of state (EOS) of neutron star by mass measurement \citep[e.g.][]{dpr+10, afw+13, of16}, and construct pulsar-based time-scale \citep{hgc+20}. 


\begin{table*}
  \centering
  \caption{Parameter ranges for known binary pulsars with various companion types in the Galactic field and the classification criteria. }
  \label{tab:classify}
  \footnotesize
  \tabcolsep 1.0mm
  \renewcommand{\arraystretch}{0.9}
\begin{tabular}{lccccl}
\hline\hline
Companion type&  $P$   & $P_b$  & $m_{\rm c,med}$ & $e$ & Coarse classification Criteria  \\
              & (ms)   & (days) & ($M_\odot$)     &     &           \\
\hline
He-WD        & [1.74, 834.84] & [0.10, 944.64] & [0.07, 0.48]  & [1.2$\times 10^{-7}$, 0.14] & $m_{\rm c,med}>0.08M_\odot$ \& $ P<10$ms \& \\
& & & & & [ ($m_{\rm c,med}<0.356M_\odot$ \& $P_b<100$d) \\
& & & & & or ($m_{\rm c,med}<0.5M_\odot$ \& $P_b>100$d) ]  \\
CO/ONe-WD    & [2.91, 1066.37]& [0.19, 95.26$^\dagger$]& [0.33$^\dagger$, 1.58] & [6.9$\times 10^{-7}$, 0.66] & $m_{\rm c,med}>0.356M_\odot$ $\& P_b<100$d \\
& & & & & $\& ~ e <0.05$ \\
Ultra light object (UL)   & [1.41, 520.95] & [0.06, 10.59]  & [0.0009, 0.097]  & [2.6$\times 10^{-6}$, 4.5$\times 10^{-3}$] & 
$m_{\rm c,med}<0.08M_\odot $ \\
MS star (unevolved) & [47.76$^\ddagger$, 763.93] & [95.17, 16800] & [1.08, 18.50] & [0.08, 0.96]      & $m_{\rm c,med}>1.0M_\odot$ $\& P_b>80$d   \\
& & & & & $\&~\rm e>0.05$ $\& P>30$ms\\
MS star (evolved)  & [1.61, 14.25]  & [0.08, 1.1\add{$^*$}]  & [0.11, 0.52]   & [4.2$\times 10^{-6}$, 2.1$\times 10^{-4}$] &  $0.1M_\odot<m_{\rm c,med}<0.52M_\odot$  $\&~ P_b<1.1$d   \\
& & & & & $\& ~ \rm e < 0.001$ $\& P<30$ms \\
& & & & &  $\& \rm Eclipse$ \\
Neutron star (NS)    & [16.96 2773.46] & [0.078, 45.06] & [0.76, 1.74]   & [0.06, 0.83]                 & $m_{\rm c,med}>0.76M_\odot$ \& $P_b<80$d    \\
& & & & &  $ \& ~ \rm e >0.05$ \\ 

\hline
\end{tabular}
\tablecomments{The about parameter spaces are rough indicators for classification of binary types. However, there have been some exceptions. For example, $\dagger$: J0823+0159 has a CO WD companion with an orbital period of 1232.4 days and a median mass of 0.226$M_\odot$ \citep{Koester+2000A&A...364L..66K}, $\ddagger$: J1903+0327 has a MS companion, a spin period of 2.150ms and an orbital period of 95.2 days \citep{Freire+2011MNRAS.412.2763F}, *: J1417$-$4402 has a red-giant companion with an orbital period of 5.4 days \citep{Strader+2015ApJ...804L..12S}. }
\end{table*}

Among 3724 known pulsars \citep{mhth05}, 421 pulsars are in binary systems. Most of the pulsars in binary systems are millisecond pulsars (MSPs) with spin periods of $P\le30$ ms and spin-down rate $\dot{P}\le10^{-17}$. These binary MSPs are generally believed to be spun up by accreting material from their companion \citep[e.g.][]{acrs82,bv91}, so they are called as ``recycled pulsars". During recycling, the magnetic field strength of pulsars are reduced and their orbits are circularized. Companions of these binary pulsars can be neutron stars (NS), white dwarfs (WD), main sequence stars (MS), ultra-light companion (UL) such as brown dwarf or white dwarf remnant or planet with mass $m_c \le 0.1M_{\sun}$. 
The spin periods $P$, pulsar ellipses, orbital periods $P_b$, orbital eccentricity $e$ and median companion masses $m_{\rm c,med}$ derived from mass functions can be used as criteria to classify companion types, see Table~\ref{tab:classify}. For example, some MSPs in compact orbits ($P_{\rm b}<1$ day) with a low-mass non- or semi-degenerate companion are ``spiders", in which the pulsar winds are evaporating the companion. Many of them show the eclipses of pulsar signals. They are further divided into ``black-widows" if the companion mass is less than $\sim 0.1M_{\sun}$, or ``redbacks"  if the companion mass is greater than $\sim 0.1M_{\sun}$ and such companions are probably evolved MSs \citep{rob13,ccth13}.


Binary systems have diverse formation history. It is generally accepted that double neutron star systems originate from high-mass X-ray binaries (HMXB) via a common envelope (CE) and spiral-in process \citep[e.g.][]{tv06}. Pulsars with CO/ONe white dwarf companions originate from intermediate-mass X-ray binaries \citep[IMXBs, e.g.][]{tlk12}, and the ones with He white dwarf companions originate from low-mass X-ray binaries \citep[LMXBs, e.g.][]{ps88,itl14}. The final orbital configuration and stellar masses are diverse, depending on the companion mass and metallicity, the initial orbital separation as well as the evolution stage of the donor star when Roche-Lobe Overflow (RLO) initiates, for example, the Case A RLO on the main sequence stage, the Case B RLO at the red giant branch (RGB) stage with hydrogen shell burning, the Case C RLO for the asymptotic giant branch (AGB) stage with helium shell burning \citep[e.g.][]{tv23}. Hence, the spin period, period derivative, orbital period, orbit eccentricity, and masses of pulsar and its companion are important fossil records for the original binary evolution \citep[e.g.][]{co84, phi92}. Based on observations of binary pulsars, the evolution theories of stellar binaries have been developed and examined \citep[e.g.][]{tlk12, itl14, cthc21}.

Binary systems with compact orbits generally exhibit a number of relativistic effects which can be quantified by Post-Keplerian (PK) parameters. With two PK parameters, masses of the pulsar and the companion as well as the inclination angle of the orbit can be determined by resorting to a theory of gravity, e.g. General Relativity (GR) \citep[e.g.][]{of16}. By combing these precisely measured pulsar masses, underlying distribution of neutron star masses can be determined \citep{ato+16}. Masses of the most massive neutron stars pose important constraints on the equation of state (EOS) of neutron stars, e.g. PSRs J0348$+$0432, J0740$+$6620 and J1614$-$2230 \citep[e.g.][]{dpr+10, afw+13, of16}. The so measured companion masses can be employed to test the theories of binary evolution \citep{pk94,ts99}. Combination of more than two PK parameters allows for the tests of GR and other theory of gravity in the strong-field regime with high precision \citep{wt84,ksm+06}. The violation of strong equivalence principle can also be tested by constraining the universality of free fall or gravitational dipole radiation \citep{fwe+12,vcf+20}. Moreover, relativistic deformation of the orbit \citep{wh16, cck+18} and relativistic spin-orbit coupling or named as Lense-Thirring process \citep{cck+18, hkw+20, vb20} can also be detected from the change of orbital inclination.

Finding and precisely timing new binary systems, especially those with compact orbits, exceptional orbital configurations or the most massive pulsars, will improve the existing tests of gravity, the constraints of EOS and and the understanding the evolution channels of stellar binaries. There have been many efforts devoted to precise measurements of the binary nature of the systems \citep[e.g.][]{drc+21,ksk+21,mzl+23}, and most of them are follow-up observations of new pulsars discovered in large pulsar survey projects.

\begin{table}[t]
  \centering
  \caption{FAST projects for collecting data for binary pulsar timing in this paper.}
  \label{tab:Project}
  \footnotesize
  \renewcommand{\arraystretch}{0.7}
\begin{tabular}{lrc}
\hline\hline
Project ID & PI  & Observation hours \\
\hline
ZD2020$\_$2  & J.L. Han  &  350+  \\
ZD2021$\_$2  & J.L. Han  &  350+  \\
ZD2022$\_$2  & J.L. Han  &  350+  \\
ZD2023$\_$2  & J.L. Han  &  350+  \\
ZD2024$\_$2  & J.L. Han  &  350+  \\
PT2020\_0136 & P.F. Wang & 30.0   \\
PT2021\_0037 & T. Wang   & 42.0  \\
PT2021\_0126 & P.F. Wang & 40.0  \\
PT2022\_0047 & W.Q. Su   & 70.0  \\
PT2022\_0158 & Z.L. Yang & 20.0  \\
PT2022\_0159 &   T. Wang & 20.0  \\
PT2022\_0174 & P.F. Wang & 20.0  \\
PT2022\_0178 & Z.L. Yang & 15.0  \\
PT2023\_0084 & Z.L. Yang & 18.9  \\
PT2023\_0085 & P.F. Wang & 19.5  \\
PT2023\_0143 & Z.L. Yang & 15.8  \\
PT2023\_0162 & W.Q. Su   & 11.5  \\
PT2023\_0190 & P.F. Wang &  7.6  \\
PT2023\_0193 & Z.L. Yang & 11.5  \\
PT2023\_0195 & Z.L. Yang & 11.2  \\
PT2024\_0007 & P.F. Wang & 23.3   \\
PT2024\_0020 & Z.L. Yang & 31.9 \\
PT2024\_0026 & P.F. Wang & 25.0   \\
PT2024\_0200 & Z.L. Yang & 13.1 \\
PT2024\_0224 & W.C. Jing & 5.6  \\
PT2024\_0231 & Z.L. Yang & 5.4  \\
\hline
\end{tabular}
\end{table}

We are carrying out the Galactic Plane Pulsar Snapshot (GPPS) survey \citep{hww+21} by using the the Five-hundred-meter Aperture Spherical radio Telescope \citep[FAST,][]{Nan+2006ScChG..49..129N,Nan+2011IJMPD..20..989N}, with a goal to discover pulsars within the Galactic latitude of $\pm 10^\circ$ of the FAST visible sky area. Up to now, we have discovered 751 pulsars \footnote{http://zmtt.bao.ac.cn/GPPS/GPPSnewPSR.html}  \citep{hzw+25}.
Among them about 160 pulsars show binary features, such as prominent acceleration in the discovery diagram or a significant variation of barycentric periods in the confirmation observation. 
We ascertain their binary nature with a few follow-up observations, and get the orbital parameters determined for about 3/4 of them. More follow-up observations of these binary pulsars have been done by several applied FAST projects, see Table~\ref{tab:Project}. Combining all data, we get timing solutions for some of them (see Table~\ref{tab:PreKep} and Table~\ref{tab:ephem_appendix} in appendix). The first pulsar discovered by the GPPS survey, PSR J1901+0658 (gpps0001), is turned out to be a double neutron star system \citep{shy+24}. The 190th GPPS pulsar, PSR J1953+1844 (gpps0190), is a binary with the shortest orbital period of only 53 minutes \citep{plj+23}, probably a descendant of an ultracompact X-Ray binary \citep{yhj+23}. PSR J1928+1815 (gpps0121) is an eclipsed millisecond pulsar in a compact orbit with an orbital period of 3.6 hours, standing as evidence for the common envelope phase \citep{yhz+25}. We get timing solution for 6 millisecond pulsars in compact orbits with massive white dwarf companions \citep{yhw+25}. 
Here, we present the results for 116 new binary pulsars, as listed in Table~\ref{tab:PreKep}. In Section 2, we briefly describe the FAST observations and data reduction procedures. The Keplerian solutions and timing solutions are presented in Section 3 together with companion classification. Conclusions and further discussion are given in Section 4.

\section{FAST observations and data reduction}

\begin{figure}
  \centering
  \includegraphics[angle=0,height = 0.25\textheight] {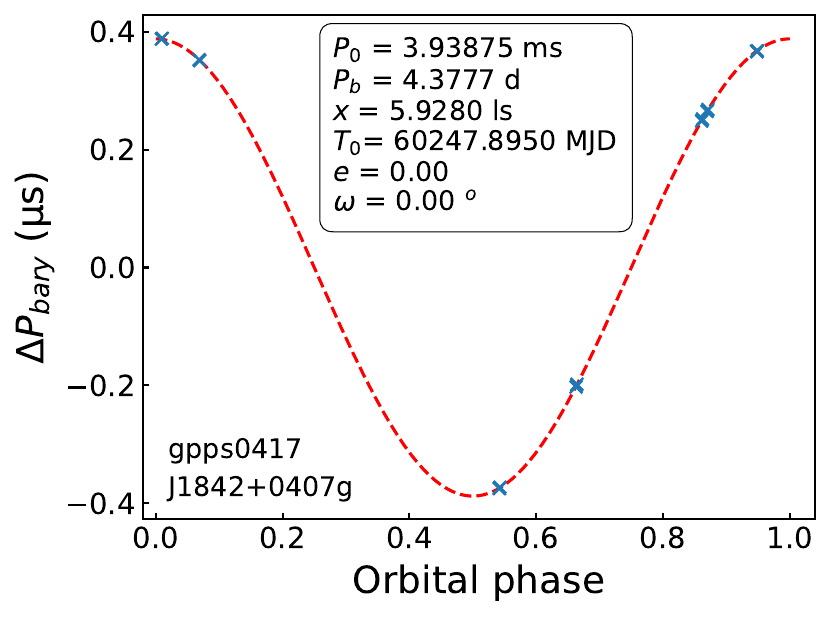}
  \caption{Variation of the barycentric periods for PSR J1842+0407g across the orbit phase. The observed barycentric periods are marked by ``x" after the spin period $P_0$ is subtracted. The error-bars are marked but too small to see for most data. Dashed line is the best-fit by using the preliminary Keplerian model with orbital parameters ($P_b$, $x$, $T_0$, $e$ and $\omega$) listed inside the panel. The orbital phase is referred to the periastron of the orbit. Plots for 78 newly discovered pulsars by the FAST GPPS survey are given in Fig.~\ref{fig:Pbary_appendix} in the Appendix.
  }
  \label{fig:Pbary}
\end{figure}

All FAST Observations used in this paper (see Table~\ref{tab:Project}) have been carried out by using the 19-beam L-band receiver. In the GPPS survey observations (ZD2020\_2 to ZD2024\_2), the snapshot observation mode has been used to cover a hexagonal sky area of 0.1575 square degrees by using 76 beams \citep[see Fig. 4 in][]{hww+21}. The follow-up observations have been carried out by using the tracking mode and  also the 19-beam L-band receiver, including these verification observations made by the ZD202x\_2 or the targeted observations by other free-applied projects (PT202x\_0yyy).

The 19-beam L-band receiver works at a central frequency of 1250~MHz with a bandwidth of $\sim$ 500~MHz \citep{jth+20}. Radio signals from two orthogonal linear polarizations, X and Y, are sampled, channelized and correlated (XX, YY, Re[$X^*Y$], Im[$X^*Y$]) in digital backends for each beam. The data stream are stored in \textsc{``search mode" psrfits} files with 2048 spectral channels and a time resolution of 49.152 $\mu s$ in our observations.

The GPPS survey observations take 5 minutes for each pointing  \citep{hww+21}. The verification observations of a pulsar take 15 minutes for the targeted objects by using the central beam (M01) of the 19-beam L-band receiver if it has been detected in any survey beam. In tracking observations the data from all 19 beams are also recorded. Sometimes we do detect some other pulsars from away beams \citep{hzw+25}. In general, at the beginning or the end of each observation session, the periodic calibration noise signals are often turned on and off for 40 second or 1 minute or 2 minutes, which are used for the flux and polarization calibrations during off-line data processing.

\movetabledown=0mm
\startlongtable
\begin{longrotatetable}
 commands
\tablecomments{Pulsar name with the GPPS number in the bracket; right ascension (RA, in hh:mm:ss.s); declination (Dec, in dd:mm:ss.s),  dispersion measure (DM, in pc~cm$^{-3}$);  pulsar spin period $P$ (in millisecond); orbital period $P_b$ (in days);  projected semi-axis $x$ (in light second);  
orbital eccentricity $e$;  mass function $f$; the minimum and median companion mass $m_{\rm c,min}$ and $m_{\rm c,med}$ of the companion estimated by assuming the orbit inclination angle of $i=90^\circ$ or $60^\circ$ together with the pulsar mass $m_p=1.35 M_{\odot}$; Companion types: He-white dwarf (He WD), CO/ONe white dwarf (CO/ONe WD), neutron star (NS), main sequence star (MS), ultra light object or palent (UL); Notes: Eclipse for eclipsing pulsar.}
\tablerefs{[0]: this work;  [1]: \citet{yhw+25};  [2]: \citet{shy+24};  [3]: \citet{yhz+25}; [4]: \citet{plj+23}. }
\end{deluxetable*}
\end{longrotatetable}


\begin{table}
  \centering
  \caption{Measured and derived parameters of PSR J0528+3529, as one example of 29 pulsars with phase-connected timing solutions presented in Table~\ref{tab:ephem_appendix}.}
  \label{tab:ephem}
  \tabcolsep 2.0mm
  \renewcommand{\arraystretch}{0.7}
\begin{tabular}{lc}
\hline\hline
Pulsar name\dotfill      & J0528+3529                     \\
\hline
GPPS name \dotfill       & gpps0537  \\
MJD range      \dotfill  & 59930-60625          \\
Dat Span (yr) \dotfill   & 1.9                  \\
Number of TOAs \dotfill  & 23                   \\
Ref. epoch (MJD)\dotfill & 60000                 \\

\hline
\multicolumn{2}{c}{Measured quantities} \\
\hline

Right ascension: RA (hh:mm:ss)\dotfill       & 05:28:28.7978(1)      \\ 
Declination: DEC (dd:mm:ss)\dotfill          & +35:29:36.81(3)   \\ 
Dispersion measure: DM (cm$^{-3}$pc)\dotfill & 111.837(6)           \\ 
Pulse frequency: $\nu$ (s$^{-1}$)\dotfill    & 12.78221930378(2) \\ 
First derivative $\nu$: $\dot{\nu}$ ($10^{-16}\, \rm Hz \, s^{-1}$)\dotfill& $-$1.202(4)    \\ 
Residual ($\mu s$)\dotfill                   & 6.037                 \\
EFAC\dotfill                                 & 0.77                     \\
EQUAD\dotfill                                & 0.0                     \\
Reduced $\chi^2$  \dotfill                   & 0.99                   \\

\hline
\multicolumn{2}{c}{Binary parameters} \\
\hline

Binary model\dotfill                        & DD                         \\
Orbital period: $P_{\rm b}$ (d)\dotfill     & 11.7261813(4)             \\ 
Projected semi-major axis: $x$ (lt-s)\dotfill & 31.43468(2)             \\ 
Periastron passage time: $T_0$ (MJD)\dotfill & 59994.190822(8)          \\ 
Orbital eccentricity: $e$ \dotfill           & 0.2901088(10)             \\ 
Longitude of periastron: $\omega$ ($\deg$) \dotfill & 184.9436(2)                             \\ 
Advance rate of $\omega$: $\dot{\omega}$ ($\deg/yr$) \dotfill & 0.0072(3)   \\ 

\hline
\multicolumn{2}{c}{Derived quantities} \\
\hline
Galactic longitude: $l$ ($\deg$) \dotfill    & 172.52387(3)          \\
Galactic latitude: $b$ ($\deg$) \dotfill     &   0.46700(2)          \\
YMW17$^1$ distance: $D_{\rm YMW}$ (kpc) \dotfill & 1.933            \\
NE2001$^2$ distance: $D_{\rm NE2001}$ (kpc) \dotfill & 2.950         \\
Spin period: $P$ (ms) \dotfill               & 78.23367572047(12)     \\ 
Derivative of $P$: $\dot{P}$ ($10^{-21}\, \rm s \, s^{-1}$) \dotfill & 736(2) \\   
Characteristic age: $\tau$ (Gyr) \dotfill    & 1.686                   \\
Surface magnetic field: $B_{\rm surf}$ ($10^{8}$G) \dotfill  & 76.762   \\
\hline
\end{tabular}
\tablecomments{
Ephemeris obtained based on the DE440 solar system model \citep{DE440}, Barycentric Dynamical Time (TDB) units, and TT(TAI) clock.  Distances estimated by YMW16 model \citep{ymw17} or NE2001 \citep{cl02}. }
\end{table}

\begin{figure}
  \centering
  \includegraphics[width = 0.45\textwidth] {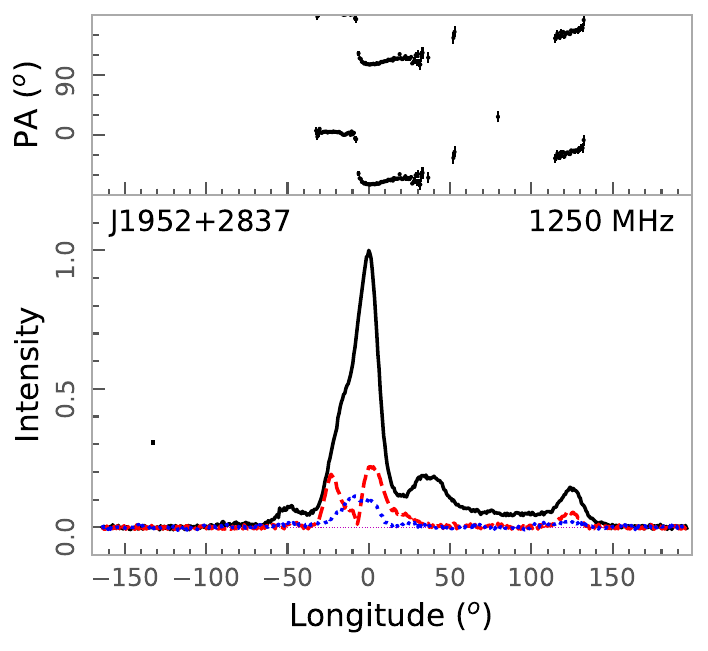}\\
  \caption{Integrated pulse profile of PSR J1952+2837.
  The total intensity, linear and circular polarization are represented by solid, dashed and doted lines in the bottom sub-panel. The left-hand circular polarization is defined to be positive. The bin size and $3\sigma$ are marked inside the sub-panel, here $\sigma$ is the standard deviation of off-pulse bins. In the top panel, dots with error-bar are measurements of polarization position angles for linear polarization intensity exceeding 3$\sigma$ line. The position angles are corrected to infinite frequency by discounting Faraday rotation. Polarized pulse profiles of the 29 pulsars are shown in Figure~\ref{fig:profs_appendix}. 
  }
  \label{fig:profs_example}
\end{figure}

\subsection{Data reduction}

With the initial period and DM obtained from the GPPS survey, we fold the survey data and initial follow-up FAST observation data by using \textsc{dspsr}\footnote{\url{http://dspsr.sourceforge.net/}} \citep{vb11}. The optimal barycentric period $P_{\rm bary}$ for each observation is first searched from a set of trial periods around its nominal value by using the \textsc{pdmp} tool from \textsc{psrchive}\footnote{\url{http://psrchive.sourceforge.net}} \citep{hvm04}. For a binary pulsar, the observed $P_{\rm bary}$ varies due to the Doppler effect caused by orbital motion \citep[e.g.][]{fkl01}. Its initial orbital parameters are obtained by grid searching over the two dimensional parameter space of the orbital period $P_{\rm b}$ and the epoch of passage of periastron $T_0$, together with fitting for the projected semi-major axis $x$ and spin period $P_0$. These parameters are further refined with \textsc{fitorbit}\footnote{https://github.com/vivekvenkris/fitorbit}, during which the orbital eccentricity $e$ and longitude of periastron $\omega$ might also be obtained. With this approach, preliminary orbital parameters are obtained for all the binaries systems, as shown in Figure~\ref{fig:Pbary} as an example. 
By combing orbital period $P_{\rm b}$ and projected semi-axis $x$, one can get the mass function $f$,
\begin{equation} 
f(m_p,m_c)=\frac{(m_c \sin i)^3}{(m_p+m_c)^2}=\frac{4 \pi^2 x^3}{T_\odot P_{\rm b}^2}.
\label{eq:massfunc}
\end{equation}
Here, $m_p$ and $m_c$ are the masses for pulsar and the companion, $i$ is the inclination angle of binary orbit, $T_\odot=G M_\odot/c^3=4.925490947 \mu s$ with $G$ the gravitational constant and $c$ the speed of light. Then the median companion masses can be obtained assuming $m_p=1.35$ M$_\odot$ and $\sin i=60^\circ$.
%

\begin{table*}
  \centering     
  \caption{Polarization profile parameters of 29 new binary pulsars from added FAST observations based on pulsar ephemeris. Numbers in brackets are uncertainties on the last digit. }
  \label{table:pol}
  \tabcolsep 8pt
  \small
  \renewcommand{\arraystretch}{0.8}
  \begin{tabular}{crrrrrrrhhhh}
    \hline\hline
PSR        & P    & $W_{50}$  &  $W_{10}$ &  L/I & V/I  & $|V|$/I & RM      & $\alpha_0$ & $\beta_0$ & Features     \\
           & (ms) & ($^{\circ}$)&($^{\circ}$)& (\%) & (\%) & (\%)    &($\rm rad/m^2$)& ($^{\circ}$) & ($^{\circ}$) &       \\
(1)        &  (2)  &  (3)   &    (4)    &   (5)    & (6) & (7)  & (8) & (9)  &  (10) & (11)  \\
\hline
J0528+3529 & 78.233 & 6.6(7) &  12.9(7) & 10.2(31) & -5.0(31) &  6.3(31) & -88(5)    & & & ot?  \\
J0622+0339 &  8.771 & 9.4(14)&   -      & 49.7(38) &-13.4(36) & 14.9(36) &  28.9(15) & & &     \\
J1838+0024 &  5.087 & 92.2(28)&   -      & 20.2(34) &  1.7(34) &  5.9(34) &  52(9)    & & &     \\
J1840+0012 &  5.338 &193.7(14)& 239.9(15)& 36.2(31) &  0.9(30) &  6.9(31) &  23.4(6)  & 131$_{-8}^{+43}$ & -22$_{-21}^{+21}$ &S,w  \\
J1844-0128 & 29.142 & 18(6)   &   -      &   -      & -3(7)    &  4(7)    &   -       & & &     \\
J1844+0028 &  3.571 & 61(3)   &   -      &  7.4(32) &  9.2(32) & 10.3(32) & -32(8)    & & &     \\
J1845+0201 &  4.309 &139(3)   & 156(3)   & 15.4(33) & -1.4(32) &  5.1(32) &  20(6)    & & &     \\
J1845+0317 &  1.851 & 63.9(28) & -       & 80.1(43) & -1.6(38)  & 7.9(38)  & 20.2(16) & & &     \\
J1857+0642 &  3.530 & 42.0(7) & 173.4(7) & 34.1(30) & -8.6(30) & 12.9(30) & -30.7(11) & 108$_{-6}^{+17}$ & -21.3$_{-12.7}^{+7.3}$ & S,ot   \\
J1903+0839 &  4.621 & 41.5(7) & 216.2(7) & 21.1(30) &  5.9(30) & 10.0(30) & 247.1(13) & & &ot,ip \\
J1904+0553 &  4.907 & 40.9(7) & 138.4(7) & 12.8(30) & 10.6(30) & 12.3(30) & -67.0(16) & & &ot   \\
J1905+0649 & 27.464 & 20(3)   &   -      &   -      &  6(5) &  8(5) &   -       & & &     \\
J1908+1036 & 10.690 &  5.5(7) &  26.1(7) & 71.3(31) & -1.0(30) &  1.3(30) & -27.2(8)  & & &hiL  \\
J1911+1253 & 27.238 & 25.6(7) &  38.8(7) &  5.3(31) &  4.2(31) & 11.8(31) & 224(7)    & & &     \\
J1912+1416 &  3.166 & 50.8(14)&  82.6(14)& 15.1(31) & -7.2(31) & 10.9(31) & 214.7(24) & & &     \\
J1916+0740 & 11.219 & 91.8(14)& 255.9(15)& 15.9(31) &-10.6(31) & 12.9(31) & 609(3) & 171$_{-1}^{+8}$ & -66$_{-23}^{+62}$ & S,ot,w \\
J1917+0615 &  3.967 & 73.5(14)&   -      & 18.6(32) &-11.5(31) & 14.9(31) &  11.4(18) & & &     \\ 
J1917+1259 &  5.637 & 20.6(14)&   -      &  9.8(33) &  4.0(33) & 10.9(33) & 247(14)   & & &     \\
J1918+0621 &  2.103 & 14.3(14)&  34.5(14)& 25.2(30) & 26.6(30) & 27.3(30) & -88.7(8)  & & &     \\
J1924+1342 &  5.721 & 69.8(28)&   -      & 14.9(33) &-12.6(31) & 16.7(31) & 103(13)   & & &     \\
J1930+1403 &  3.209 & 25.1(14)&  51.9(14)& 42.5(31) &  2.2(31) &  2.4(31) &  59.4(7)  & 62$_{-60}^{+116}$ & -26$_{-63}^{+25}$ & S    \\
J1932+2121 & 14.244 & 12.1(7) &  36.6(7) & 16.1(30) & -8.4(30) &  9.5(30) &  99.5(10) & & &ip   \\
J1936+2035 & 32.927 & 17.1(14)&   -      & 64.4(38) & -0.7(37) &  7.0(37) &  12.7(12) & & &c    \\
J1938+2302 & 52.761 & 11.2(7) &   -      &  7.1(33) & -2.2(33) &  4.0(33) &  66(8)    & & &     \\
J1943+2206 &  4.681 & 20(3)   &   -      & 20(4)    &  5(4)    & 11(4)    &-165(25)   & & &     \\
J1946+0904 & 25.772 & 30.7(7) &  64.4(7) & 19.8(30) &  4.1(30) &  5.0(30) & -105.3(6) & 18$_{-16}^{+100}$ & -9.3$_{-32}^{+8}$ & S, ot   \\
J1947+2011 &  8.177 &  9.1(7) &  27.5(7) &  7.8(32) & 11.7(32) & 13.6(32) & -48(8)    & & &     \\
J1952+2837 & 18.020 & 21.1(7) & 159.8(7) & 21.4(30) &  9.7(30) & 11.2(30) & -42.5(5)  & & &ot   \\
J2018+3518 & 31.316 & 20.3(28)&   -      & 42.2(40) &  5.7(35) &  3.9(35) &-332(7)    & & &c    \\
\hline

  \end{tabular}
  
\tablecomments{   
  Columns (1)-(2): pulsar name; spin period,
Columns (3)-(8): the profile properties: pulse widths $W_{50}$ and
$W_{10}$ at 50\% and 10\% the peak intensities, degree (and
uncertainty on the last digit) of linear, circular and absolute
circular polarization $L/I$, $V/I$ and $|V|/I$; FAST measured RM (and uncertainty), i.e. $\rm RM=RM_{obs}-RM_{ion}$,  
%
%
%
}
\end{table*}

Some binary pulsars are monitored for years after initial discoveries by the GPPS survey. Further timing analysis is performed in the following steps. The ELL1 or DD model is first employed to model the orbital motion, depending on if $x e^2$ is much less or larger than the uncertainty of the time of arrival. Because one can not reliably define the time and location of pariastron for a nearly circular orbit, the covariance can be avoided by the Laplace-Lagrange eccentricity parameterizations by using $\epsilon_1=e \sin \omega$ and $\epsilon_2=e \cos \omega$ \citep{lcw+01}. The ELL1 model is parameterized by $P_{\rm b}$, $x$, $T_{\rm asc}$, $\epsilon_1$ and $\epsilon_2$. Here, $T_{\rm asc}=T_0-\omega P_{\rm b}/2\pi$ represents the epoch of the ascending node. The DD model is parameterized by $P_{\rm b}$, $x$, $T_{0}$, $e$ and $\omega$. These five Keplerian parameters of ELL1 or DD together with previously estimated pulsar position ($\alpha$ and $\delta$), rotation period $P$ and dispersion measure (DM) form the initial pulsar ephemeris. 

With the initial ephemeris, we fold the data to form \textsc{``fold mode" psrfits} archive files with \textsc{dspsr}. After removing radio-frequency interference (RFI) using \textsc{paz} and \textsc{psrzap}, one can integrate all subintegrations and all channels using \textsc{pam} in the \textsc{psrchive} tool and get a noise-free standard profile template by using \textsc{paas}. Then, we obtain TOAs from cross-correlating the template with all observed pulse profiles by using \textsc{pat}. With the initial ephemeride and TOAs, phase coherent timing solution is obtained by determining the global rotational count through mapping the gaps between adjunct observations with \textsc{dracula} \citep{fr18}.

Using the phase-connected ephemeris, the recorded data are de-dispersed and folded again. They are calibrated in polarization following the procedures described by \citet{whx+23}. The whole frequency channels are summed to one for data of every 5 minutes. New TOAs are extracted again. To account for the influence of white noise, ToA uncertainties are scaled by EFAC and quadratically added with EQUAD with the help of the efacEquad plugin of tempo2 \citep{hem06}. The ephemeris is finally refined to get a better parameter estimation with $\chi^2 \sim 1$. One example is given in Table~\ref{tab:ephem}, and the ephemeris of 29 pulsars in Table~\ref{tab:ephem_appendix}.

\begin{figure*}
  \centering
  \includegraphics[bb =  5 40 965 165, clip, width=0.95\textwidth] {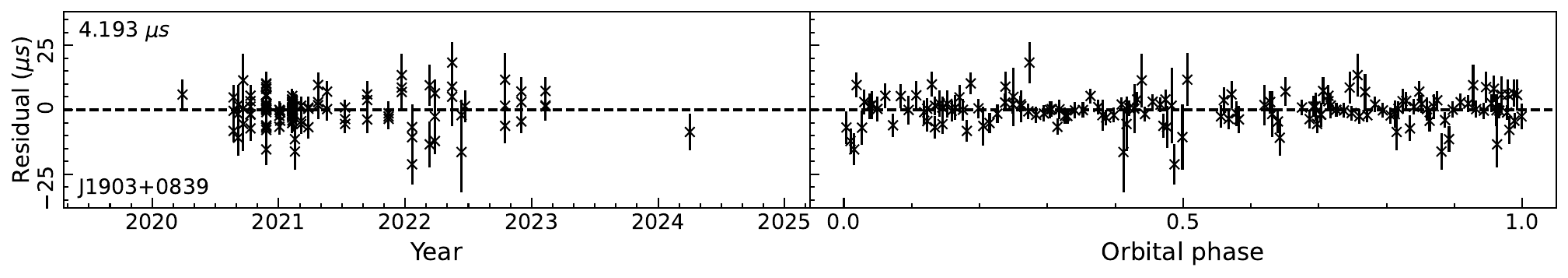} 
  \includegraphics[bb =  5 5 965 165, clip, width=0.95\textwidth] {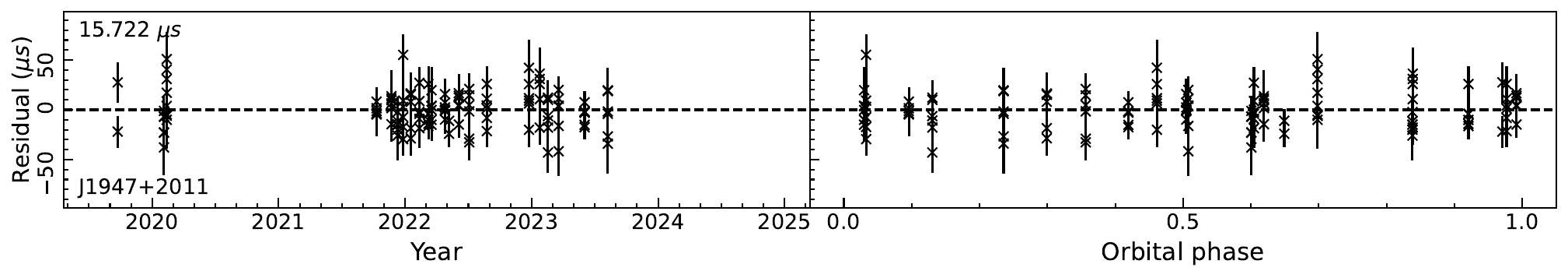} 
  \caption{Timing residuals of two example pulsars PSR J1903+0839 and J1947+2011. {\it Left panels:} Residuals versus observation epochs. The weighted root-mean-square residual of each pulsar is indicated in the top left corner of the panel. {\it Right panels:} Residuals versus orbital phase. The orbital phases are referred to ascending node or periastron depending on the binary model of each pulsar. Timing residuals of the 29 binary pulsars are shown in Figure~\ref{fig:Res_appendix}. }
  \label{fig:Res}
\end{figure*}

\subsection{Pulsar polarization profiles}

Polarized pulse profile represents the mean radiation feature of each pulsar. It is obtained by integrating the emission of tens of thousands of individual pulses. Data from multiple timing observations are then combined to form the integrated pulse profiles, as shown in Figure~\ref{fig:profs_example} for PSR J1952+2837 as an example. Polarization profiles for 29 pulsars are shown in  Figure~\ref{fig:profs_appendix}. The profile width at 50\% and 10\% the peak intensity, the fractional linear, circular and absolute circular polarization are measured and listed in columns (3) to (7) of Table~\ref{table:pol}. The rotation measures (RM) are listed in column (8). 

These pulsars are generally MSPs with wide profiles. The widths of their profiles range from $12.9^\circ$ to $255.9^\circ$ at $10\%$ the peak intensity. Profiles of PSRs J1840+0012 and J1916+0740 are extreme wide, emissions of which extend to more than 180$^\circ$. PSRs J1903+0839 and J1932+2121 exhibit inter pulse emissions that separate by about 180$^\circ$ from the main pulses. PSRs J1840+0012, J1857+0642, J1916+0740, J1930+1403 and J1946+0904 have S-shaped position angle variations. PSRs J0528+3529, J1857+0642, J1903+0839, J1904+0553, J1916+0740, J1946+0904 and J1952+2837, exhibit orthogonal modes manifesting as $90^\circ$ position angle jumps. The leading component of PSRs J1857+0642 and J1916+0740, the inter pulse of PSR J1903+0839 and the profiles of PSRs J1845+0317 and J1908+1036 are highly linearly polarized. PSRs J1936+2035 and J2018+3518 are affected by interstellar scattering, which results in scattering tails of profiles together with flat PAs. These diverse polarization properties resemble those categories reported in \citet{whx+23}. 

\section{Detailed Results of various binary pulsars}

More than 160 binary pulsars have been found by the FAST GPPS survey. We have got 9 pulsars published \citep{shy+24, plj+23, yhj+23, yhw+25, yhz+25}, as mentioned at the end of introduction. For 6 binary pulsars, PSRs J0653+0443, J1852$-$0044, J1856$-$0039, J1921+1631, J1922+1511 and J1933+2038, we have obtained the time solutions and are preparing independent papers. 
Here we list 116 pulsars in Table~\ref{tab:PreKep}. The table consists of two parts, the first for pulsars with phase-connected timing solution, and the second part for pulsars without solutions. There are 38 pulsars in the first part, including the 9 published pulsars for completeness. 
For these 38 pulsars, we listed in the first part of Table~\ref{tab:PreKep} their names, right ascension, declination and dispersion measure in columns (1), (2), (3) and (4). Spin period $P$ obtained from discovery is refined from modeling the orbital motion, as listed in column (5). Columns (6), (7) and (8) are for orbital period $P_{\rm b}$, projected semi-axis $x$, and orbital eccentricity $e$ obtained by modeling the Keplerian orbit. Mass functions of these binary systems are listed in column (9) of Table~\ref{tab:PreKep}. By assuming
$i=90^\circ$ or $60^\circ$ and $m_p=1.35 M_\odot$, the minimum and median companion mass are estimated, as listed in columns (10) and (11). With the spin period, orbital period, eccentricity and the rough companion mass, the companion type can be roughly estimated, as listed in column (13) of Table~\ref{tab:PreKep}.
The new ephemeris with measured and derived parameters for 29 pulsars are given in details in Appendix Table~\ref{tab:ephem_appendix} and one example is shown in Table~\ref{tab:ephem}. The timing residuals are shown for two pulsars in Figure~\ref{fig:Res} and for all 29 pulsars in Figure~\ref{fig:Res_appendix}, which are plotted along the observation epochs and versus the orbital phase. Timing residuals range from 1.053 to 174.563 $\mu s$.

\begin{table}
  \centering
\renewcommand\arraystretch{0.8}
  \caption{Population and companion statistics for pulsar binaries.}
  \label{tab:Comp} 
\begin{tabular}{lccc|cc}
\hline\hline
Binary pulsars      & \multicolumn{3}{c}{Known}& \multicolumn{2}{c}{FAST GPPS}  \\
\hline
Total               & \multicolumn{3}{c}{420} & \multicolumn{2}{c}{116}   \\
\hline
Location            & GF  & GC & EC & GF & GC \\
                    & 315 & 104 & 1 & 115 & 1 \\                    
\hline
He-WD              & 129 & 23 & - &  58  & - \\
CO/ONe-WD          &  39 &  3 & - &  25  & - \\
Ultra light object &  42 & 26 & - &  11  & 1 \\
Main sequence star &  22 &  5 & 1 &   5  & - \\
Neutron star       &  17 &  2 & - &   3  & - \\
Giant star         &   1 &  - & - &   -  & - \\
Helium star        &  -  &  - & - &   1  & - \\
uncertain          &  65 & 45 & - &  12  & - \\
\hline
\end{tabular}
\tablecomments{GF, GC and EG stand for Galactic field, Globular clusters and extra galaxy, respectively. The known pulsar binaries are from ATNF pulsar Catalogue 2.4.0 \citep{mhth05}. It incorporates 3 triple systems with two having He-WD and one having UL companions. }
\end{table}

\begin{figure}
  \centering
  \includegraphics[width = 0.45\textwidth] {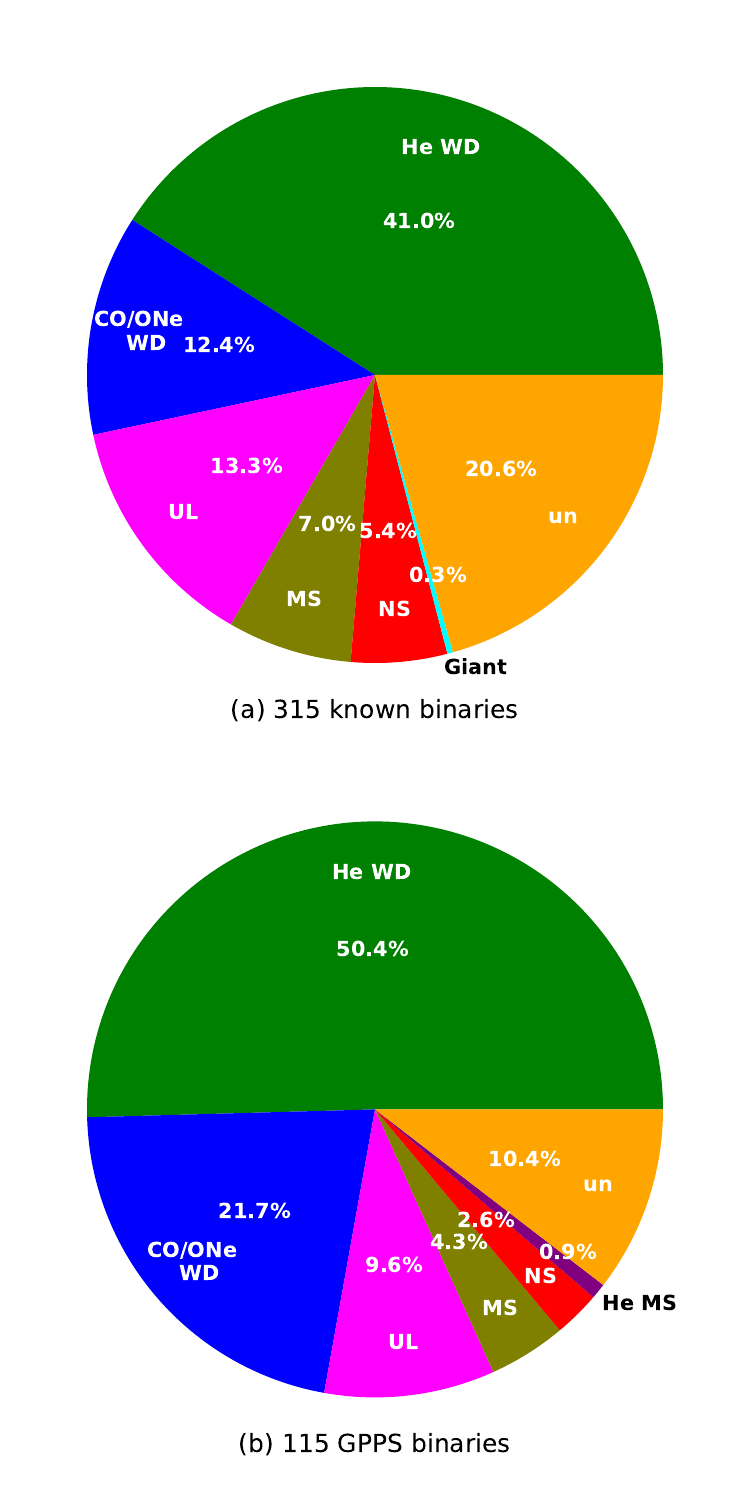}\\
  \caption{Fractions of binary pulsars with different types of companions in the Galactic field. The top panel is for 315 known binaries and the bottom panel for the 115 GPPS discovered ones. Their numbers are listed in Table~\ref{tab:Comp}.}
  \label{fig:binary_pie}
\end{figure}

\begin{figure*}[t]
  \centering
    \includegraphics[angle=0,width = 0.497\textwidth] {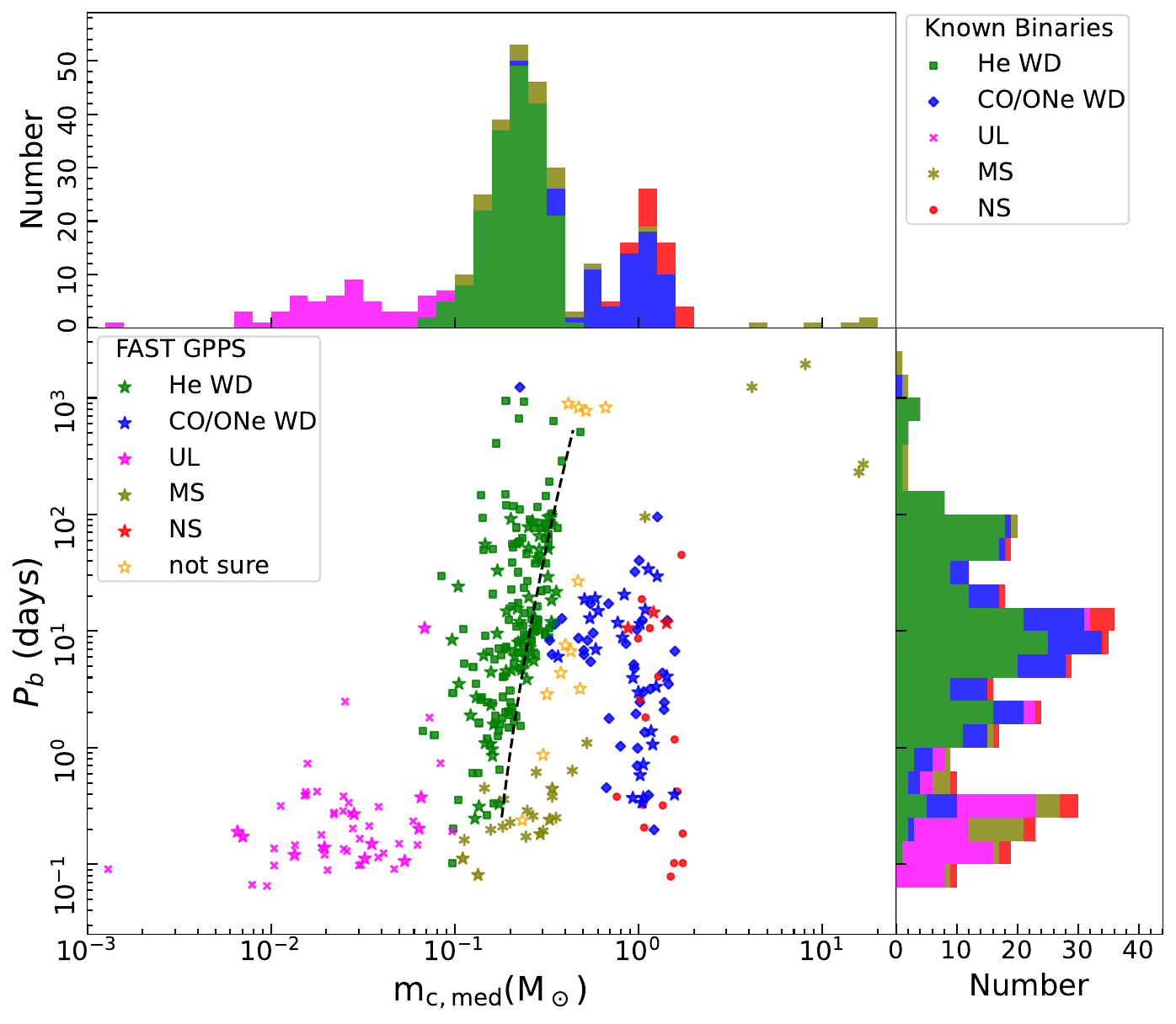} 
    \includegraphics[angle=0,width = 0.497\textwidth] {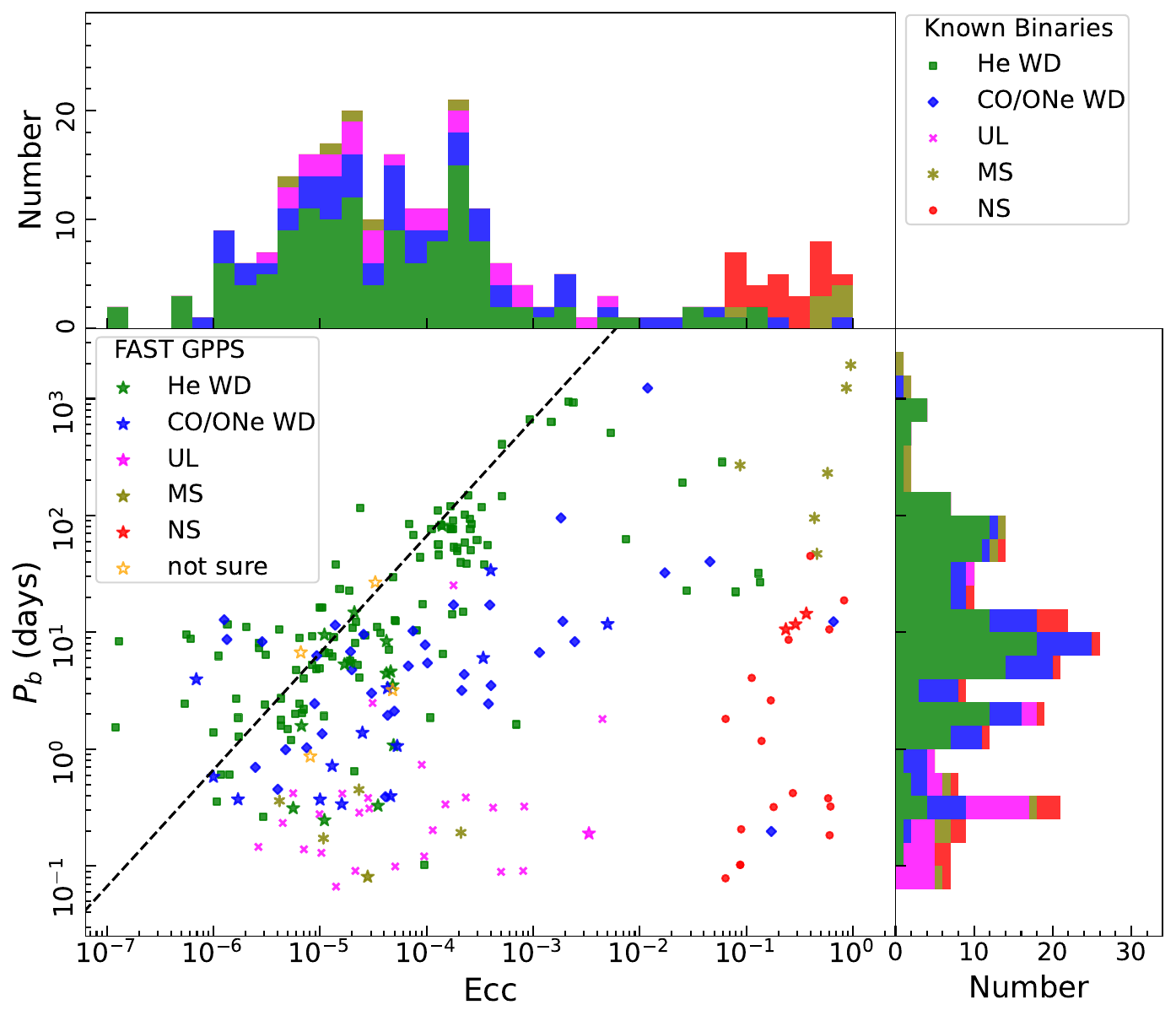} 
  \caption{Binary parameters are clustered for various companion types. Binary parameters of previously known pulsars are taken from the ATNF puslar catalogue     \citep[https://www.atnf.csiro.au/people/pulsar/psrcat/][]{mhth05}, plus the newly discovered binary pulsars by the FAST GPPS survey. In the main panel of the left plot, green squares, blue diamond, magenta crosses, olive asterisks and red dots represent the known He-WD, CO/ONe WD, ultra light (UL), main sequence star (MS) and   neutron star (NS) companions, respectively. The binary pulsars reported in this work are represented by stars with colors for various companion types as the known ones and the the orange for uncertain companion type. The dashed line is for neutron star - He-WD binaries predicted by \citet{ts99}. Histograms of the orbital period and companion mass are shown in the right and top panels for the sum of both the known and the new GPPS binaries with different types of companions. In the main panel of the right plot, same data point for the orbital period versus the orbit eccentricity. The dashed line represents the correlation between orbital period and eccentricity for MSP- He-WD systems as proposed by \citet{phi92}. Histograms of the orbital period and eccentricity are shown in the right and top panels for the sum of known pulsars and GPPS pulsars.
  }
  \label{fig:Pb_Mc_Ecc}
\end{figure*}

For the other 78 pulsars, we have not yet got the timing solution from the limited number of observations, but we get the currently best-fitted preliminary Keplerian parameters obtained from available FAST observations made by the FAST GPPS survey or follow-up tracking observations in applied projects. Their positions have not been well determined yet, so their temporal names have``g" at the end. 

The companions of binary pulsars are diverse (see Table~\ref{tab:Comp}). The ranges of spin period ($P$), orbital period ($P_b$), the estimated companion mass ($m_{\rm c,med}$) and orbital eccentricity ($e$) are listed in Table~\ref{tab:classify} for the binary pulsars with different types of companions in the Galactic field. Criteria for their classification is demonstrated in the last column. 
Companion type of a binary pulsar is determined based mainly on the measured $P$, $\dot{P}$, $P_{\rm b}$, $m_{\rm c,med}$ and $e$ for a given binary system, as well as its distribution in the $P_b$ v.s. $m_{\rm c,med}$ and $P_b$ v.s. $e$ diagrams, as shown in Figure~\ref{fig:Pb_Mc_Ecc}.  

See Table~\ref{tab:Comp} for the numbers of the GPPS binary pulsars with various companions. Their fractions are shown in Figure~\ref{fig:binary_pie}. It is apparent that the GPPS binaries have a significantly larger fraction of He-WDs and CO-WDs compared with the known population.

In the following we discuss the GPPS binary pulsars according to their probable companions.

\subsection{Pulsars with He-WD companions}

The pulsar He-WD binary systems are generally believed to be formed from the LMXBs via the Case A RLO which results in a fully recycled pulsar in an orbital period shorter than one day, or via the Case B RLO that results in a fully or partially recycled pulsar with an orbital period in the ranging from one to a thousand days \citep[e.g.][]{wrs83,ts99,prp02, tv23}. The pulsar He-WD binaries might also be evolved from the IMXBs when the Case A RLO is not initiated too late during the main-sequence evolution of the donor star \citep[e.g.][]{prp02}. Binary pulsars evolving in this channel are fully recycled and have orbital period between 3-20 days \citep{Tauris+2011ASPC}. 

\begin{figure}
  \centering
  \includegraphics[width = 0.47\textwidth] {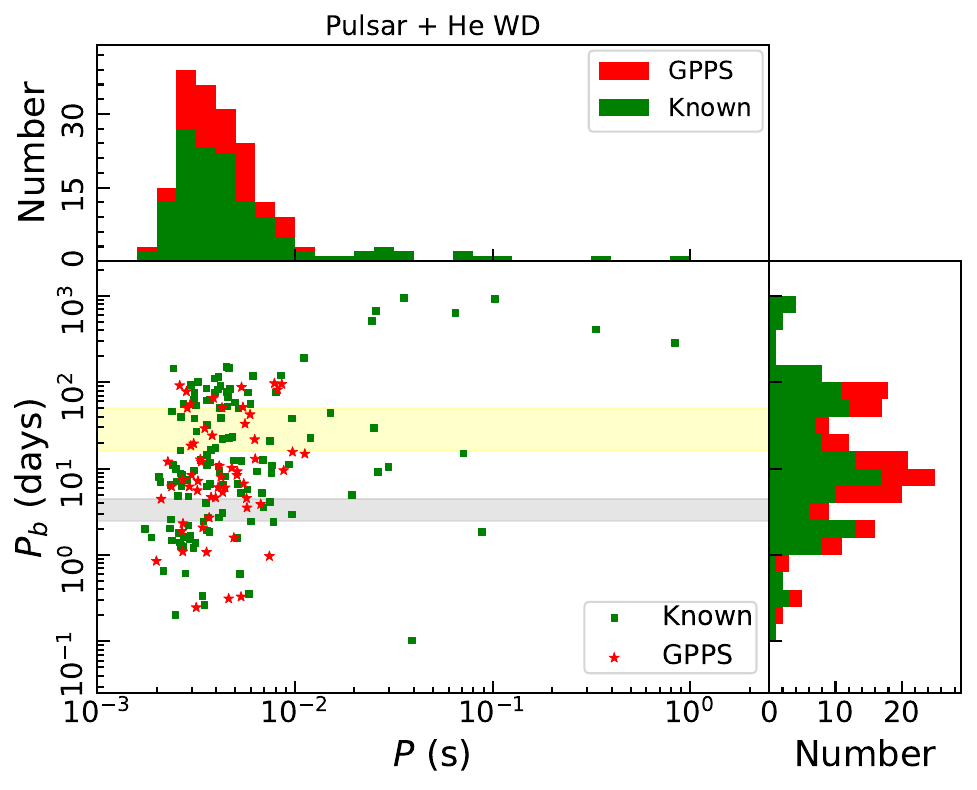} 
  \caption{The distribution of pulsar periods and orbit periods for binary pulsars with He-WD companions. In the main panel, green dots represent the 129 known binary pulsars with He-WD companions in the Galactic field, the stars stand for the 58 GPPS pulsars with He-WD companions. The histograms on the top and right show the distributions of pulsar spin period and orbital period.}
  \label{fig:PvsPb_He}
\end{figure}

Pulsars with He-WD companions are generally fully recycled and hence have spin periods $P<10$~ms with a lognormal distribution with the most probable spin period of about 3.5 ms, as shown in Figure~\ref{fig:PvsPb_He}. The orbital periods range from 0.1 day to several hundred days. 
According to the criteria listed in Table~\ref{tab:classify} and the one suggested by \citet{tlk12}, the companion of a pulsar with spin period less than 10 ms is most likely a He-WD if the median mass is in the range of $0.08 M_{\odot}<m_{\rm c, med}< 0.356 M_{\odot}$ for a system with an orbital period of $P_{\rm b} < 100$ days, or the median mass in the range of $0.08 M_{\odot}<m_{\rm c, med}< 0.5 M_{\odot}$ for a system with an orbital period of $P_{\rm b} > 100$ days. 

We get 58 GPPS binary pulsars that most likely have He-WD companions. We have obtained timing solution for 14 such pulsars, but not yet for another 44 pulsars (see Table~\ref{tab:PreKep}). Among the GPPS pulsars, 5 pulsars, PSRs J1840+0012, J1903+0839, J1912+1416, J1829$-$0235g and J1908$+$0705g, have orbital periods shorter than one day and exhibit no eclipsing as shown from observations. They are most likely formed via the Case A RLO from LMXBs. The other 53 MSPs have orbital periods longer than one day, and are most likely formed via the Case B RLO from LMXBs. 

During the evolution, the recycling process results in two fossil relations for the orbital parameters for MSPs if they have a low mass companion with an orbital period $P_{\rm b} \ge 1$ day. One is $P_{\rm b}$ versus $m_c$ \citep{rw71,ts99,itl14, ant14}, as shown in Figure~\ref{fig:Pb_Mc_Ecc}. 
The GPPS pulsars, e.g., PSRs J0622+0339, J1844+0028, J1904+0553, J1916+0740, J1917+0615, J1918+0621, J1930+1403 and J1947+2011, are around the relation, which is an indication for the companion type.

The other fossil relation is $P_{\rm b}$ versus $e$ for He-WD binaries \citep{phi92}, as shown in Figure~\ref{fig:Pb_Mc_Ecc}. Positive correlations between $P_{\rm b}$ and $e$ are evident for pulsars with He companions with correlation coefficient of 0.66. This correlation is related to tides resulting from density fluctuations in the convective envelope, and systems with wider orbits during the mass transfer prevent perfect circularization \citep{tv23}. The GPPS pulsars, e.g., J1857+0642, J1943+2206 and J1947+2011, are consistent with the relation. 

The distribution of orbital periods was noticed to have two gaps around $P_{\rm b}\sim 25-50$ days \citep[e.g.][]{tau96,tkr00} and around $P_{\rm b}\sim 2.5-4.5$ days \citep{hwh+18}. With the newly discovered GPPS binary pulsars, we see the peaks are enhanced, the gaps are confirmed with the lower boundary for the large gap extending from 25 to about 16 days, as shown in Figure~\ref{fig:PvsPb_He}.

\begin{figure}
  \centering
  \includegraphics[width = 0.47\textwidth] {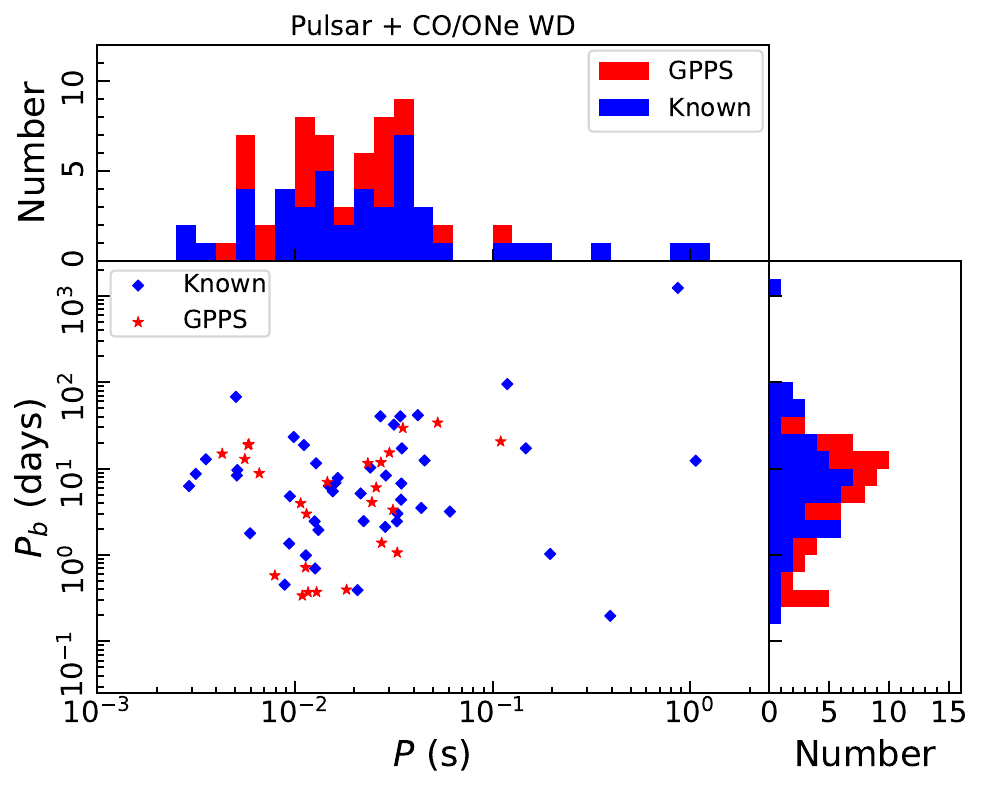} 
  \caption{Distribution of pulsar spin periods and orbit periods of binaries with CO/ONe WD companions. There are 39 known systems, plus 25 GPPS binary pulsars with a CO/ONe WD companion.}
  \label{fig:PvsPb_CO}
\end{figure}

\subsection{Pulsars with CO/ONe-WD companions}

A pulsar CO/ONe WD binary system can be formed from an IMXB via the Case A RLO in small orbits that results in a fully recycled pulsar with an  orbital period of 3-20 days \citep{prp02, tlk11}, or via an early Case B RLO in a wider orbit that results in a partial recycled pulsar with an orbital period of 3-50 days\citep{prp02}, or via the late Case B or Case C RLO and a common envelope in very wide orbits that result in a partial recycled pulsar with CO/ONe companion with an orbital period less than 20 days \citep{ijc+13, tv23}. In addition, there are also systems that might be formed from LMXBs via the late Case B RLO \citep{ts99}, which leads to a slowly spinning pulsar in an extremely wide orbit with orbital periods $\gtrsim800$ days. The NS in such a system is only mildly recycled. Another possibility is that in a HMXB the star with a higher initial mass evolves into a massive WD instead of an NS due to mass ratio reversal
\citep{Kaspi+2000ApJ...543..321K}. The NSs in these systems are non-recycled and their orbit are eccentric.

The companion of a pulsar is likely a CO/ONe WD when the pulsar has an orbital eccentricity $e<0.05$, an orbital period $P_{\rm b}<100$ days, together with the companion having the median mass $m_{\rm c, med}>0.356 M_{\odot}$, as set in Table~\ref{tab:classify} and suggested by \citet{tlk12}. 

\begin{figure}
  \centering
  \includegraphics[angle=0,width = 0.45\textwidth] {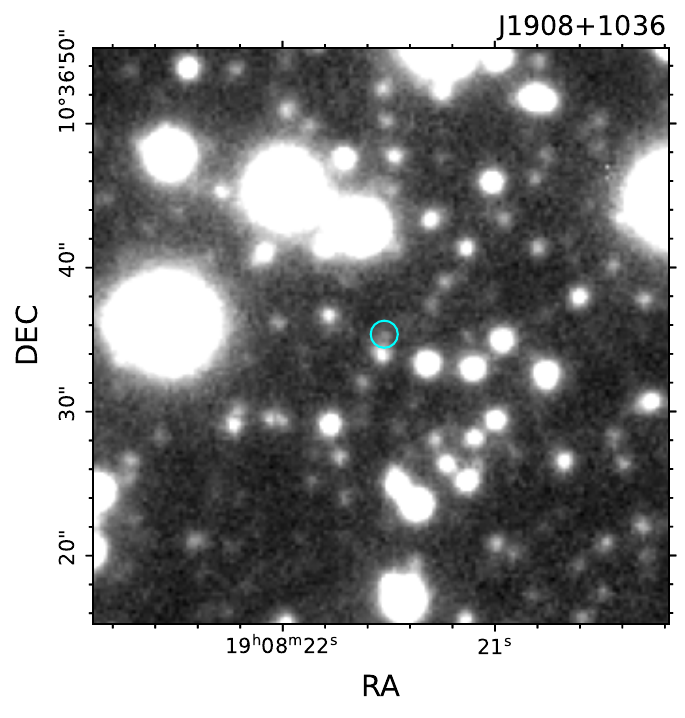}
  \caption{Optical images around the companion of PSR J1908+1036 from Pan-STRARRS1 \citep{Chambers+2016arXiv161205560C}. The cyan circle is centered at the pulsar with a radius of one arcsecond.}
  \label{fig:optical}
\end{figure}

We have 25 binary pulsars discovered in the GPPS survey which probably have CO/ONe WD companions, as listed in Table~\ref{tab:PreKep}. Timing solutions have been obtained for 13 of them, with 6 ones reported in \citet{yhw+25}. Figure~\ref{fig:PvsPb_CO} shows the distribution of these pulsars in $P$ vs $P_{\rm b}$ diagram. It is apparent that the binary pulsars with CO companions have a broad range of spin periods from as low as 2.9 ms to about 1.1 s, and have a most probable orbital period of about 10 days.

Among these systems, PSR J1908+1036 has the smallest eccentricity among all the pulsar CO-WD binary systems, though with large measurement uncertainty. The pulsar has a small DM of 10.91 $\rm pc~ cm^{-3}$, which indicates that it is a nearby system. We do find the optical counterpart of the companion simply by inspecting manually of the images from the Pan-STRARRS1 image cutout server \footnote{http://ps1images.stsci.edu/cgi-bin/ps1cutouts} with stack image at $grizy_{P1}$ bands, as shown in Figure~\ref{fig:optical}. The detailed evolution history needs to be further investigated.

\subsection{Pulsars with ultra-light companions}

Some pulsars have an UL companion with a median mass $m_{\rm c, med}<0.08 M_{\odot}$ \citep{mhth05, tlk12}. The UL companions are generally formed via the Case A RLO of LMXBs \citep{tv23}. Magnetic braking takes away their orbital momentum leading to compact orbits. Some of them, such as PSR J1953+1844 \citep{yhj+23}, undergo evolution from ultra-compact X-ray binaries (UCXBs), then to accreting X-ray millisecond pulsars (AXMSPs), and finally to binary millisecond pulsars with compact orbit together with UL companions \citep{prp02,vnv+12,Guo+2024MNRAS}.

\begin{figure}
  \centering
  \includegraphics[bb=42 9 420 735, clip, width=0.48\columnwidth] {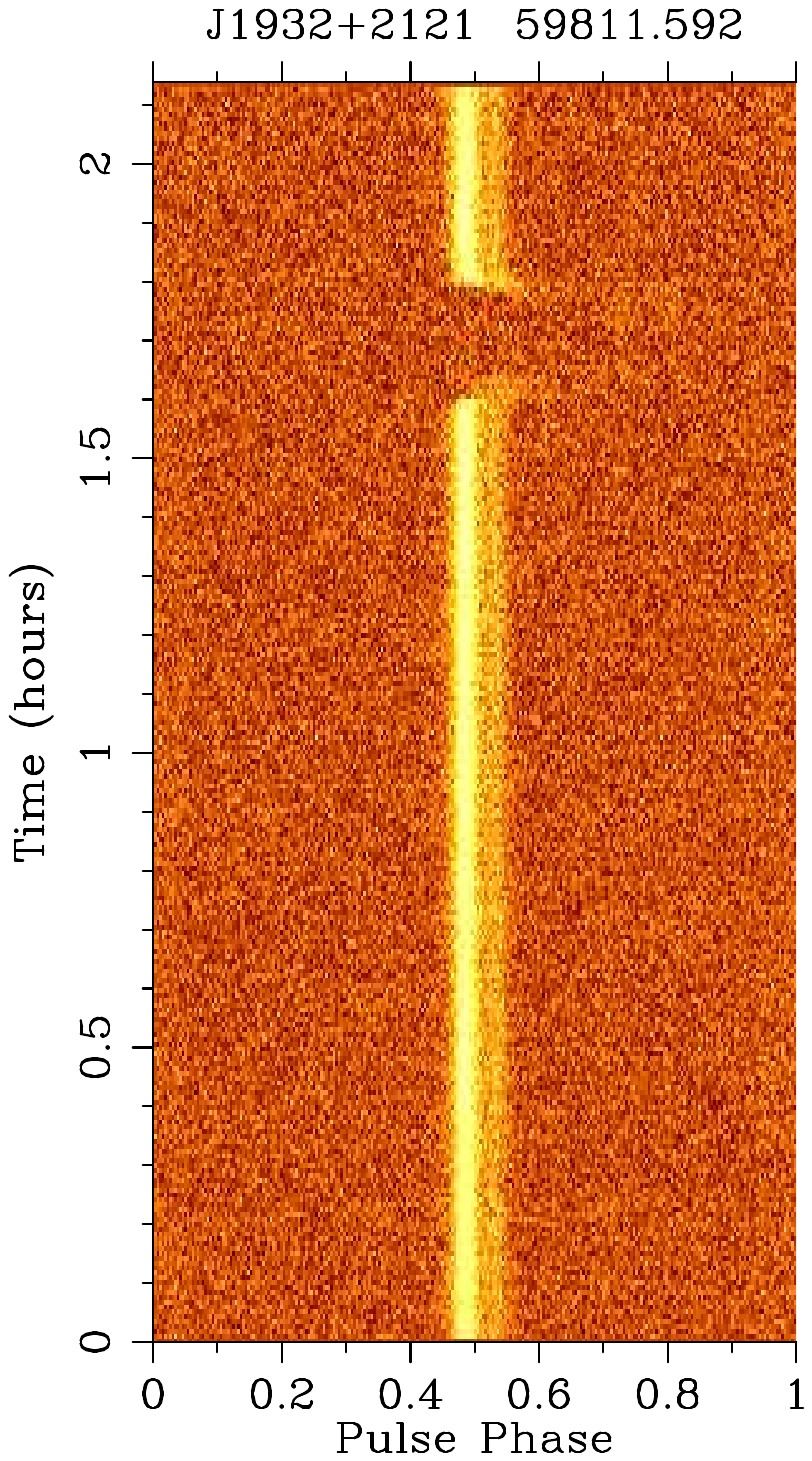}
  \includegraphics[bb=42 9 420 735, clip, width=0.48\columnwidth] {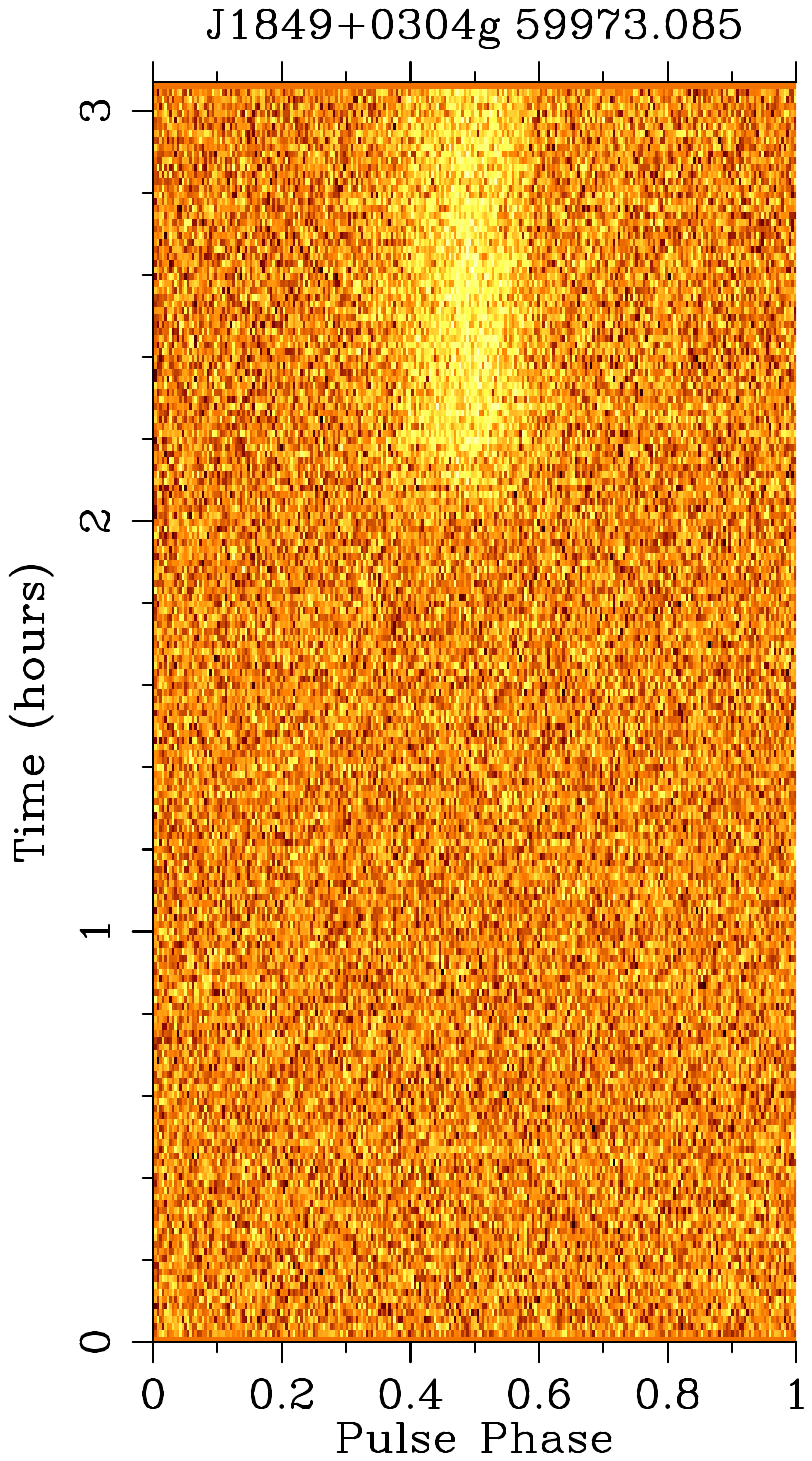}\\
  \includegraphics[bb=42 9 420 355, clip, width=0.48\columnwidth] {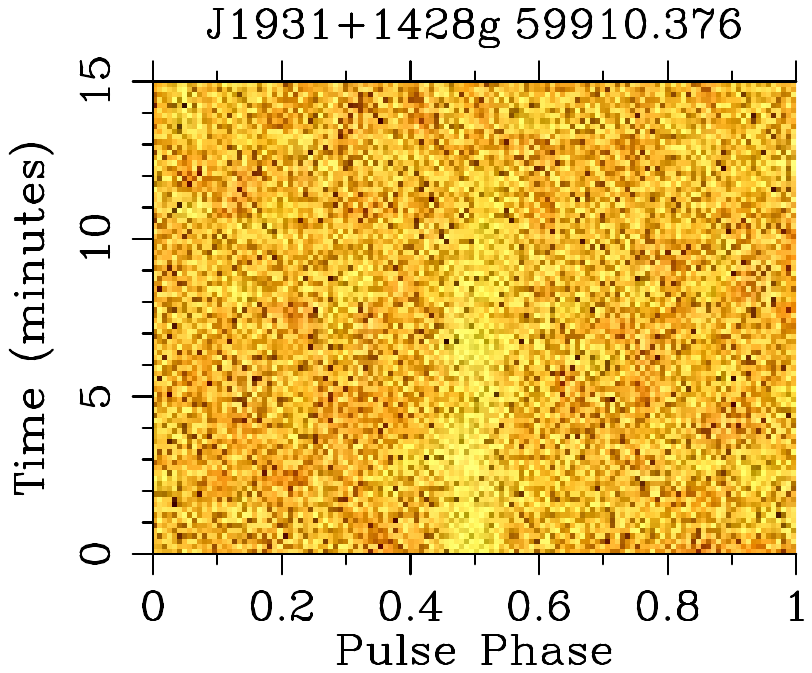}
  \includegraphics[bb=42 9 420 355, clip, width=0.48\columnwidth] {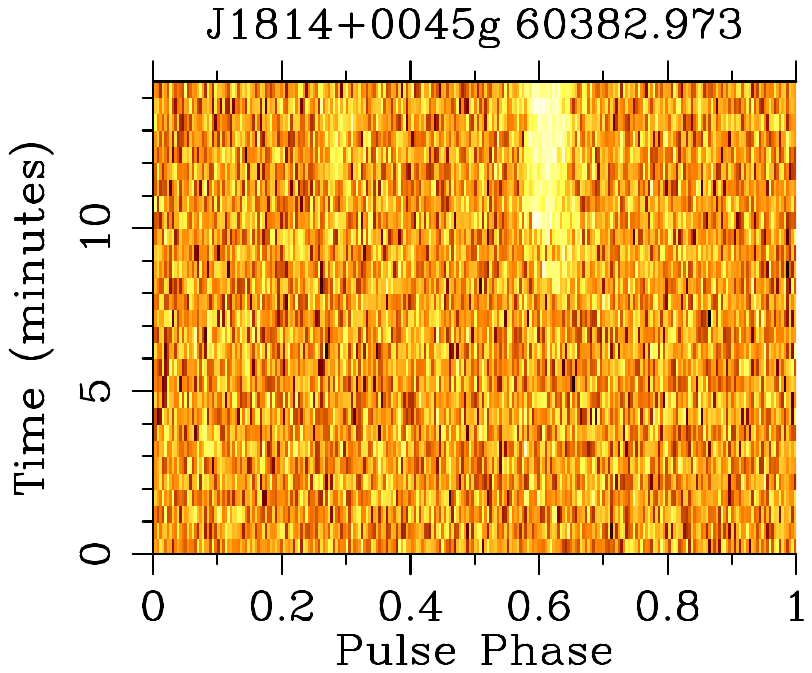}\\  
  \caption{Eclipses of four binary pulsars. PSRs J1932+2121 and J1849+0304g are observed for 2.1 and 3.1 hours, PSRs J1931+1428g and J1814+0045g are observed for 15 minutes each.}
  \label{fig:eclipse}
\end{figure}

12 GPPS discovered pulsars, PSRs J0541+2959g, J1814+0045g, J1830$-$0106g, J1838+1507g, J1847$+$0342g, J1856$+$1000g, J1911+1206g, J1919+0126g, J1953+1006g, J2003+3032g, J1845+0317 and J1953+1844, likely have UL companions. These pulsars are fully recycled. PSRs J1814+0045g and J1953+1006g are black widows exhibiting eclipsing, whose companions have median masses of 0.064 and 0.013 $M_\odot$. Eclipse around the egress of PSR J1814+0045g is detected in a short observation, as shown in Figure~\ref{fig:eclipse}. While the signals are fully eclipsed for PSR J1953+1006g during observations. Others are black widow candidates. Companions of PSRs J1845+0317 and J1911+1206g are the lightest and are most likely planets.

\subsection{Pulsars with main sequence star companions}

A small number of pulsars have MS star companions with the median mass $m_{\rm c, med} > 0.5 M_{\odot}$ and orbital period $P_{\rm b} > 50$ days \citep{tlk12}. Some redback systems with $P_{\rm b} \lesssim 1$ day and $m_{\rm c}\sim0.1-0.4 $ M$_\odot$ might also have low-mass MS companions that experienced irradiation-induced mass loss.

We have 5 GPPS binary pulsars which likely have MS companions, PSRs J1849+0304g, J1859+0313g, J1919+1502g, J1931+1428g and J1932+2121. They are redbacks, and have a companion with a median mass from $0.110 M_\odot$ to $0.339 M_\odot$ in compact orbits and exhibit eclipses. The eclipses of PSRs J1932+2121, J1849+0304g and J1931+1428g are shown in Figure~\ref{fig:eclipse}, as revealed by FAST observations. 
The FAST observation of PSR J1932+2121 exhibits a full eclipse lasting for about 10 minutes that is about 8.6\% of the orbital phase. During the ingress and egress, pulsar emission is gradually delayed due to the extra DM contributed by the eclipsing material. The egress of eclipse has been also observed for PSR J1849+0304g. The eclipse around ingress is detected in some short observations of PSR J1931+1428g. Eclipses of PSRs J1859+0313g and J1919+1502g result in non-detection of pulsar emission in several observations. 

The eclipses can be observed at multiple wavelengths and are desired for careful investigation to understand the out-flowing material from the companion stars \citep[e.g.][]{pod91}, the companion properties and the binary orbit \citep[e.g.][]{myc18,llm19, dyc+23}, which is hard to include in this paper.

\subsection{Pulsars with NS companions}

The double neutron star binary systems are generally formed from HMXBs through the Case BB RLO following a CE phase, which leads to mild or marginal recycling for the first-born NSs \citep{tkf+17}. As summarized by \citet{shy+24}, there have been about 28 double neutron star systems, 23 DNS in the Galactic field (including 5 DNS candidates: PSRs J1753$-$2240, J1755$-$2550, J1759+5036, J1906+0746, and J2150+3427) with an orbital period in the range from 0.078 to 50 days and an eccentricity in the range from 0.064 to 0.828 \citep{tlk12}. 

 \begin{figure}
   \centering
   \includegraphics[width = 0.46\textwidth] {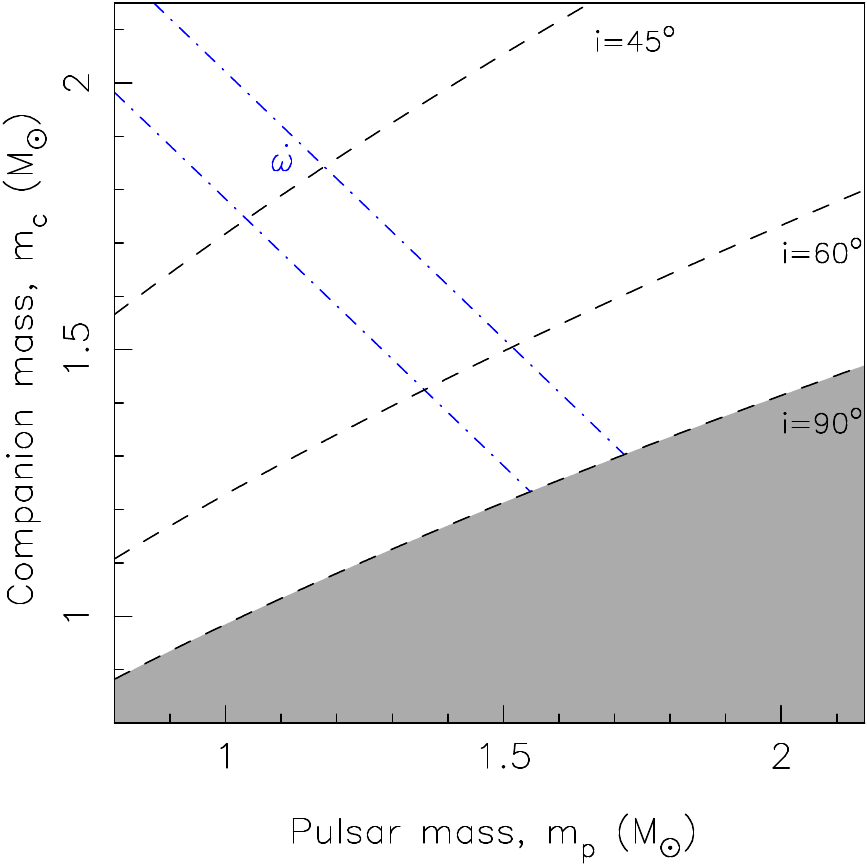}\\
   \caption{Mass-mass diagram of the double neutron star system PSR J0528+3529. The gray area represents the excluded parameter spaces from its mass function, with the boundary defined by an inclination angle of $i=90^\circ$. The area between two dash-doted blue lines is the possible parameter space, as constrained by the measured $\dot{\omega}$ within $\pm1\sigma$.
   }
   \label{fig:m1m2}
\end{figure}

In the discovery of the FAST GPPS survey, three binary pulsars, PSRs J0528+3529, J1844-0128 and J1901+0658, are most likely in DNS systems. PSR J1901+0658 is the first GPPS discovered pulsar, and was reported in \citet{shy+24}. 

PSR J0528+3529 has a spin period of 78.2 ms and a period derivative of $7.36\times10^{-19} $s~s$^{-1}$. It is in an elliptic orbit with a period of 11.73 days and an eccentricity of 0.29. The ephmeris is presented in Table~\ref{tab:ephem}. Its companion is estimated to have a median mass of 1.42 $M_\odot$. The relatively large companion mass together with the large eccentricity indicates that the companion is most likely a neutron star. Our timing observations demonstrate that its periastron advances with $\dot{\omega}$ of 0.0072(3) $\rm deg/yr$ (see Table~\ref{tab:ephem}). According to general relativity, the rate of advance of the periastron is described by \citep{bt76, dd85},
\begin{equation}
  \dot{\omega}=  3T_\odot^{2/3}\left(\frac{P_{\rm b}}{2\pi}\right)^{-5/3}\frac{1}{1-e^2} (m_p+m_c)^{2/3}.
  \label{eq:omegadot}
\end{equation}
With which, total mass of the system is estimated to be 2.90(12) M$_\odot$. The possible mass spaces for both the pulsar and its neutron star companion is shown in Figure~\ref{fig:m1m2}. 

PSR J1844-0128 has a spin period of 29.1 ms, and is in an eccentric orbit with a period of 10.6 days and an eccentricity of 0.235. Its companion is estimated to have a median mass of 0.876 $M_\odot$, as listed in Table~\ref{tab:PreKep}. Its periastron advance is marginally detect with $\dot{\omega}$ of 0.0059(18) $\rm deg/yr$, which indicates a total mass of 1.7(8) M$_\odot$.

\subsection{Pulsars with a helium MS companion}

PSR J1928+1815 is in an compact circular orbit with a companion with a  median mass of $1.395 M_\odot$ and shows the eclipse near the conjunction phase. \citet{yhz+25} suggested the companion to be a He main-sequence star. 

\subsection{Pulsars with undetermined companions}

The companion nature of a binary system discussed above is inferred based on previous knowledge. However there are too many evolution channels, each with very uncertain parameters. There are some ambiguous parameter space between the NS He-WD binary and the NS CO-WD binary, or between the NS He-WD binary and the NS-MS binary, or between the NS CO-WD binary and the DNS binary. Such an ambiguity is mainly caused by the unknown orbital inclination angle and the lack of companion observations at other wave bands. Among newly discovered binary pulsars by the FAST GPPS survey, 11 pulsars may have a companion of either a He-WD or CO-WD, and one pulsar has a companion of either He-WD or MS star.   

PSRs J1838+0028g, J1842+0407g, J1857+0642, J1908+0949, J1917+1259 and J1943+2206 are fully recycled millisecond pulsars in orbits with periods of 3-27 days. The median masses of companion are in the range of 0.32 to 0.48 $M_\odot$. Such systems are typical descendants from the IMXB Case A evolution channel, and their companions are either CO-WDs or He-WDs \citep{Tauris+2011ASPC}.  

PSRs J1900+0213g, J1910+0423g, J1917+1046g and J1923+2022g are typical descendants of LMXBs but in orbit with periods as long as almost 900 days. It is then difficult to assess the companion to be a He-WD or CO-WD according to \cite{ts99}. PSR J1952+2837 has a companion mass large and beyond that predicted by the $P_{\rm b}$ versus $m_{\rm c}$ relationship for the typical He-WDs. Moreover, unlike typical descendants of LMXBs, PSR J1952+2837 is mildly recycled with a spin period of 18 ms which probably evolve from an IMXB. The median companion mass is 0.30 M$_\odot$, which cannot be a CO-WD unless its orbital inclination is sufficiently low. But if the companion is a He-WD, the formation channel for such a system is unclear.

PSRs J1857-0125g has an orbital period of 0.24 days and a median companion mass of $m_{\rm c, med}=0.23 \rm M_{\odot}$. Such a low-mass companion might be a He-WD or an MS exhibiting as redback systems with feedback from the neutron star \citep{ccth13,ssc+19}. Further observations are desired to determine the companion nature by uncovering the possible pulsar eclipses and orbital period variations.

\begin{table*}
  \centering
  \caption{Kinematic corrections of $\dot{P}$ and relavant parameters for 4 pulsars with proper motion measurement.}
  \label{tab:Pdot}
  \small
  \tabcolsep 1.2mm
  \renewcommand{\arraystretch}{0.7}
\begin{tabular}{crrrrrrrr}
\hline\hline
Pulsar & $\mu_T$       & $\dot{P}$     & $\dot{P_S}$  & $\dot{P_G}$ & $\dot{P_I}$ & $B_{\rm surf,c}$ & $\tau_c$ & $\dot{E_c}$ \\
       &($\rm mas~yr^{-1}$)&($10^{-21}s~s^{-1}$) & ($10^{-21}s~s^{-1}$) & ($10^{-21}s~s^{-1}$)& ($10^{-21}s~s^{-1}$) & ($10^8$Gs)    & (Gyr)    & ($10^{33} \rm erg~s^{-1}$) \\
(1)    & (2)           &(3)            & (4)          & (5)        & (6)         & (7)           & (8)      & (9)         \\
\hline
J1857+0642 & 5.9(9)   &   4.69 & 0.30  &   0.03 &   4.36 &  1.26 & 12.80 & 3.91  \\
J1903+0839 & 5.0(9)   &   4.67 & 1.49  &  -1.78 &   4.96 &  1.53 & 14.74 & 1.98  \\
J1904+0553 & 8.3(13)  &  12.75 & 3.65  &  -0.94 &  10.04 &  2.25 &  7.72 & 3.36  \\
J1908+1036 &12.0(22)  &  15.73 & 2.51  &  -0.03 &  13.25 &  3.81 & 12.75 & 0.43  \\
\hline

\end{tabular}
Notes: $\mu_T$: total proper motion, $\dot{P_S}$: Shklovskii effect,
$\dot{P_G}$: Galactic acceleration, $\dot{P_I}$: intrinsic period derivative, 
$B_{\rm surf,c}$: corrected surface magnetic field, $\tau_c$: corrected age,
$\dot{E_c}$: corrected energy loss rate.
\end{table*}


\begin{figure}
  \centering
  \includegraphics[angle=0,width = 0.95\columnwidth] {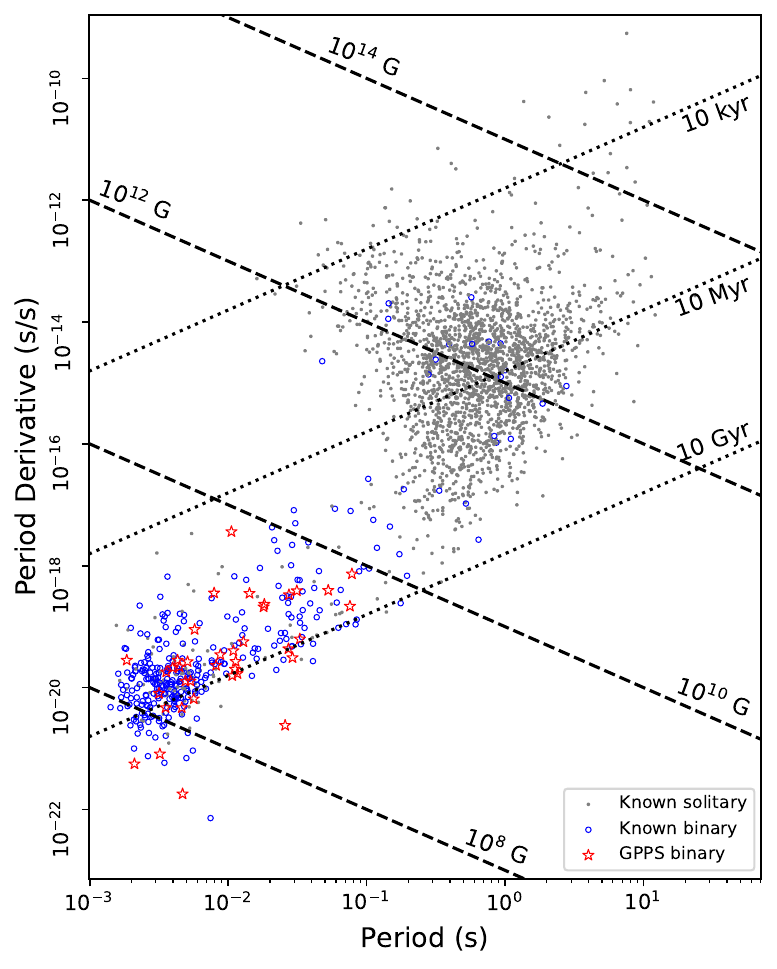}
  \caption{Binary pulsars in the period versus period derivative diagram. Gray dots represent soliday  single pulsars listed in the ATNF pulsar catalogue \citep{mhth05}. Binary pulsars are indicated by  circles. The newly discovered 38 binary pulsars by the GPPS survey which have timing solutions already are indicated by red stars. Dashed and doted lines represent constant surface magnetic field and characteristic age, respectively.}
  \label{fig:P_Pdot}
\end{figure}
\section{Conclusions and discussion}

In this work, we report 116 binary pulsars discovered by the FAST GPPS survey. We have got the timing solutions of 38 binary pulsars measured, as indicated in the distribution in the $P-\dot{P}$ diagram in Figure~\ref{fig:P_Pdot}. In addition to the timing solutions for 9 pulsars published previously, we present the measured and derived parameters for 29 binary pulsars with phase-coherent timing solutions. For the other 78 pulsars, we have got the best-fitted Kepelerian parameters of the binary orbits based on several FAST observations. 
Most of the binary systems need to be further timed in the future, so that precise spin and orbital parameters of the systems with preliminary Keplerian parameters can be obtained. The PK parameters of the relativistic systems can be measured and improved. For example, the possible Shapiro delay in tight systems requires long term monitoring with good cadence and observation sensitivity. 

With currently available measurements, we see that the pulsar binary systems we discovered in the GPPS survey have a broad range of spin periods, orbital periods and eccentricity. Among 116 pulsars, 104 of them have spin period $P<30$ ms; 30 GPPS binaries have orbital periods shorter than one day and and two of them shorter than 0.1 day. The shortest is 53 minutes for the black widow pulsar PSR J1953+1844 \citep{plj+23}. The observed $\dot{P}$ of PSR J1943+2206 is the smallest among all the binary pulsars reported in this work. The inferred surface magnetic field strength is only $2.95\times10^{7}$~G. 

Most of these binary pulsars have nearly circular orbits with $e < 0.001$, with He- or CO/ONe-WD companions in general. We noticed that three pulsars, PSRs J0528+3529, J1844$-$0128 and J1901+0658 have orbital eccentricities $e > 0.1$. Because of the large companion masses, they are very likely double NS systems, or NS-WD systems in which the NS formed after the massive WD, or a NS-MS system with a newly born NS. For evolved DNS or NS-WD systems, the large eccentricity indicates that the second formed object is most likely a compact NS if a significant eccentricity remains from the second supernova explosion \citep{tv23}. 

Among these binaries discovered by the FAST GPPS survey, 59 pulsars rotate with He-WD companions in the orbit, 24 pulsars with a CO/ONe-WD companion, 11 pulsars with an UL companion, 5 with a MS companion, 3 pulsars with a NS companion and one with a He star as companion. We have 12 pulsars with companion of unknown natures. Combing these GPPS binaries with those known ones, we get the largest binary sample in the Galactic field. The He-WD, CO/ONe WD, UL, MS and NS systems have fractions of 43.5\%, 14.9\%, 12.3\%, 6.3\% and 4.7\%, in total.

For these pulsars with timing solutions, we obtained polarization pulse profiles by combing multiple FAST observations. The Geometry parameters have been derived from the polarization position angle curves as fitted by the RVM model. Optical counterpart is found for the companion of PSR J1908+1036 from the Pan- STARRS survey. The non-detection reveals that He-WD companions are generally cool and old. Post-Keplerian parameter $\dot{\omega}$ has been measured for the double neutron star systems PSRs J0528+3529 and J1844-0128, from which the total mass of the systems are estimated to be 2.90(12)$M_{\odot}$ and 1.7(8)$M_{\odot}$. 

For very nearby pulsars, it should be noted that the observed pulsar spin down rate comprises its intrinsic term $\dot{P_{\rm I}}$ plus the contributions from the Shklovskii effect $\dot{P_{\rm S}}$ and the Galactic acceleration $\dot{P_{\rm G}}$, in the form of 
\begin{equation}
  \dot{P}=\dot{P_{\rm I}}+\dot{P_{\rm S}}+ \dot{P_{\rm G}}.
  \label{eq:pdot}
\end{equation}
The Shklovskii effect results from the increase of projected distance between pulsar and solar system barycenter when a pulsar moves, which leads to the increase of pulse period \citep{bh86}. The Shklovskii term, $\dot{P_{S}}$, as indicated in column (4) in Table~\ref{tab:Pdot}, contributes 32\% of the measured $\dot{P}$ for PSR J1903+0839 as an example. In a differential Galactic potential, the acceleration between a pulsar and the solar system barycenter, i.e., the Galactic acceleration term, can be found from \citet{nt95}. We list the influences of gravitational acceleration $\dot{P_{G}}$ listed in column (5) of Table~\ref{tab:Pdot}. They are generally negative except for PSR J1857+0642. Its influence on period derivatives is more than 38\% for PSR J1903+0839 as an example. Both the Shklovskii effect and the Galactic acceleration term can not be neglected in calculating the intrinsic $\dot{P}_{I}$. With $\dot{P}_{I}$, the surface magnetic field, characteristic age and spin-down energy loss rate are re-estimated, as listed in columns (7)-(9) of Table~\ref{tab:Pdot}.

\begin{acknowledgements}
This work made use of the data from FAST (Five-hundred-meter Aperture Spherical radio Telescope)(https://cstr.cn/31116.02.FAST). FAST is a Chinese national mega-science facility, operated by National Astronomical Observatories, Chinese Academy of Sciences. P. F. Wang is supported by the National SKA program of China (No. 2020SKA0120200), the National Natural Science Foundation of China (No. 12133004), the Chinese Academy of Science (No. JZHKYPT-2021-06) and the National Key R\&D Program of China (No. 2021YFA1600401 and 2021YFA1600400). J. L. Han is supported by the National Natural Science Foundation of China (No. 11988101 and 11833009).
\end{acknowledgements}

\appendix

\restartappendixnumbering

\renewcommand\thefigure{A\arabic{figure}}
\renewcommand\thetable{A\arabic{table}}

\section*{Tables and Figures for a large sample of binary pulsars}

In the main text of this paper, we give only one example of timing solution. Here we present the timing solutions of 29 binary pulsars. Their ephemerides are listed in Table~\ref{tab:ephem_appendix}. 
All ephemerides obtained here are based on the DE440 solar system ephemeris model, the Barycentric Dynamical Time (TDB) units, and TT(TAI) clock.  
The ephemeris items include 
standard pulsar name defined by accurate position, 
the temperate name and the GPPS discovery number, 
the MJD range and the span for FAST observation data 
and number of TOAs. 
Measured quantities include: 
right ascension (RA) in hh:mm:ss.ss, 
declination (DEC) in +/-dd:mm:ss.s,  
dispersion measure (DM) in cm$^{-3}$~pc, 
pulsar rotation frequency $\nu$ in Hz,  
first derivative of pulsar rotation frequency $\dot{\nu}$ in 10$^{-16}$~Hz per second, 
proper motion in right ascension $\mu_{\alpha} \cos \delta$ in milli-arc-second (mas) per year (if measured), 
proper motion in declination $\mu_{\delta}$ in in milli-arc-second (mas) per year (if measured),
the timing residule in microsecond ($\mu$s); 
timing residual scaling factor EFAC; 
timing residual quadratic adding factor EQUAD;
Reduced $\chi^2$ of model-fitting.
Binary parameters include orbital period $P_{\rm b}$ in days, 
projected semi-major axis of pulsar's orbit $x$ in light-second, 
the time of passing through periastron $T_0$  (if measured), 
the longitude of periastron $\omega$  (if measured), 
time of ascending node $T_{\rm asc}$ in MJD, 
first and second Laplace parameters $e_1$ and $e_2$. 
(3) Derived quantities include: 
the Galactic longitude $l$ in degree,  
the Galactic latitude $b$ in degree, 
distance estimates $D_{\rm YMW}$ and $D_{\rm NE2001}$ by using the Galactic electron density distribution models YMW17 \citep{ymw17} and NE2001 \citep{cl02} in kpc, 
the spin period of a pulsar $P$ in millisecond, 
the derivative of the spin period $\dot{P}$ in $second per second$,
characteristic age $\tau$ in Gyr, 
surface magnetic field strengt $B_{\rm surf}$ in $10^8$G, 
and the orbital eccentricity $e$ (if measured).

\begin{figure*}
  \centering
  \includegraphics[bb =  5 40 965 165, clip, width=0.95\textwidth] {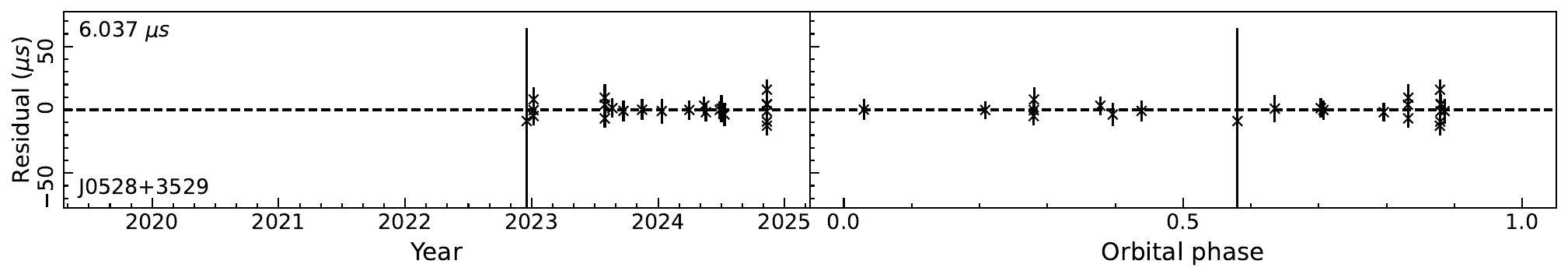} 
  \includegraphics[bb =  5 40 965 165, clip, width=0.95\textwidth] {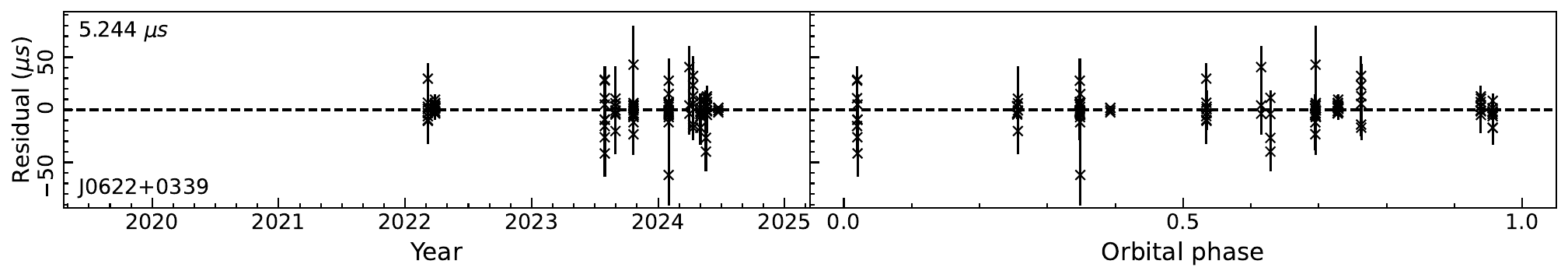} 
  \includegraphics[bb =  5 40 965 165, clip, width=0.95\textwidth] {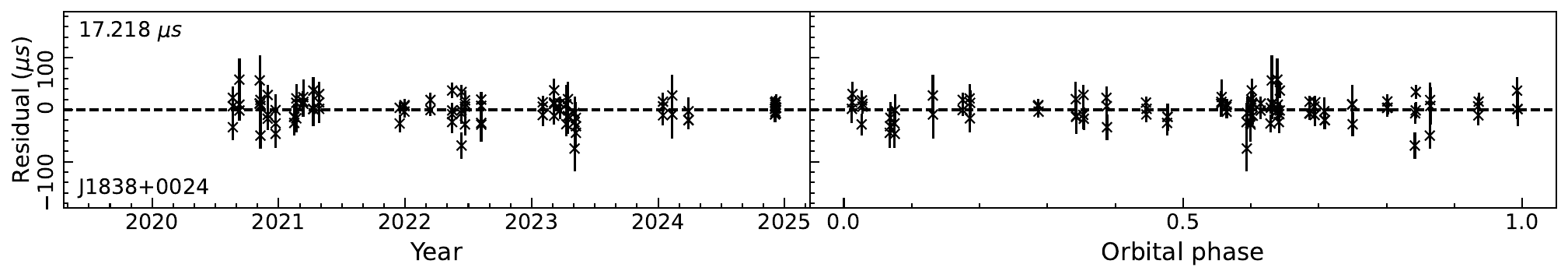}  
  \includegraphics[bb =  5 40 965 165, clip, width=0.95\textwidth] {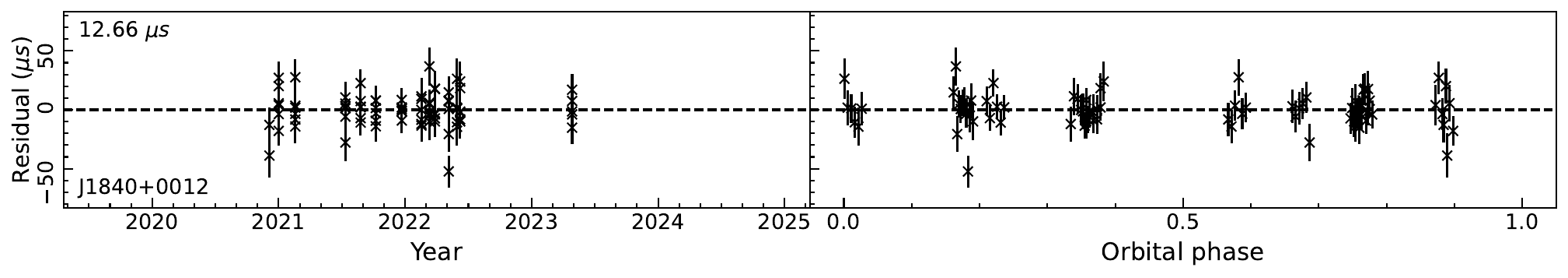}
  \includegraphics[bb =  5 40 965 165, clip, width=0.95\textwidth] {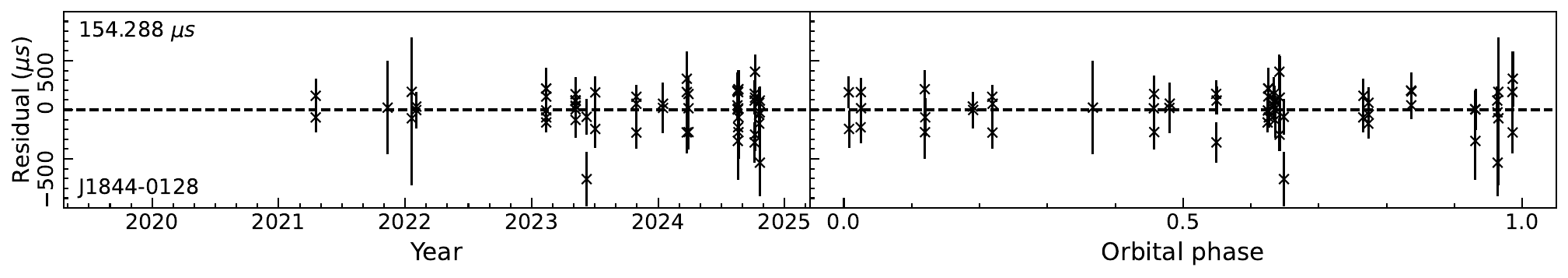}  
  \includegraphics[bb =  5 40 965 165, clip, width=0.95\textwidth] {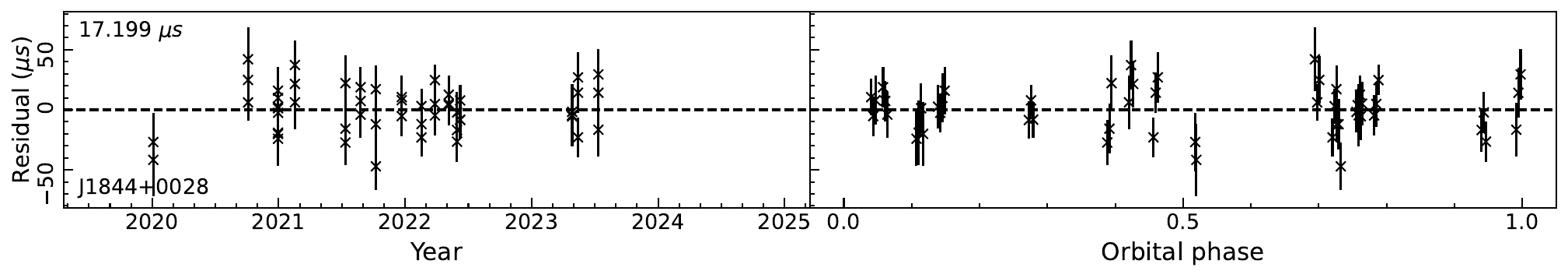}
  \includegraphics[bb =  5 40 965 165, clip, width=0.95\textwidth] {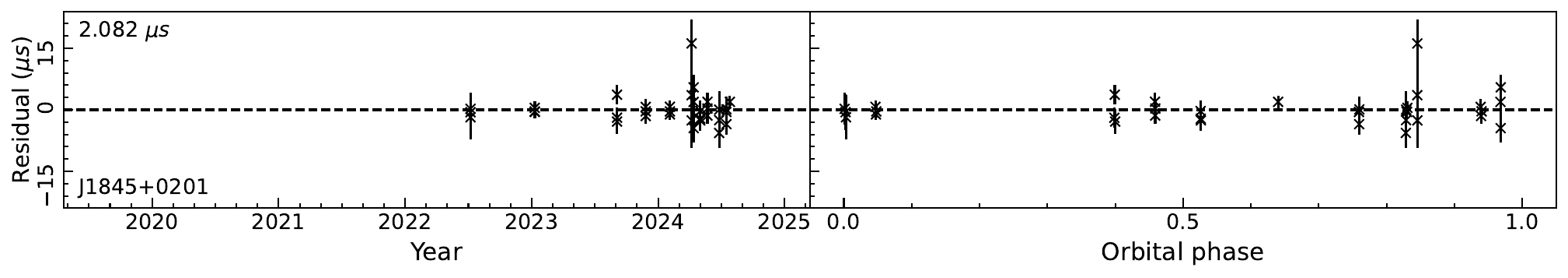}
  \includegraphics[bb =  5 40 965 165, clip, width=0.95\textwidth] {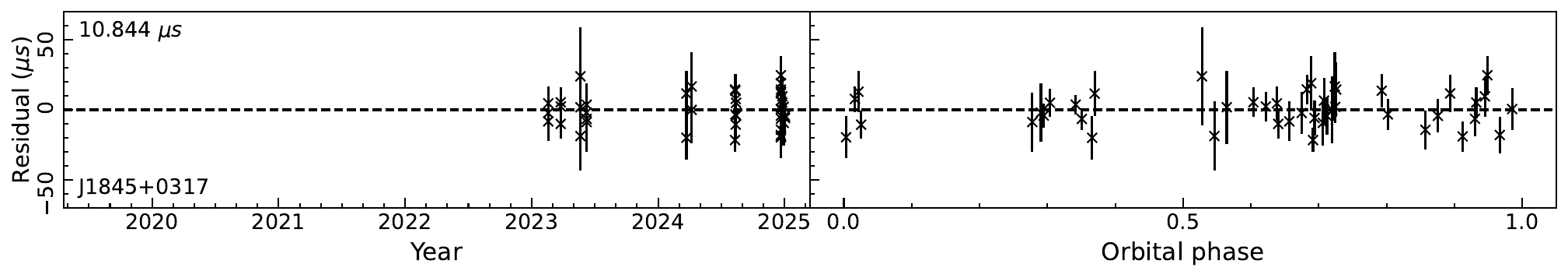}  
  \includegraphics[bb =  5 5 965 165, clip, width=0.95\textwidth] {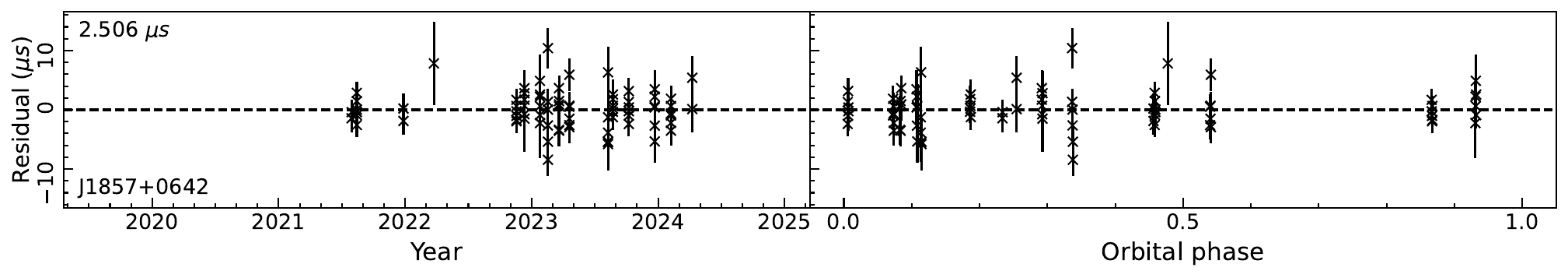}
  \caption{Timing residuals of 29 newly discovered binary pulsars by the FAST GPPS survey. {\it Left panels:} Residuals versus observation epochs. The weighted root-mean-square residual of each pulsar is indicated in the top right corner of the panel. {\it Right panels:} Residuals versus orbital phase. The orbital phases are referred to ascending node or periastron depending on the binary model of each pulsar.}
  \label{fig:Res_appendix}
\end{figure*}

Timing residuals of 29 binary pulsars are shown in Figure~\ref{fig:Res_appendix}. The left panels are for the residuals along the observation epochs, and the right panels for residuals versus the orbital phase.

For these pulsars with timing solution, we can add all FAST polarization measurements and get their polarization profiles, as shown in Figure~\ref{fig:profs_appendix}. We tried to fit the polarization angle curves with the rotating vector model \citep{rc69} for 5 pulsars, PSRs J1840+0012, J1857+0642, J1916+0740, J1930+1403 and J1946+0904.

For 78 binary pulsars without timing solutions, the measurements of barycentric periods are plotted across the orbit phase in Figure~\ref{fig:Pbary_appendix}, togethet with the currently best-fitting preliminary Keplerian model. These data are the base for further follow-up observations.

\newpage

\begin{table*}
  \centering
  \caption{Phase-coherent timing solutions of 29 binary pulsars.}
  \label{tab:ephem_appendix}
 \footnotesize
  \tabcolsep 1.0mm
  \renewcommand{\arraystretch}{0.7}
\noindent
\begin{tabular}{lccccc}
\hline\hline
Pulsar name\dotfill      & J0528+3529         & J0622+0339          & J1838+0024         & J1840+0012          & J1844-0128           \\
\hline
GPPS number \dotfill    & gpps0537            & gpps0388            & gpps0107           & gpps0146            & gpps0555            \\
MJD range     \dotfill  & 59930 -- 60625      & 59645 -- 60484      & 59082 -- 60651     & 59188 -- 60062      & 59321 -- 60604      \\
Data Span (yr) \dotfill & 1.9                 & 2.3                 & 4.3                & 2.4                 & 3.5                 \\
Number of TOAs \dotfill & 23                  & 106                 & 97                 & 78                  & 59                   \\
Ref. epoch (MJD)\dotfill& 60000               & 60000               & 59500              & 59500               & 60000              \\
\hline
\multicolumn{6}{c}{Measured quantities} \\
\hline
RA (hh:mm:ss)\dotfill   & 05:28:28.7978(1)    & 06:22:19.4862(2)    & 18:38:23.3856(2)   & 18:40:49.2066(1)    & 18:44:20.995(2)    \\
\hline
\end{tabular}
\end{table*}

\addtocounter{table}{-1}
\begin{table*}
  \centering
 \caption{\it -- continued --}
 \footnotesize
  \tabcolsep 1.0mm
  \renewcommand{\arraystretch}{0.7}
\begin{tabular}{lccccc}
\hline\hline
Pulsar name\dotfill      & J1844+0028       & J1845+0201      & J1845+0317       & J1857+0642          & J1903+0839           \\
\hline
GPPS number \dotfill     & gpps0109         & gpps0547        & gpps0566        & gpps0236            & gpps0100             \\
MJD range      \dotfill  & 58852 -- 60138   & 59768 -- 60518  & 59993 -- 60677  & 59425 -- 60409      & 58936 -- 60402      \\
Data Span (yr) \dotfill  & 3.5              &  2.1            & 1.9             & 2.7                 & 4.0                 \\
Number of TOAs \dotfill  & 52               & 34              & 44              & 79                  & 144                 \\
Ref. epoch (MJD)\dotfill & 59500            & 60000           & 60300           & 59500               & 59500               \\
\hline
\multicolumn{6}{c}{Measured quantities} \\
\hline
RA (hh:mm:ss)\dotfill    & 18:44:37.3595(2)    & 18:45:03.07708(4)  & 18:45:22.8858(3)   & 18:57:58.6789(2)    & 19:03:51.84526(5)    \\
\hline
\end{tabular}
\end{table*}

\addtocounter{table}{-1}

\begin{table*}
  \centering
  \caption{\it -- continued --}
 \footnotesize
  \tabcolsep 1.0mm
  \renewcommand{\arraystretch}{0.7}

\begin{tabular}{lccccc}
\hline\hline

Pulsar name\dotfill      & J1904+0553        & J1905+0649        & J1908+1036      & J1911+1253         & J1912+1416           \\
\hline
GPPS number \dotfill     & gpps0039          & gpps0229          & gpps0114       & gpps0181           & gpps0169             \\
MJD range      \dotfill  & 58898 -- 59753    & 58897-- 60237     & 58989 -- 60348 & 59271 -- 60348     & 59256 -- 60231       \\
Data Span (yr) \dotfill  & 2.4               & 3.7               & 3.7            & 3.0                & 2.7                  \\
Number of TOAs \dotfill  & 88                & 37                & 131            & 90                 & 44                   \\
Ref. epoch (MJD)\dotfill & 59500             & 59500             & 59500          & 59500              & 59500                \\

\hline
\multicolumn{6}{c}{Measured quantities} \\
\hline

RA (hh:mm:ss)\dotfill    & 19:04:16.82283(4)   & 19:05:05.2578(13)  & 19:08:21.5212(2)  & 19:11:28.6990(2)   & 19:12:30.51373(5)    \\
\hline

\end{tabular}
\end{table*}

\addtocounter{table}{-1}

\begin{table*}
  \centering
 \caption{\it -- continued --}
 \footnotesize
  \tabcolsep 1.0mm
  \renewcommand{\arraystretch}{0.7}

\begin{tabular}{lccccc}
\hline\hline

Pulsar name\dotfill      & J1916+0740      & J1917+0615        & J1917+1259       & J1918+0621          & J1924+1342          \\
\hline
GPPS number \dotfill     & gpps0166        & gpps0460          & gpps0012         & gpps0494            & gpps0032            \\
MJD range      \dotfill  & 59263 -- 60403  & 58713 -- 60453    & 58770 -- 59985   & 58713 -- 60344      & 58989 -- 60518      \\
Data Span (yr) \dotfill  & 3.1             & 4.8               & 3.3              & 4.5                 & 4.2                 \\
Number of TOAs \dotfill  & 118             & 39                & 79               & 94                  & 35                  \\
Ref. epoch (MJD)\dotfill & 59500           & 59500             & 59500            & 59500               & 59500                \\

\hline
\multicolumn{6}{c}{Measured quantities} \\
\hline

RA (hh:mm:ss)\dotfill    & 19:16:15.4327(7)   & 19:17:20.2887(2)  & 19:17:21.3371(2)  & 19:18:00.6061(13)   & 19:24:16.5972(4)   \\
\hline

\end{tabular}
\end{table*}

\addtocounter{table}{-1}

\begin{table*}
  \centering
  \caption{\it -- continued --}
 \footnotesize
  \tabcolsep 1.0mm
  \renewcommand{\arraystretch}{0.7}

\begin{tabular}{lccccc}
\hline\hline
Pulsar name\dotfill      & J1930+1403       & J1932+2121      & J1936+2035       & J1938+2302          & J1943+2206          \\
\hline
GPPS number \dotfill     & gpps0013         & gpps0403        & gpps0197         & gpps0392            & gpps0514            \\
MJD range      \dotfill  & 58808-60408      & 59661 -- 60254  & 58901 -- 60257   & 59654 -- 60392      & 59380 -- 60540      \\
Data Span (yr) \dotfill  & 4.3              & 1.6             & 3.7              & 2.0                 & 3.2                 \\
Number of TOAs \dotfill  & 90               & 91              & 57               & 277                 & 166                 \\
Ref. epoch (MJD)\dotfill & 59500            & 59500           & 59500            & 60000               & 59500               \\
\hline
\multicolumn{6}{c}{Measured quantities} \\
\hline
RA (hh:mm:ss)\dotfill   & 19:30:17.62052(5)  & 19:32:21.2014(2)  & 19:36:38.3072(7)    & 19:38:11.6306(4)     & 19:43:43.6485(3)     \\
\hline
\end{tabular}
\end{table*}

\addtocounter{table}{-1}
\begin{table*}
  \centering
 \caption{\it -- end --}
 \footnotesize
  \tabcolsep 1.0mm
  \renewcommand{\arraystretch}{0.7}
%
\begin{tabular}{lccccc}
\hline\hline

Pulsar name\dotfill      &   & J1946+0904      & J1947+2011      & J1952+2837       & J2018+3518                   \\
\hline
GPPS number \dotfill     &   & gpps0242        & gpps0011        & gpps0064         & gpps0393    \\
MJD range      \dotfill  &   & 59413 -- 60545  & 58749 -- 60165  & 58852 -- 60624   & 59633 -- 60045                                    \\
Data Span (yr) \dotfill  &   & 3.1             & 3.9             & 4.9              & 1.1                                            \\
Number of TOAs \dotfill  &   & 35              & 125             & 126              & 83                                             \\
Ref. epoch (MJD)\dotfill &   & 60000           & 59500           & 59500           & 60000                                          \\
\hline
\multicolumn{6}{c}{Measured quantities} \\
\hline
RA (hh:mm:ss)\dotfill   &   & 19:46:56.8707(2)  & 19:47:47.71374(9)  & 19:52:49.65344(5)  &  20:18:48.149(7)         \\
\hline
\end{tabular}

\end{table*}


\addtocounter{figure}{-1}
\begin{figure*}
  \centering
  \includegraphics[bb =  5 40 965 165, clip, width=0.95\textwidth] {pngs/J1903+0839_Residual.pdf}
  \includegraphics[bb =  5 40 965 165, clip, width=0.95\textwidth] {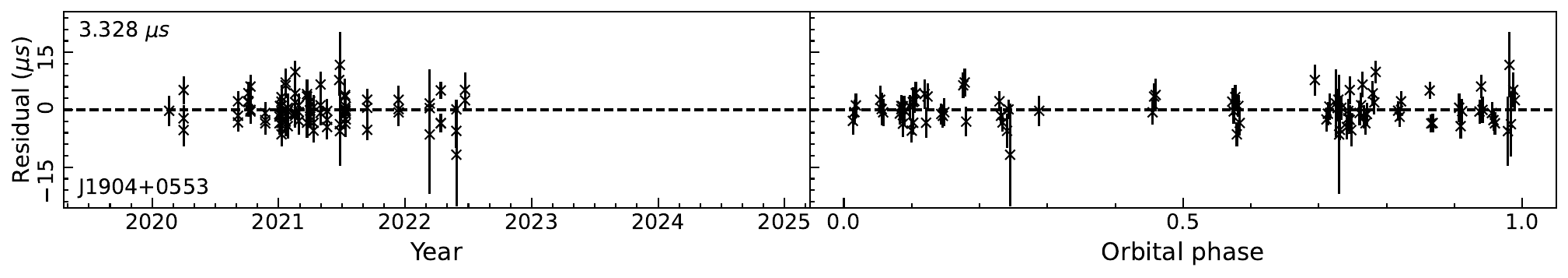}
  \includegraphics[bb =  5 40 965 165, clip, width=0.95\textwidth] {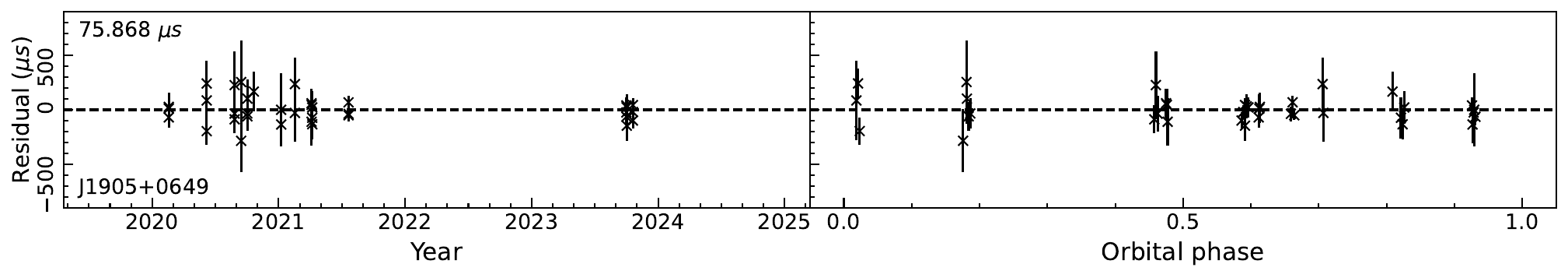}
  \includegraphics[bb =  5 40 965 165, clip, width=0.95\textwidth] {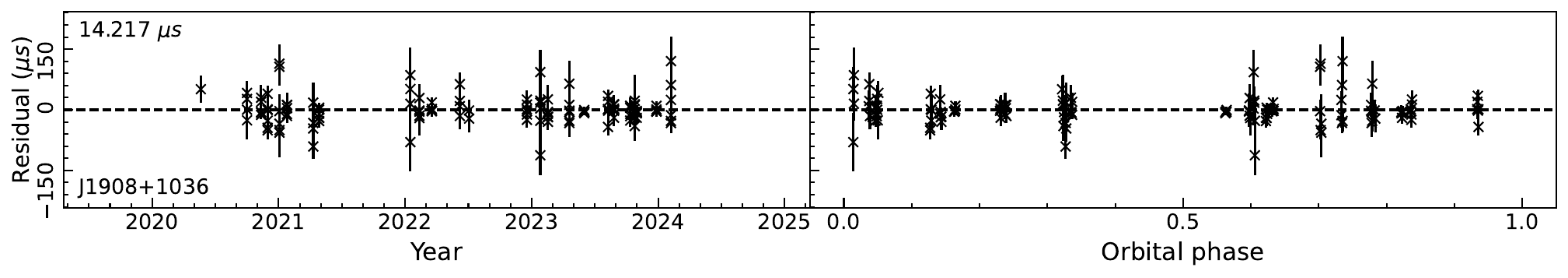}
  \includegraphics[bb =  5 40 965 165, clip, width=0.95\textwidth] {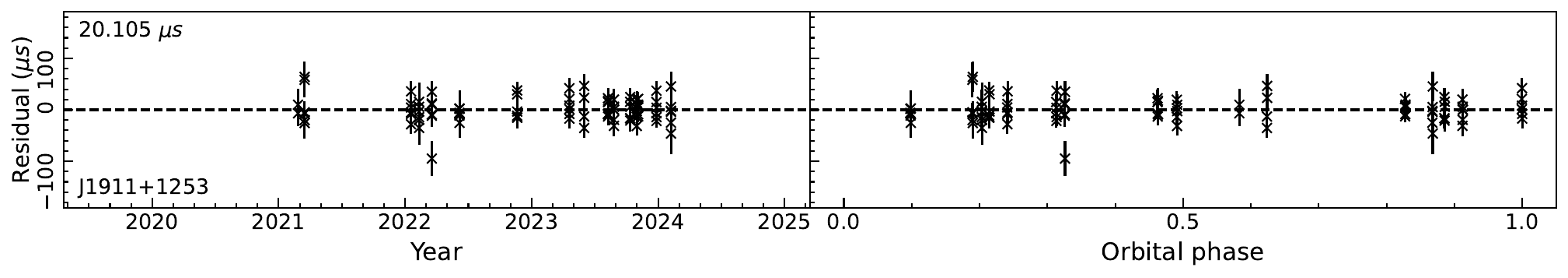} 
  \includegraphics[bb =  5 40 965 165, clip, width=0.95\textwidth] {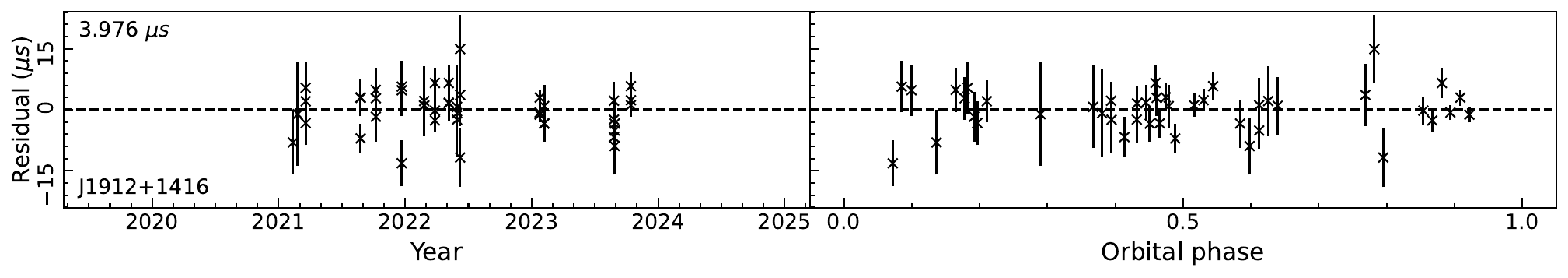}
  \includegraphics[bb =  5 40 965 165, clip, width=0.95\textwidth] {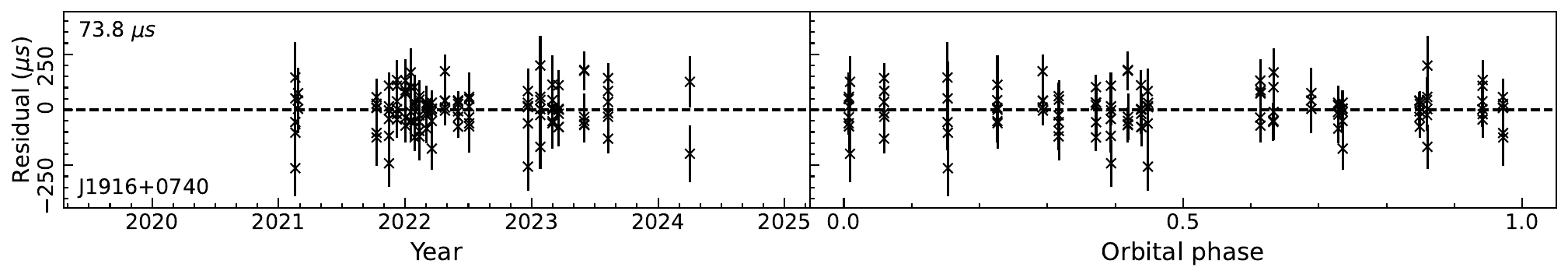}
  \includegraphics[bb =  5 40 965 165, clip, width=0.95\textwidth] {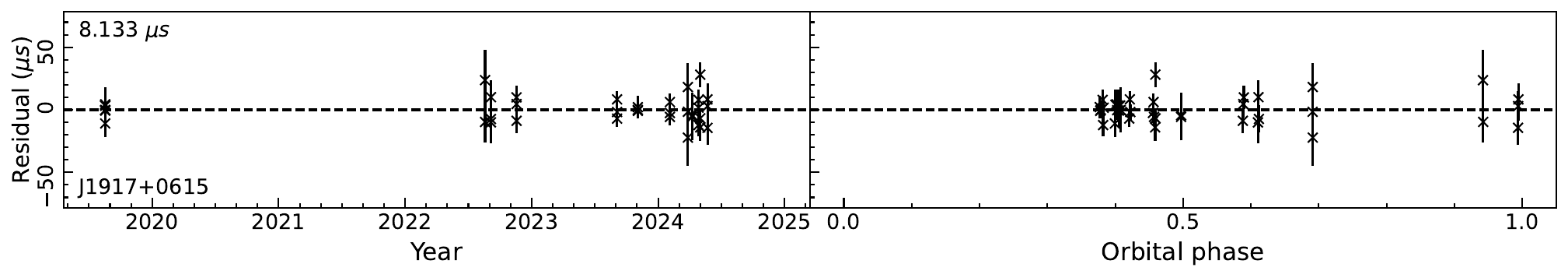}
  \includegraphics[bb =  5 40 965 165, clip, width=0.95\textwidth] {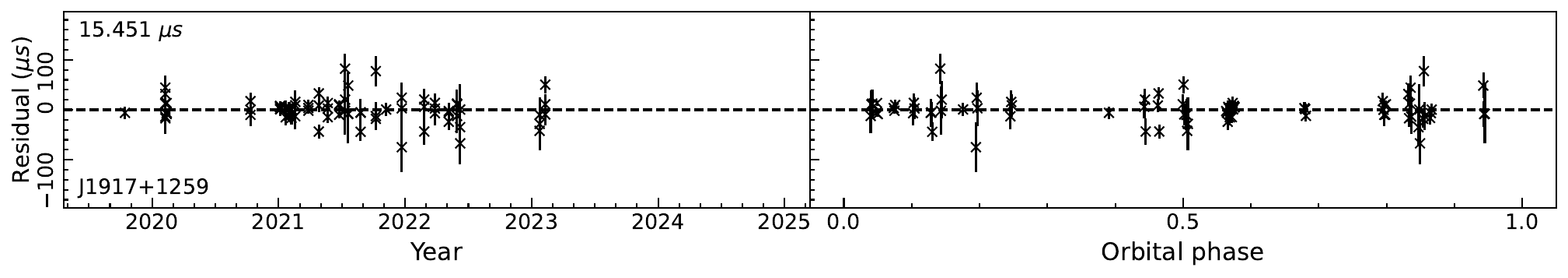}
  \includegraphics[bb =  5 5 965 165, clip, width=0.95\textwidth] {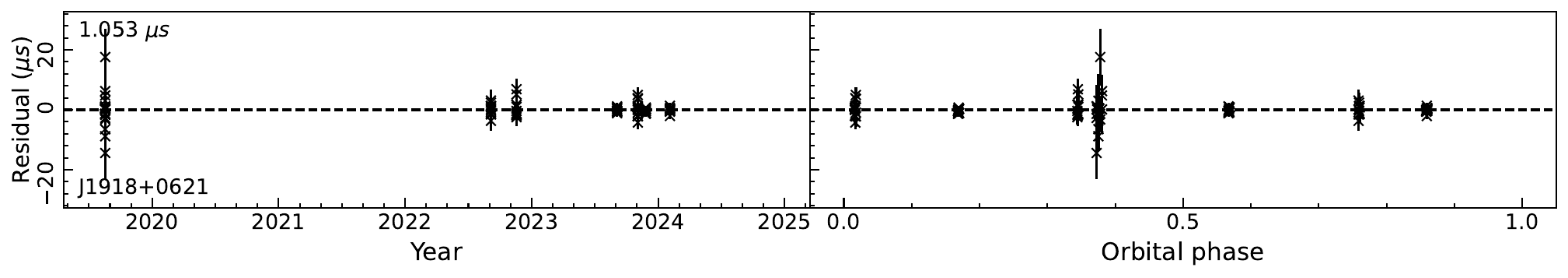}

  \caption{Continue---}
\end{figure*}
\addtocounter{figure}{-1}
\begin{figure*}
  \centering
  \includegraphics[bb =  5 40 965 165, clip, width=0.95\textwidth] {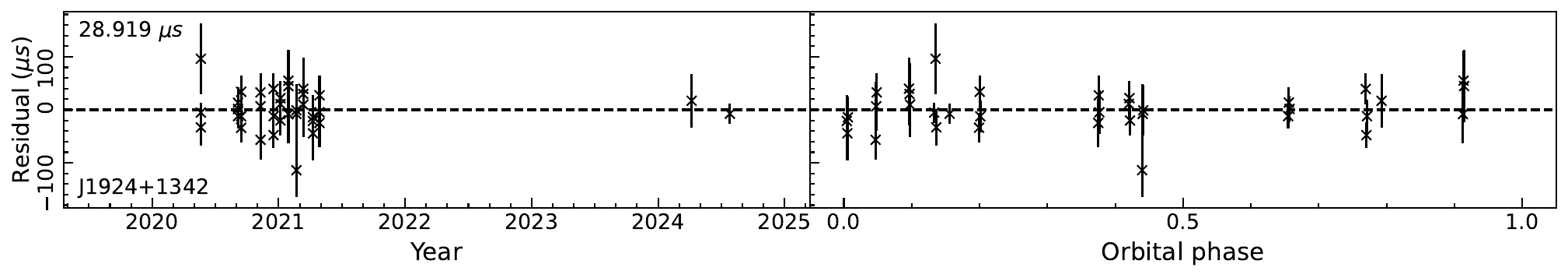}
  \includegraphics[bb =  5 40 965 165, clip, width=0.95\textwidth] {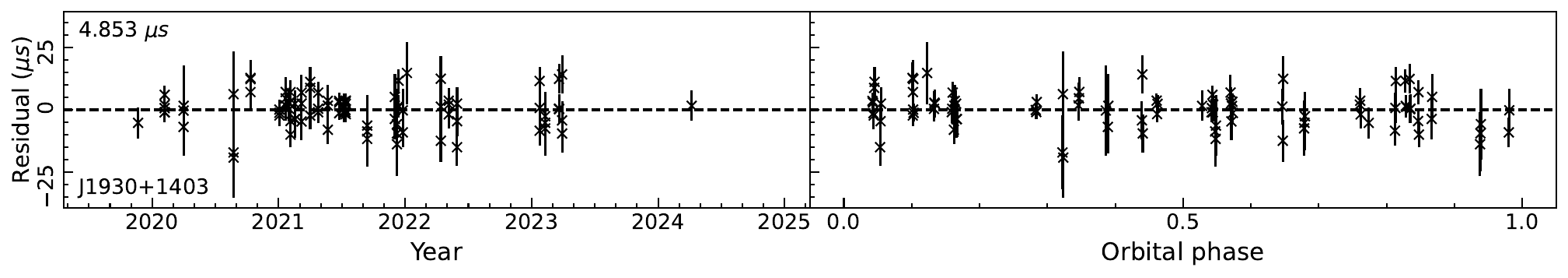} 
  \includegraphics[bb =  5 40 965 165, clip, width=0.95\textwidth] {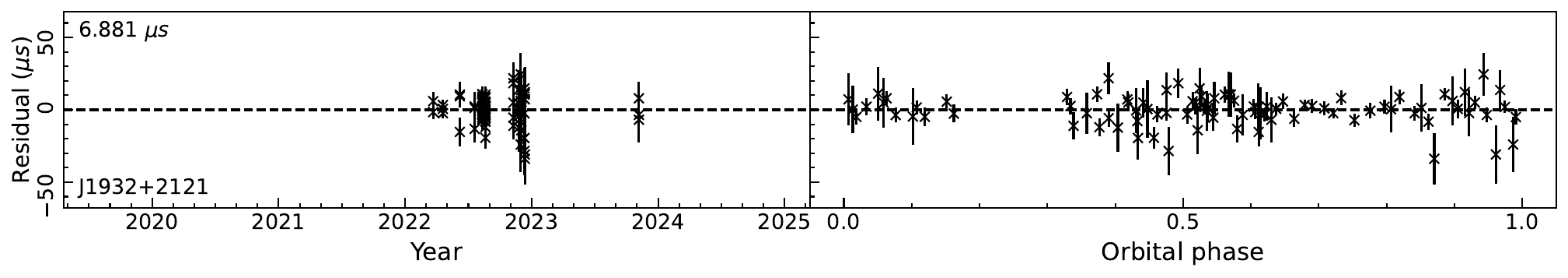}
  \includegraphics[bb =  5 40 965 165, clip, width=0.95\textwidth] {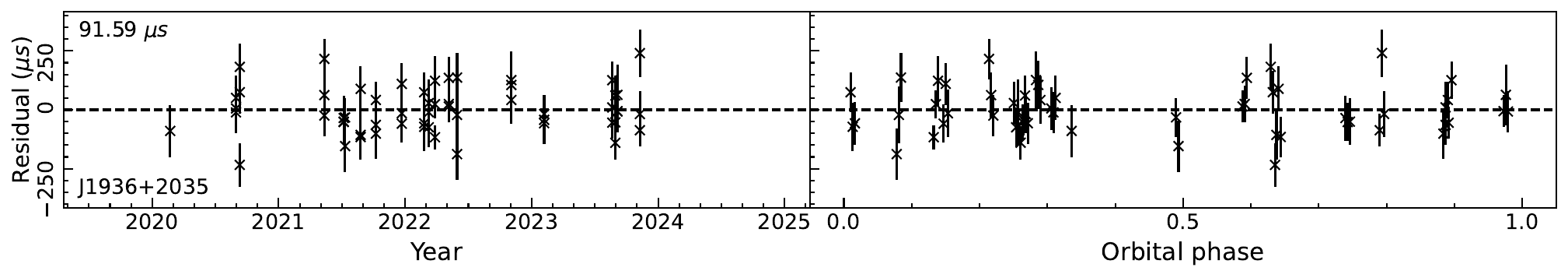}
  \includegraphics[bb =  5 40 965 165, clip, width=0.95\textwidth] {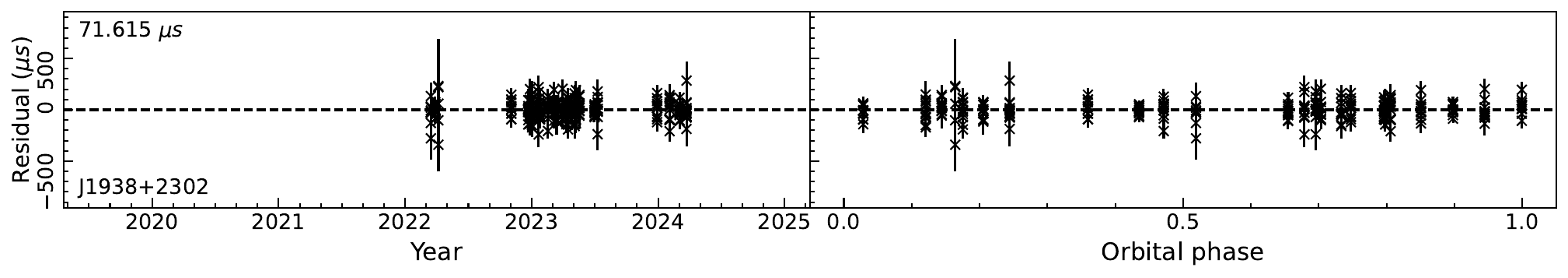}
  \includegraphics[bb =  5 40 965 165, clip, width=0.95\textwidth] {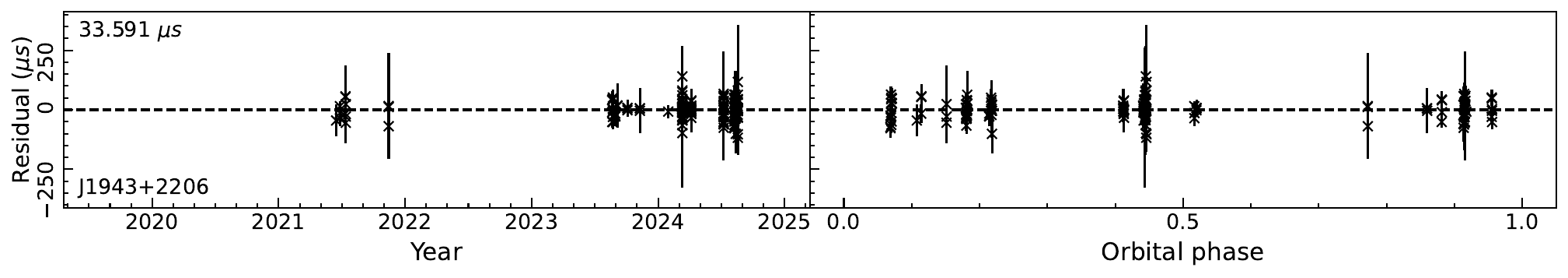}
  \includegraphics[bb =  5 40 965 165, clip, width=0.95\textwidth] {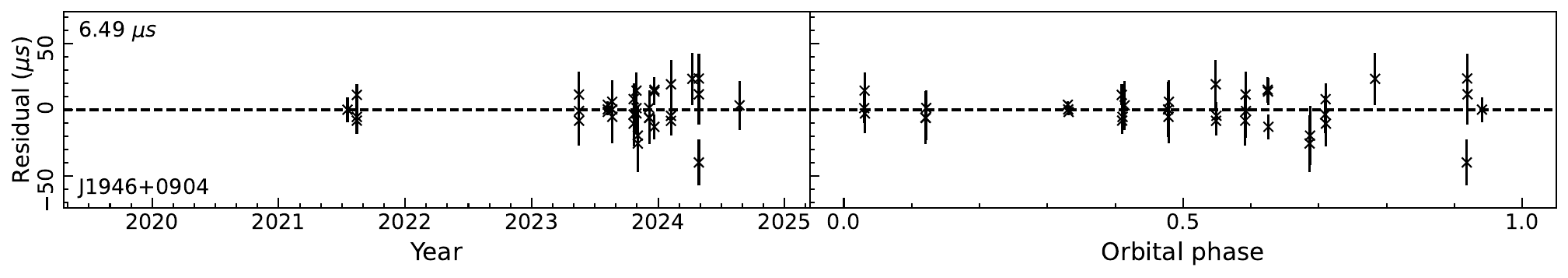}
  \includegraphics[bb =  5 40 965 165, clip, width=0.95\textwidth] {pngs/J1947+2011_Residual.pdf}
  \includegraphics[bb =  5 40 965 165, clip, width=0.95\textwidth] {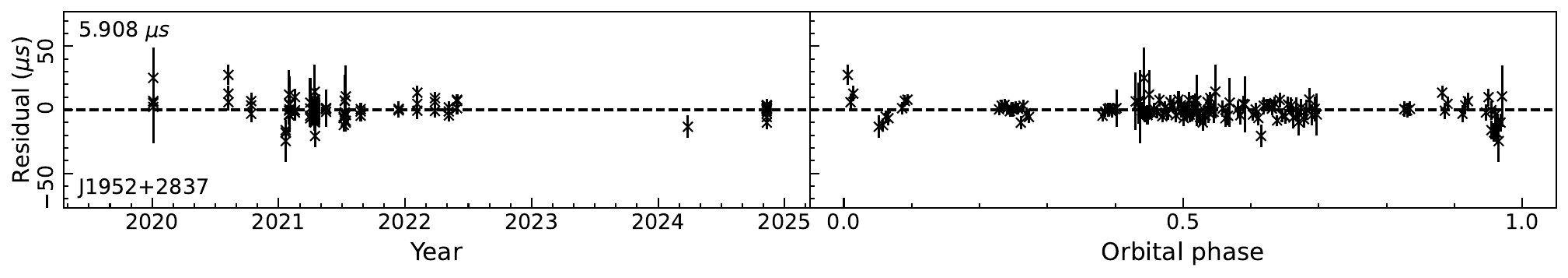}
  \includegraphics[bb =  5 5 965 165, clip,  width=0.95\textwidth] {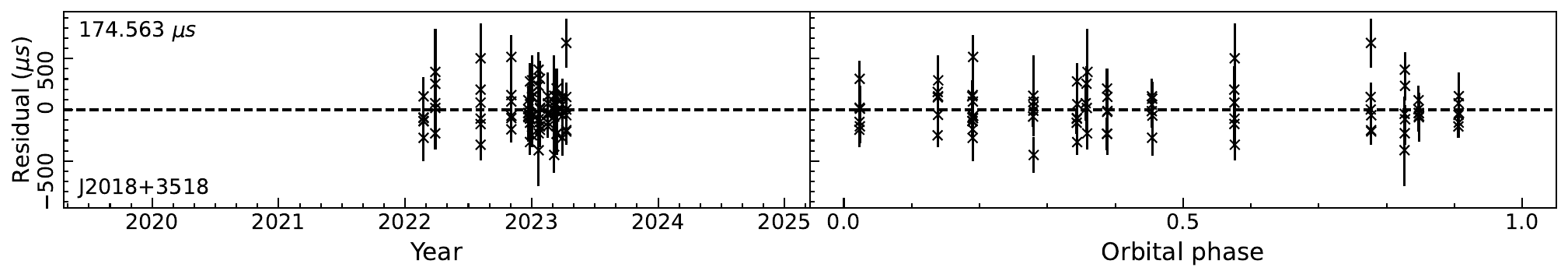}
  
  \caption{---End.}
\end{figure*}


\begin{figure*}
  \centering
  \includegraphics[width = 0.303\textwidth] {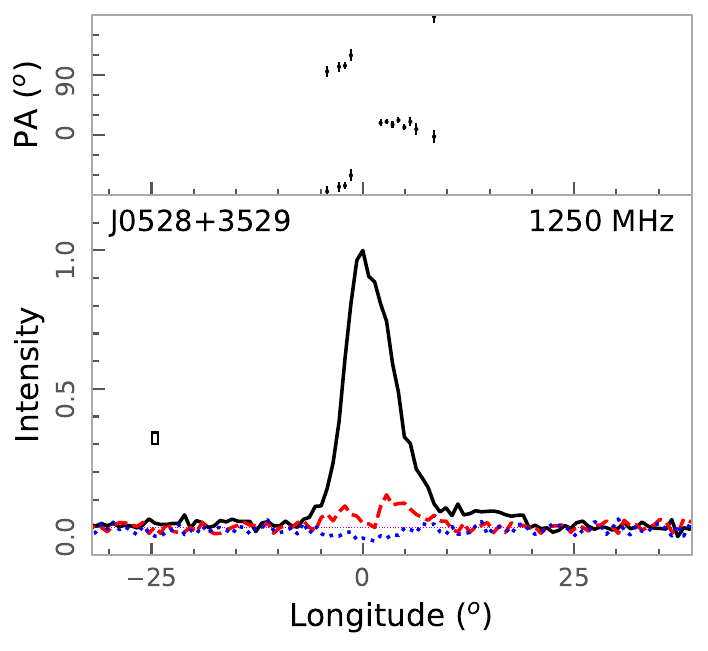}
  \includegraphics[width = 0.303\textwidth] {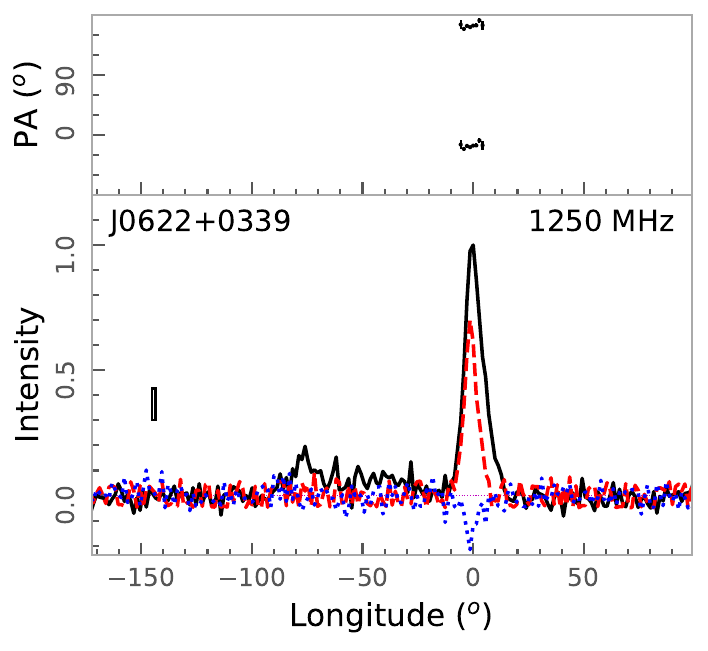}
  \includegraphics[width = 0.303\textwidth] {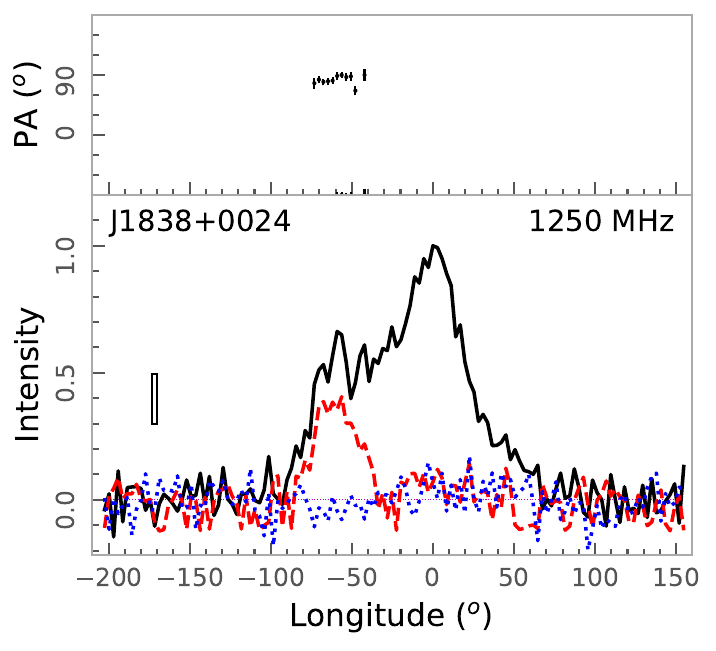}\\ 
  \includegraphics[width = 0.303\textwidth] {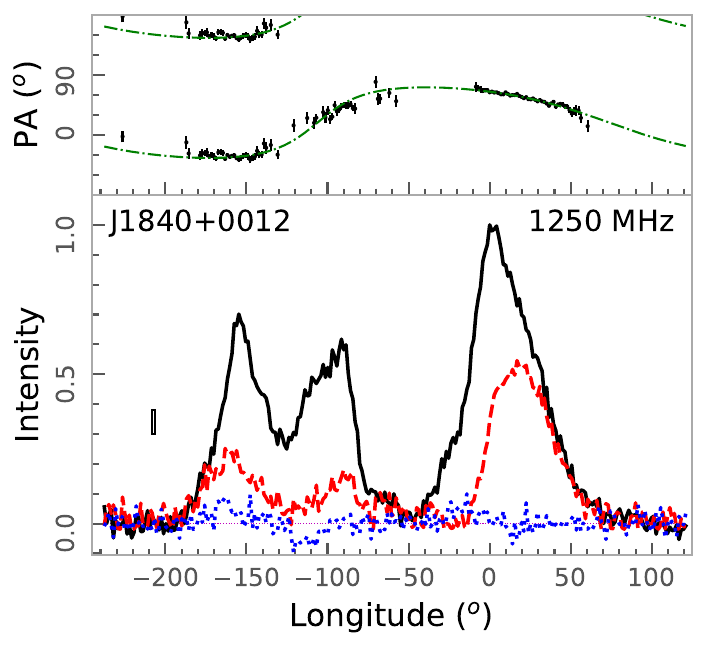}
  \includegraphics[width = 0.303\textwidth] {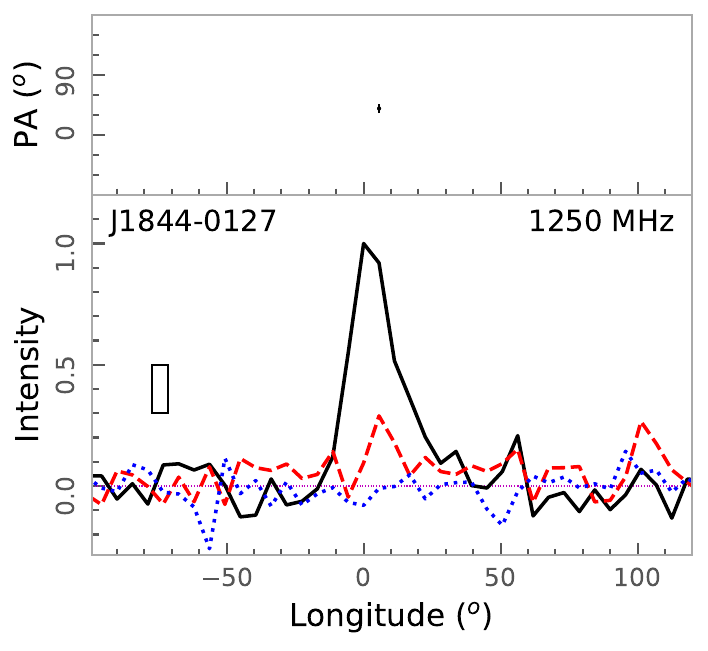}  
  \includegraphics[width = 0.303\textwidth] {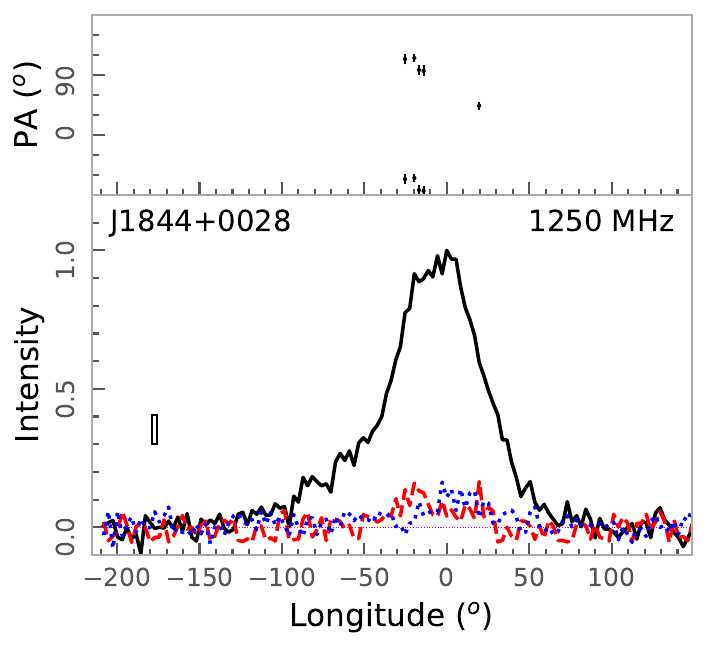}\\
  \includegraphics[width = 0.303\textwidth] {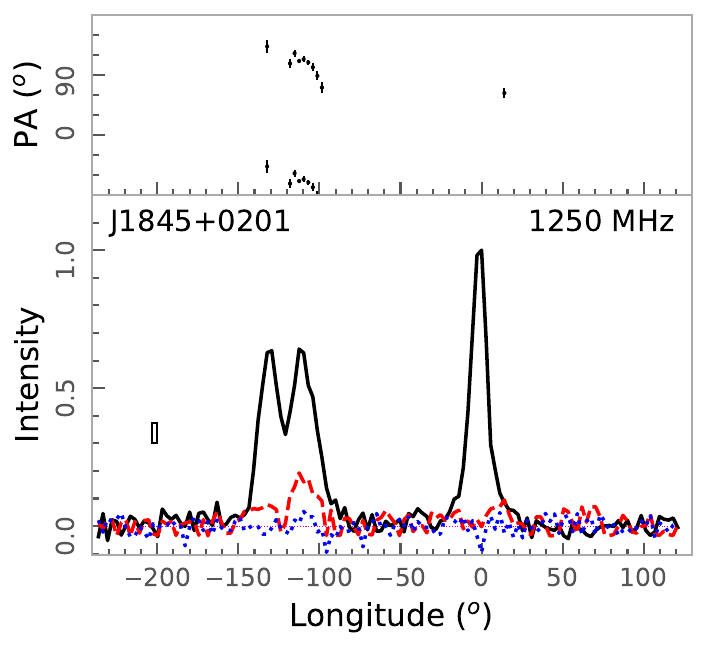}
  \includegraphics[width = 0.303\textwidth] {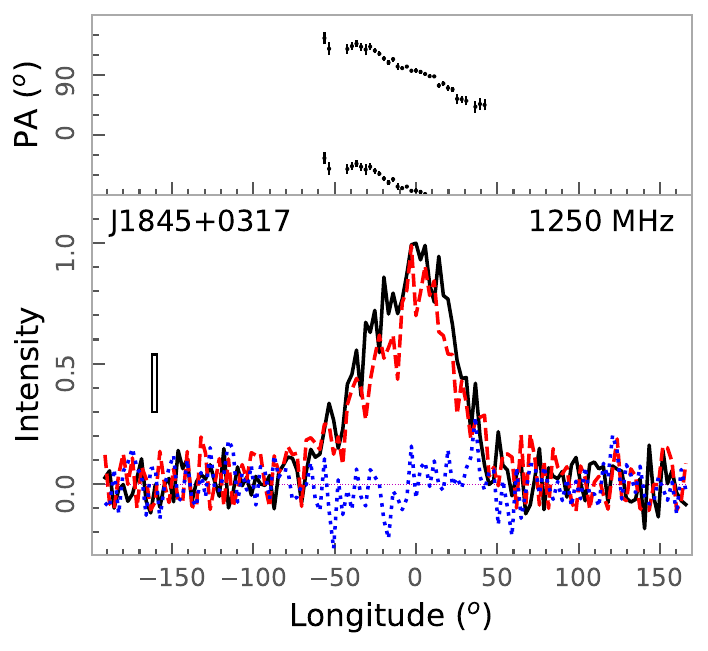}  
  \includegraphics[width = 0.303\textwidth] {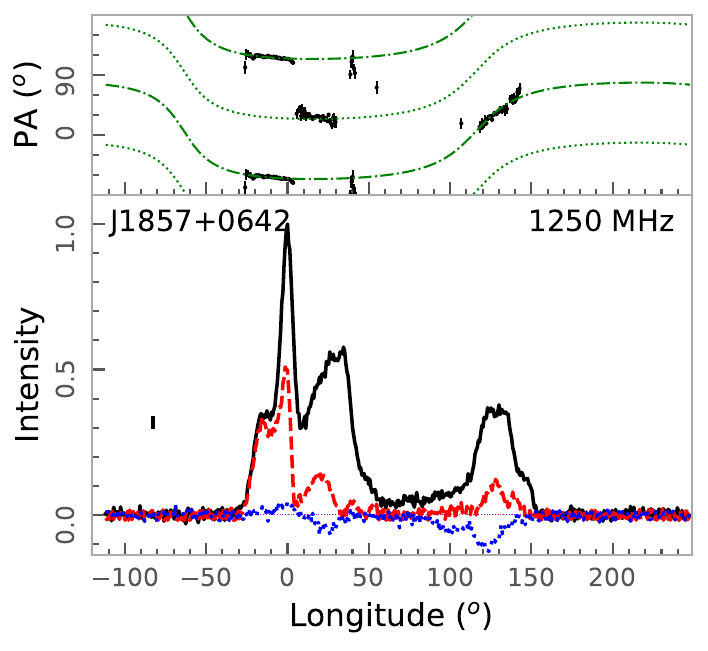}\\
  \includegraphics[width = 0.303\textwidth] {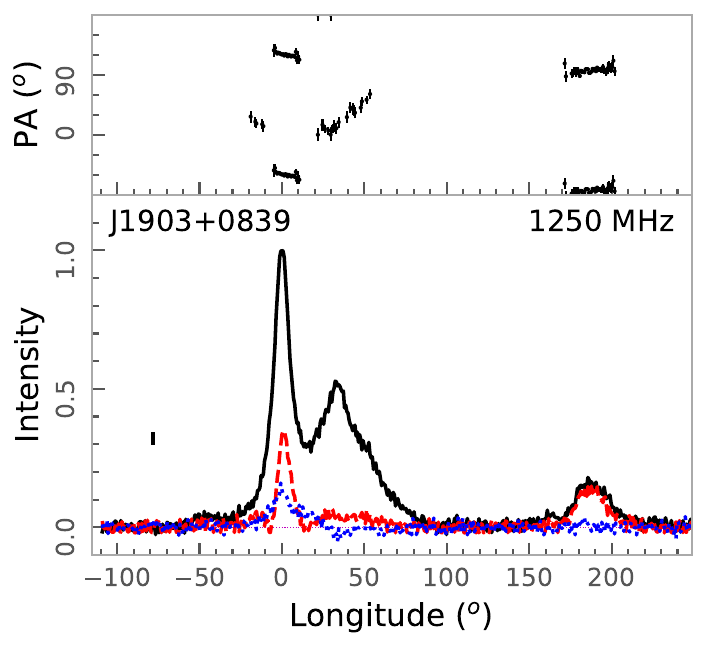}
  \includegraphics[width = 0.303\textwidth] {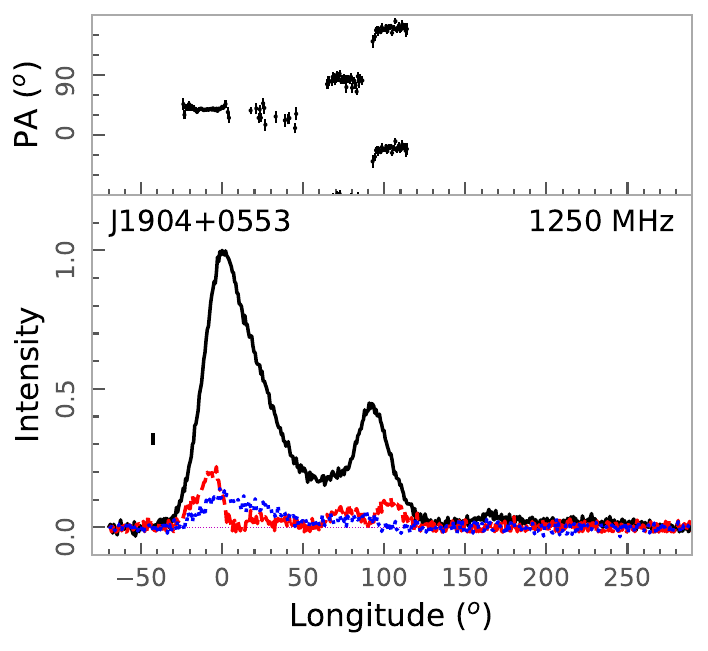}
  \includegraphics[width = 0.303\textwidth] {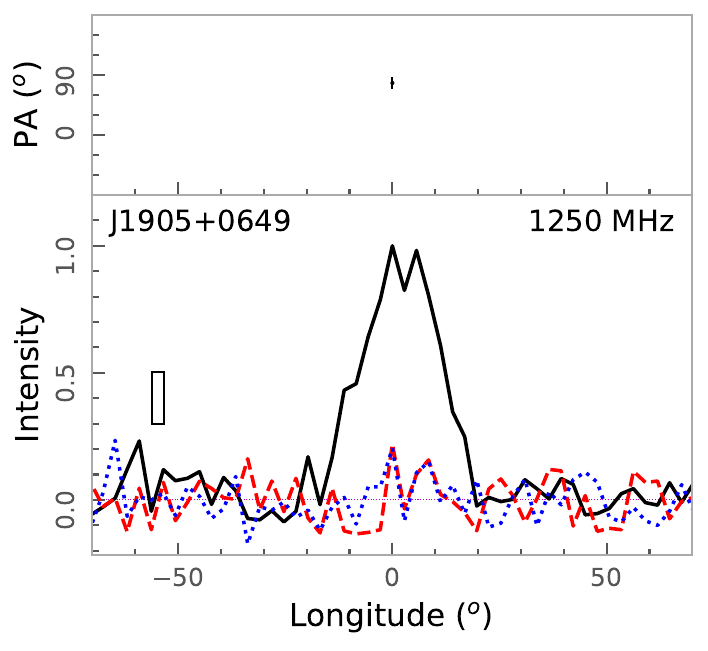}\\ 
  \caption{Integrated pulse profiles of 29 pulsars. The total intensity, linear and circular polarization are represented by solid, dashed and doted lines in the bottom sub-panel. The left-hand circular polarization is defined to be positive. The bin size and $3\sigma$ are marked inside the sub-panel, here $\sigma$ is the standard deviation of off-pulse bins. In the top panel, dots with error-bar are measurements of polarization position angles for linear polarization intensity exceeding 3$\sigma$ line. The position angles are corrected to infinite frequency by discounting Faraday rotation. We tried to fit the polarization angle curves with the rotating vector model \citep{rc69} for 5 pulsars, PSRs J1840+0012, J1857+0642, J1916+0740, J1930+1403 and J1946+0904.}
  \label{fig:profs_appendix}
\end{figure*}
\addtocounter{figure}{-1}
\begin{figure*}
  \centering
  \includegraphics[width = 0.303\textwidth] {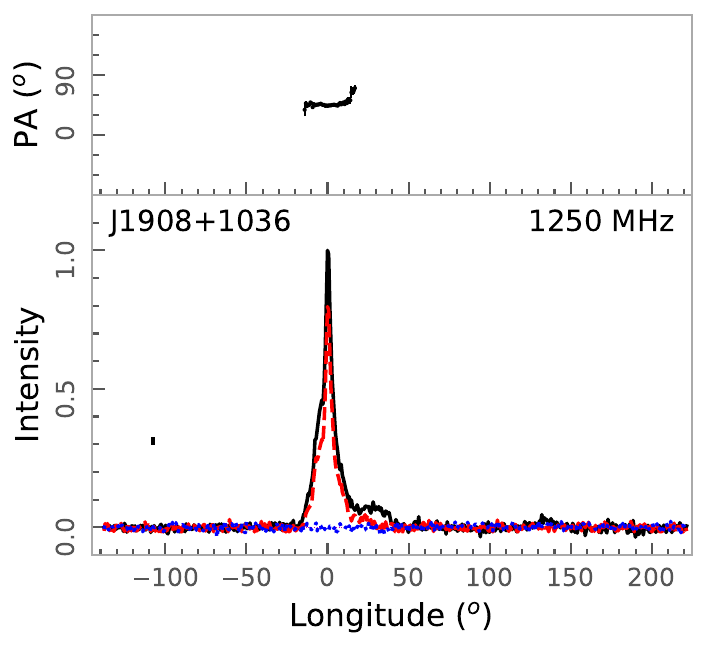}
  \includegraphics[width = 0.303\textwidth] {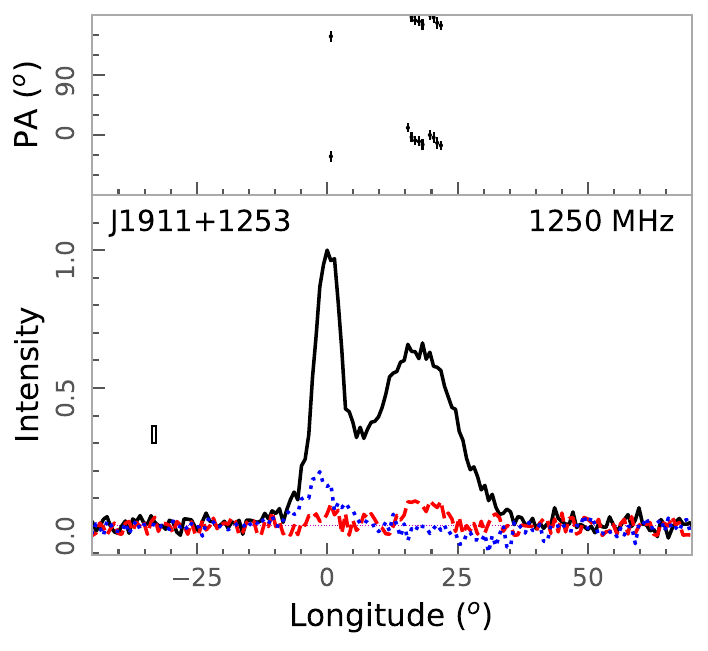}
  \includegraphics[width = 0.303\textwidth] {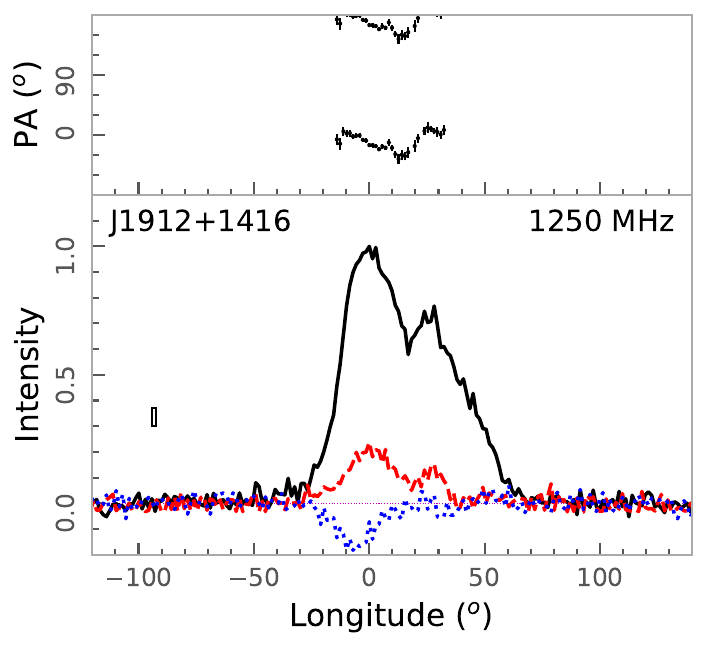}\\
  \includegraphics[width = 0.303\textwidth] {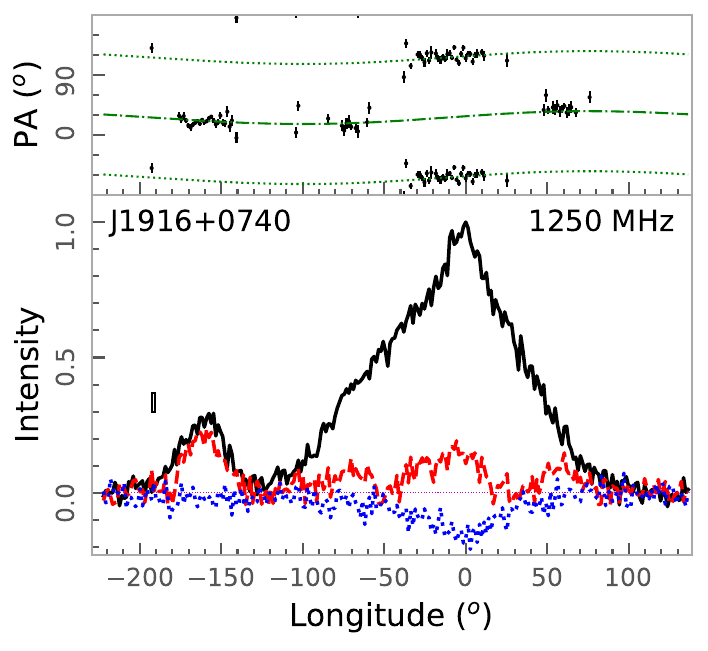}
  \includegraphics[width = 0.303\textwidth] {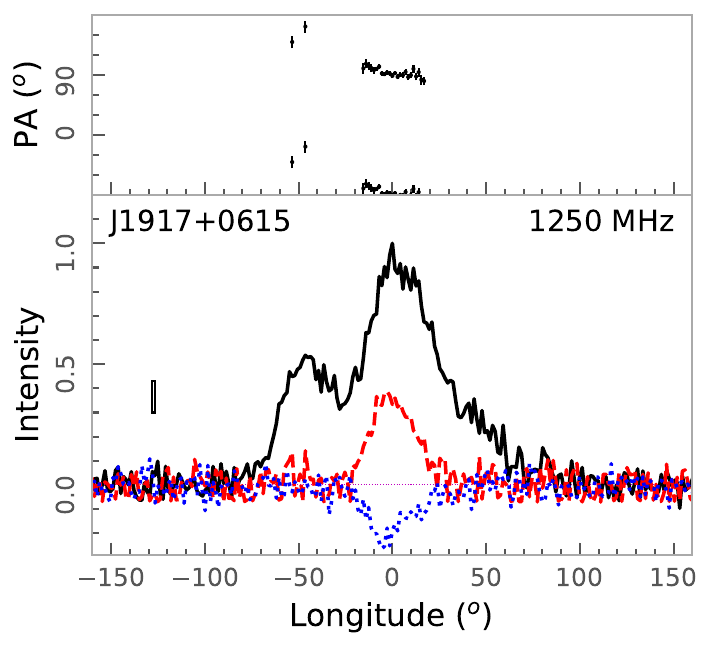}
  \includegraphics[width = 0.303\textwidth] {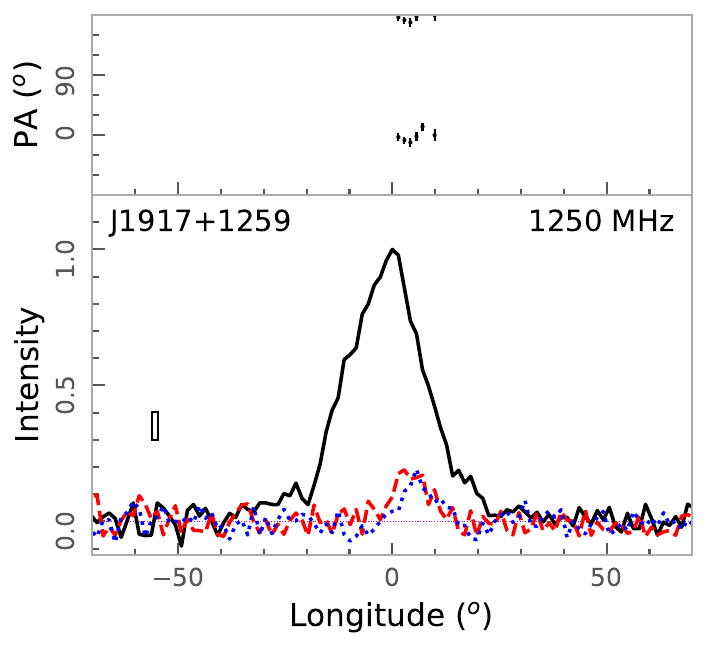} \\
  \includegraphics[width = 0.303\textwidth] {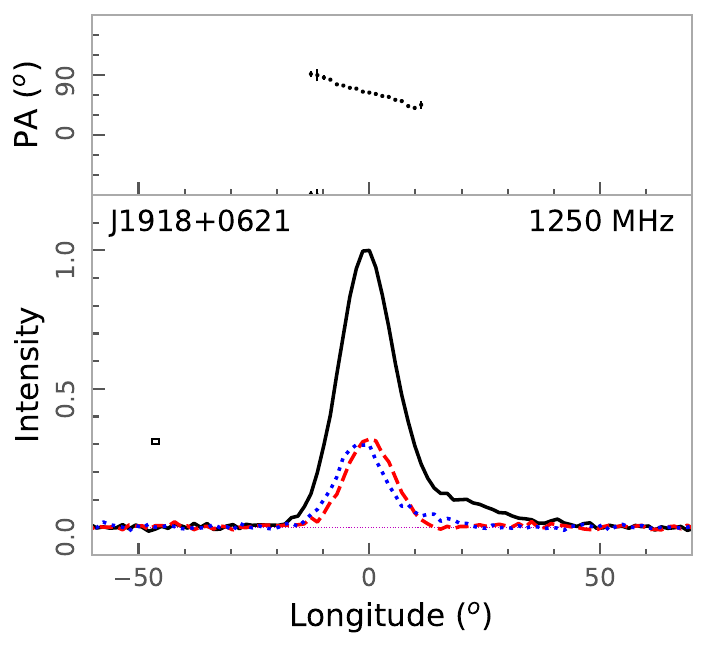}
  \includegraphics[width = 0.303\textwidth] {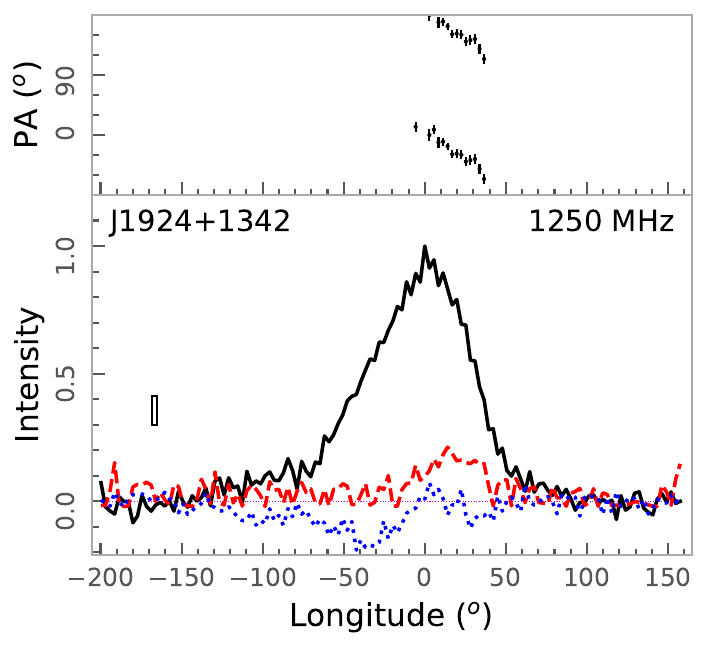}
  \includegraphics[width = 0.303\textwidth] {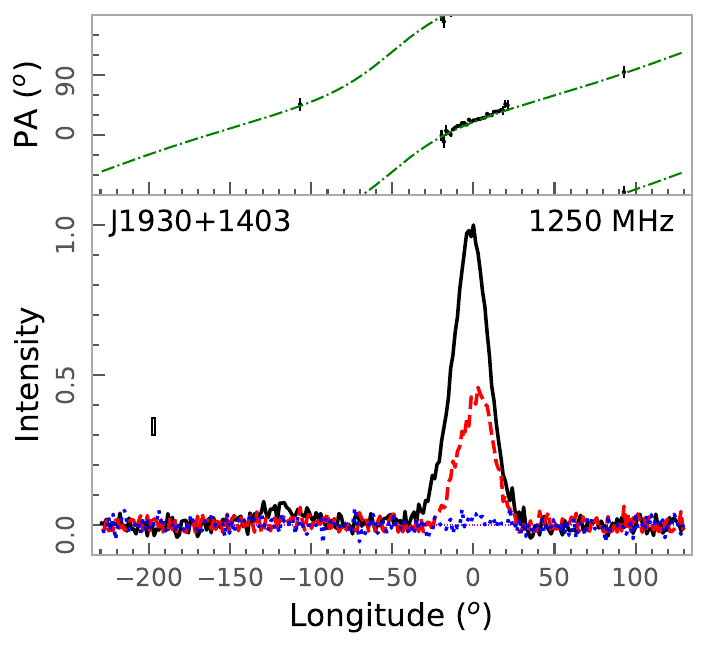}\\
  \includegraphics[width = 0.303\textwidth] {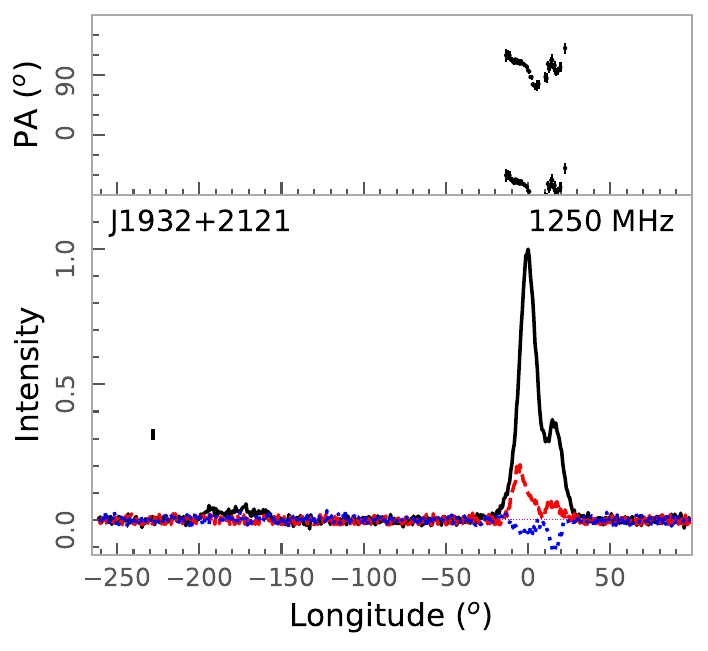}
  \includegraphics[width = 0.303\textwidth] {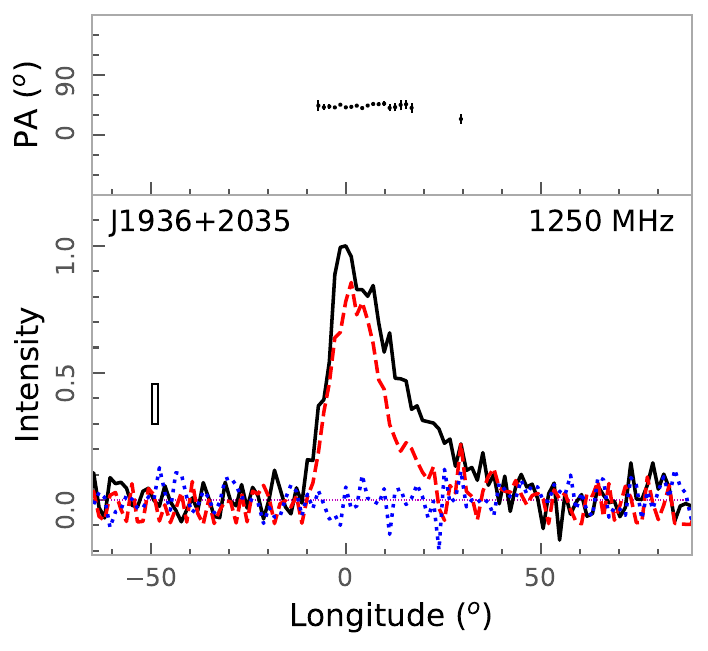}
  \includegraphics[width = 0.303\textwidth] {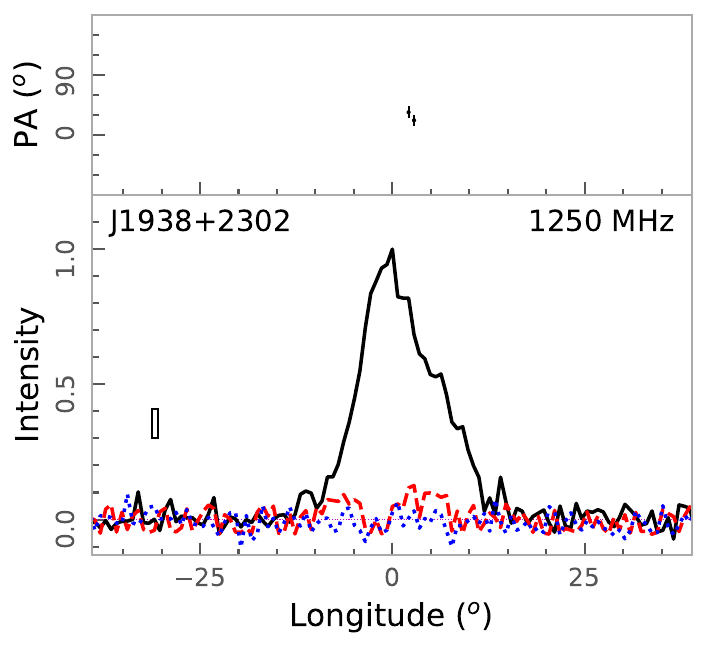}\\
  \caption{Continue---}
\end{figure*}
\addtocounter{figure}{-1}
\begin{figure*}
  \centering
  \includegraphics[width = 0.303\textwidth] {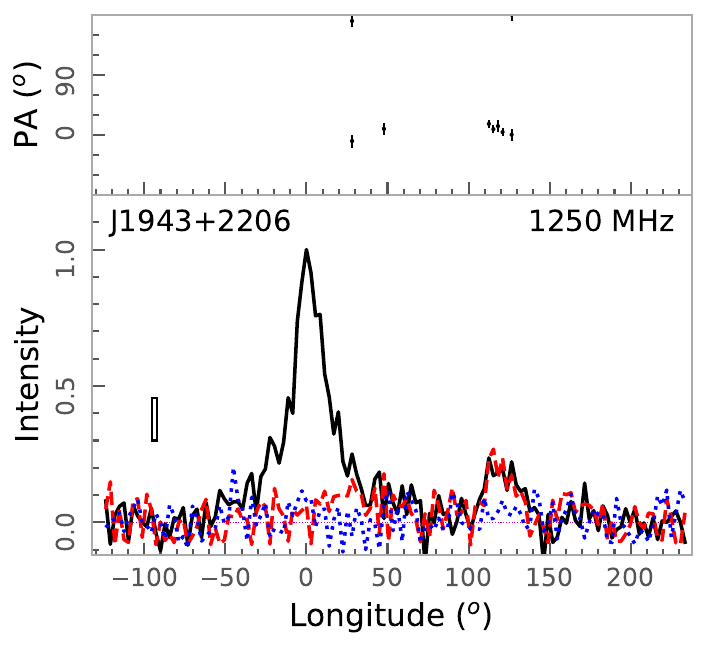}
  \includegraphics[width = 0.303\textwidth] {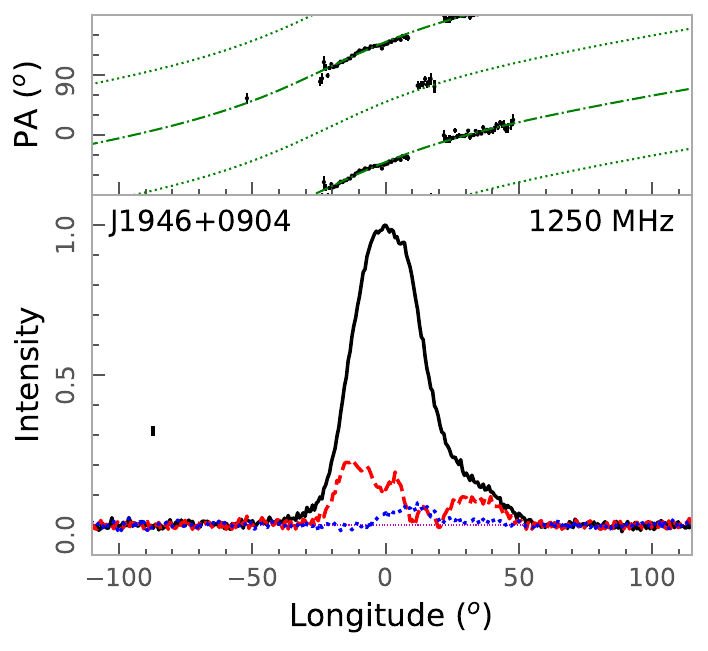} 
  \includegraphics[width = 0.303\textwidth] {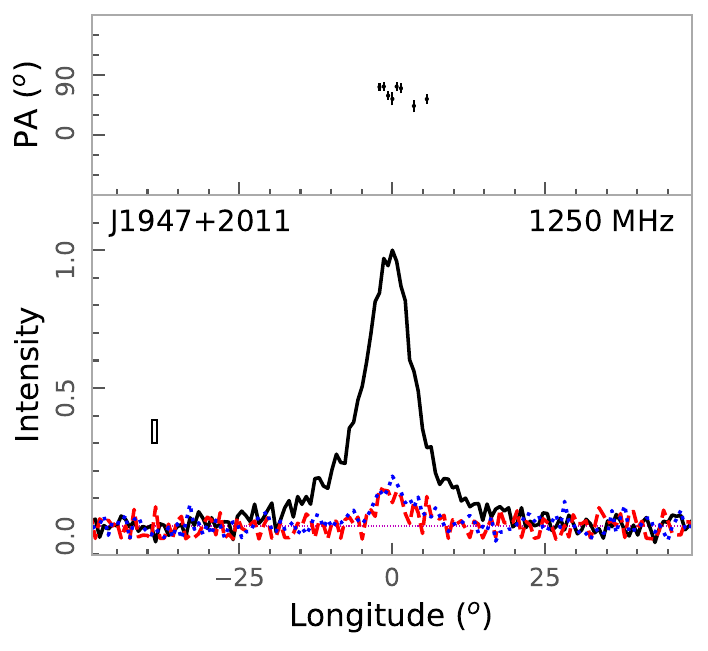} \\
  \includegraphics[width = 0.303\textwidth] {pngs/J1952+2837_IntProf.pdf} 
  \includegraphics[width = 0.303\textwidth] {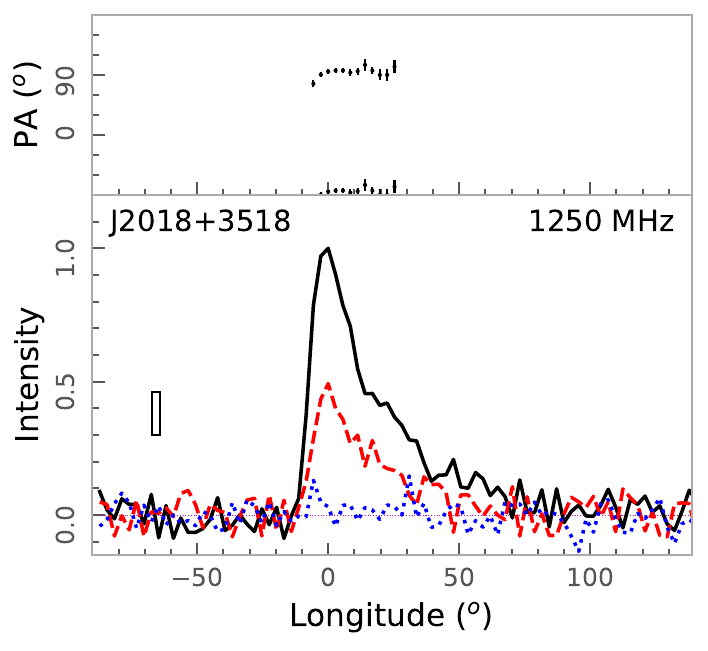}  
    \caption{---End}
\end{figure*}

\begin{figure*}
  \centering
  \includegraphics[height = 0.18\textheight] {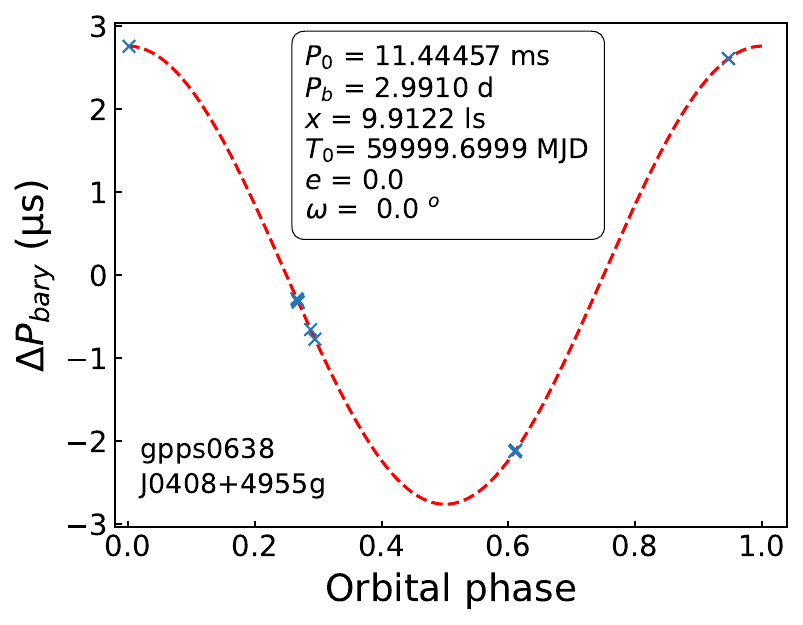}
  \includegraphics[height = 0.18\textheight] {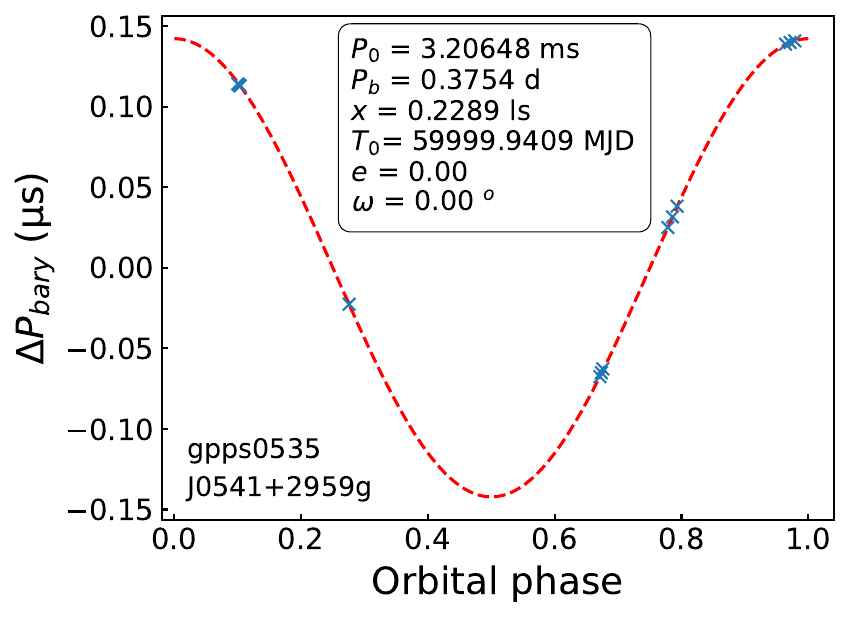}
  \includegraphics[height = 0.18\textheight] {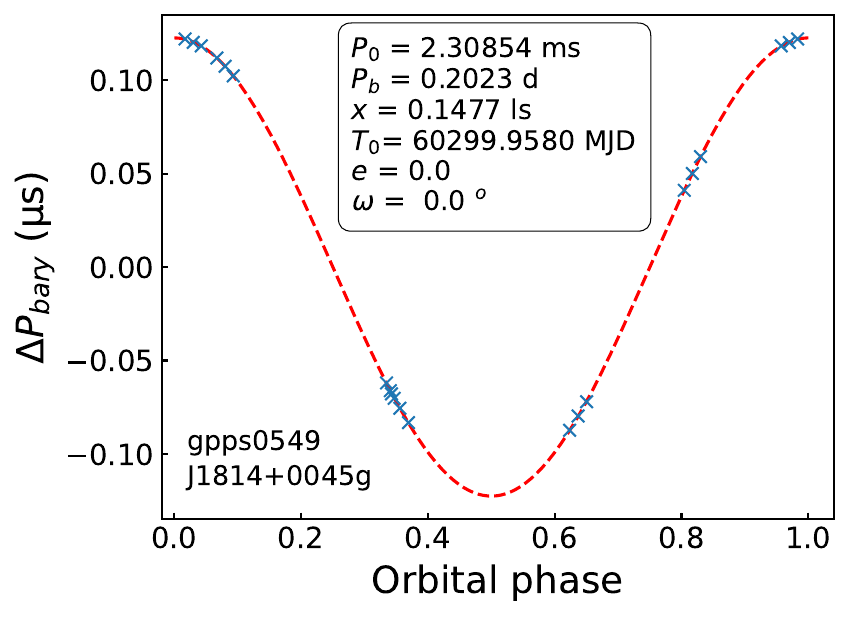}\\
  \includegraphics[height = 0.18\textheight] {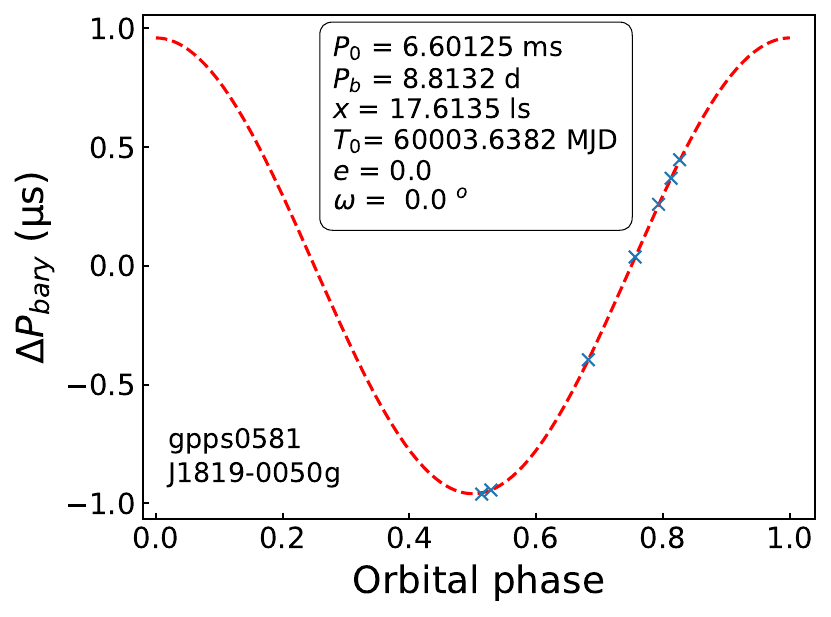}
  \includegraphics[height = 0.18\textheight] {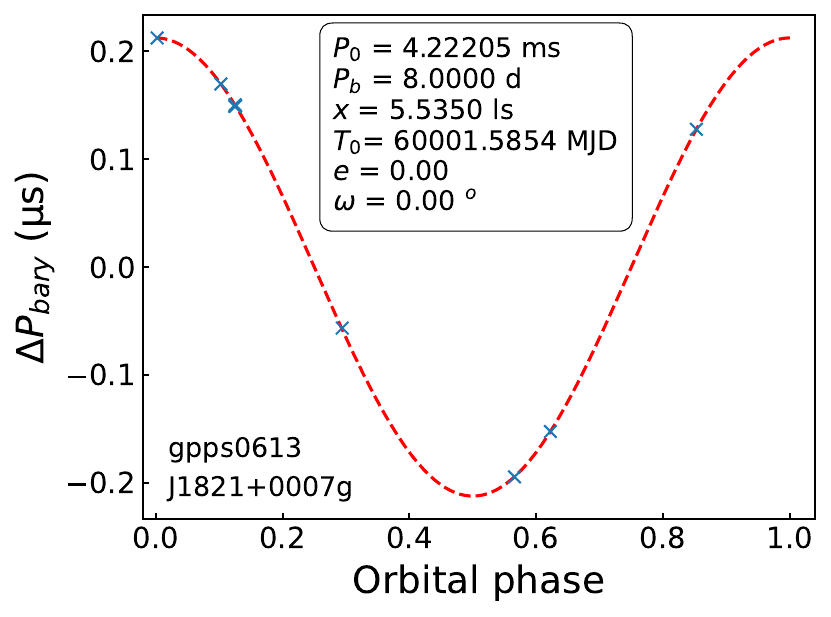}
  \includegraphics[height = 0.18\textheight] {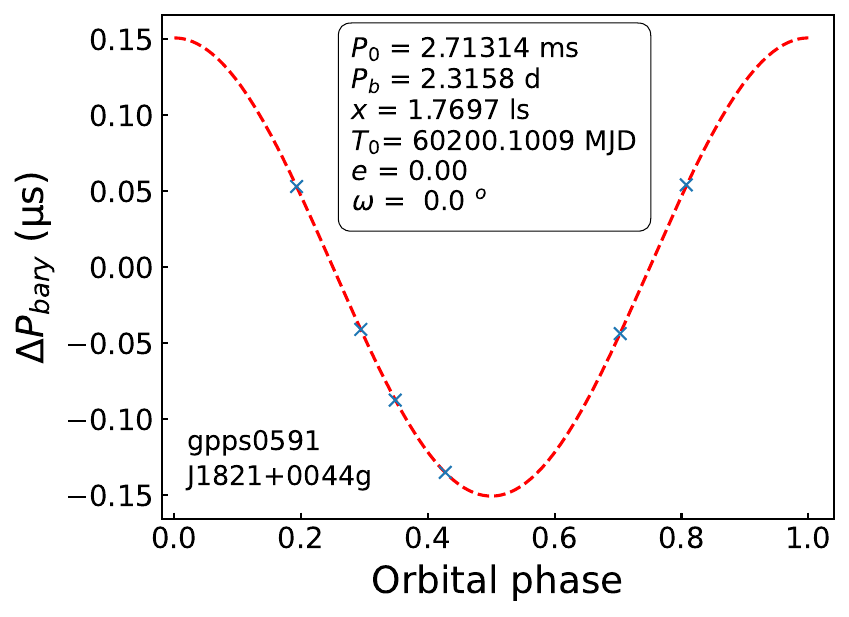}\\
  \includegraphics[height = 0.18\textheight] {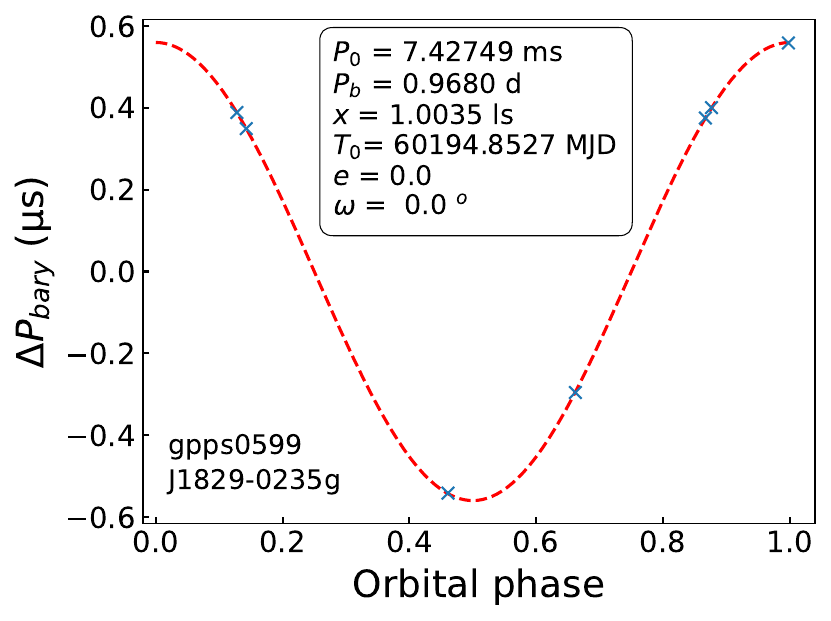}
  \includegraphics[height = 0.18\textheight] {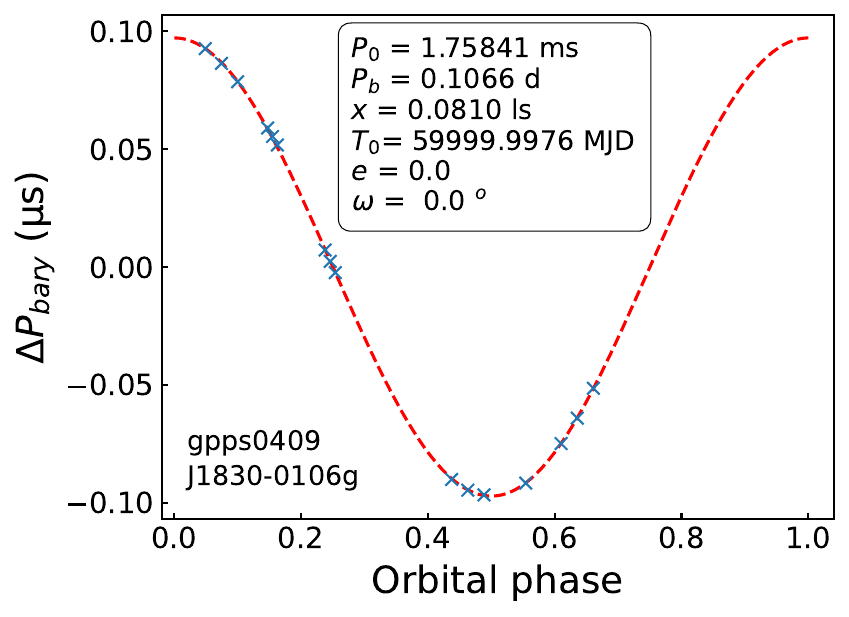}
  \includegraphics[height = 0.18\textheight] {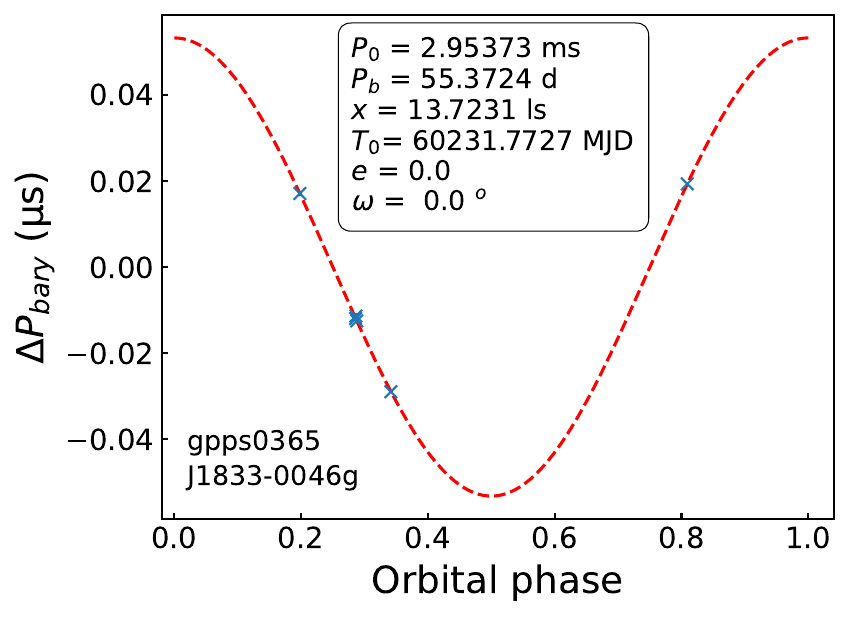}\\
  \includegraphics[height = 0.18\textheight] {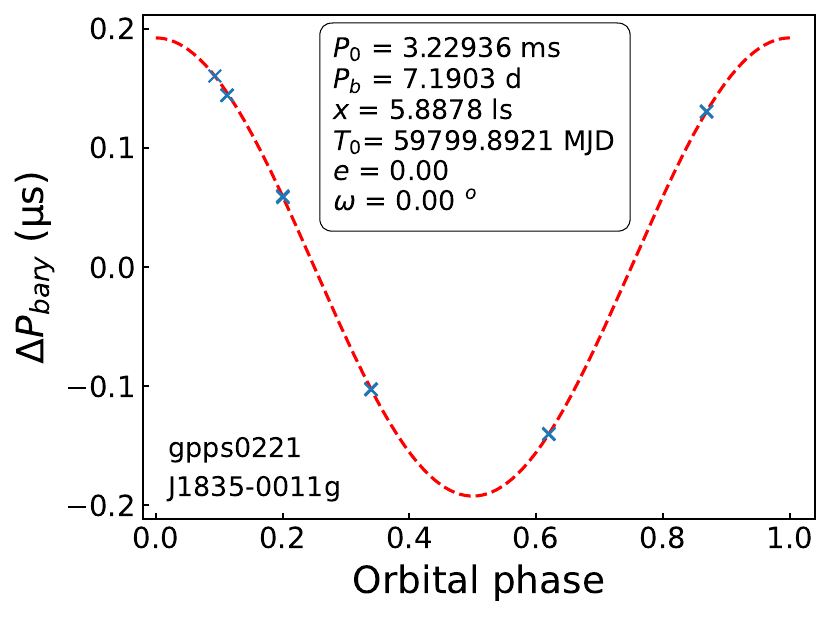}
  \includegraphics[height = 0.18\textheight] {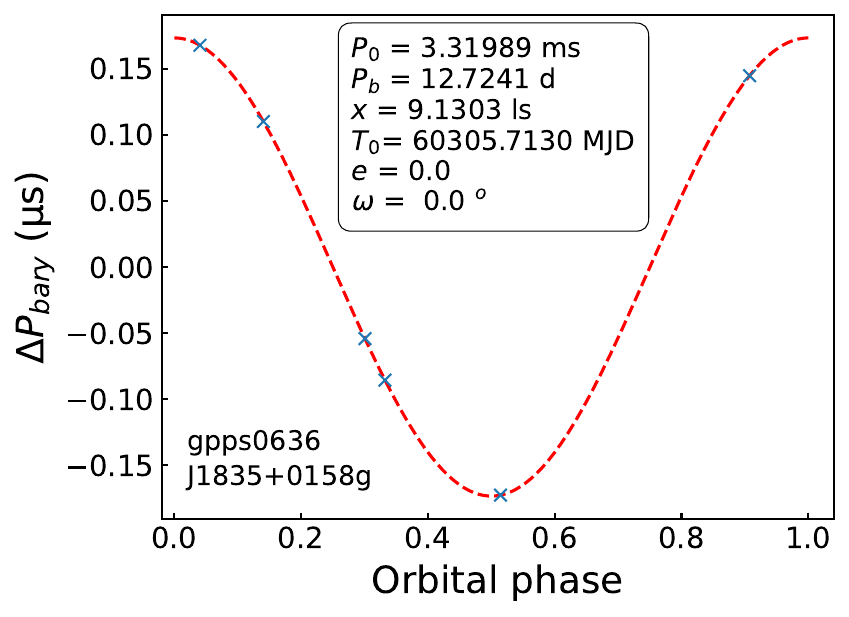} 
  \includegraphics[height = 0.18\textheight] {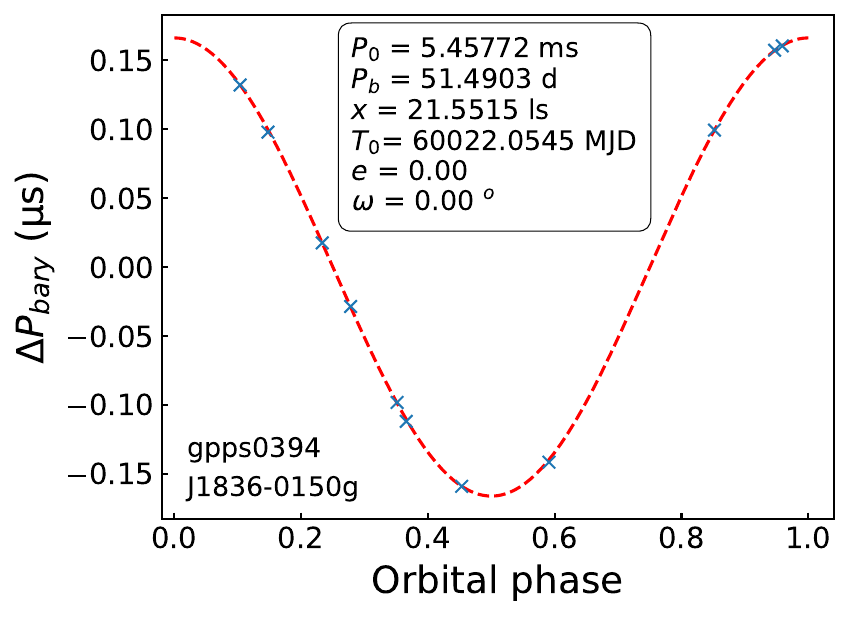}\\
  \includegraphics[height = 0.18\textheight] {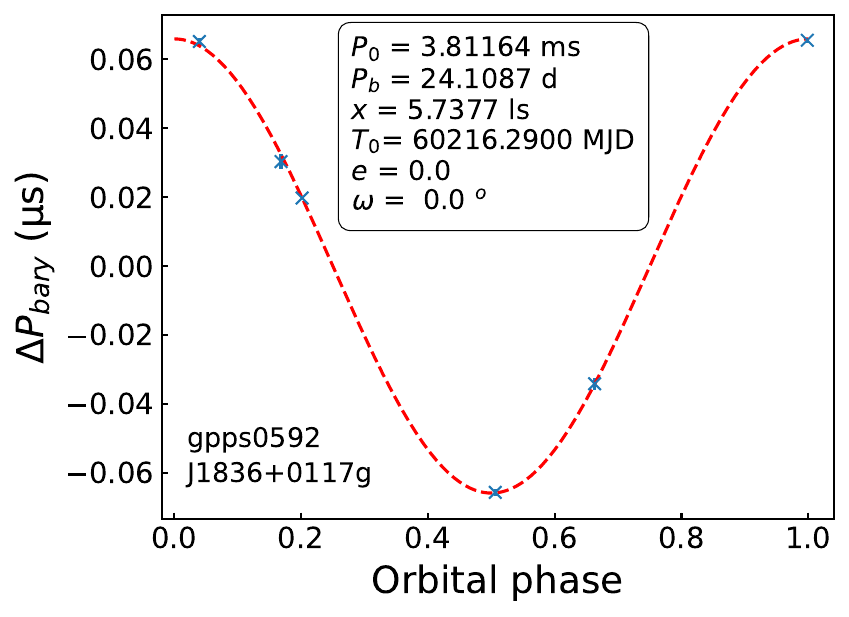}
  \includegraphics[height = 0.18\textheight] {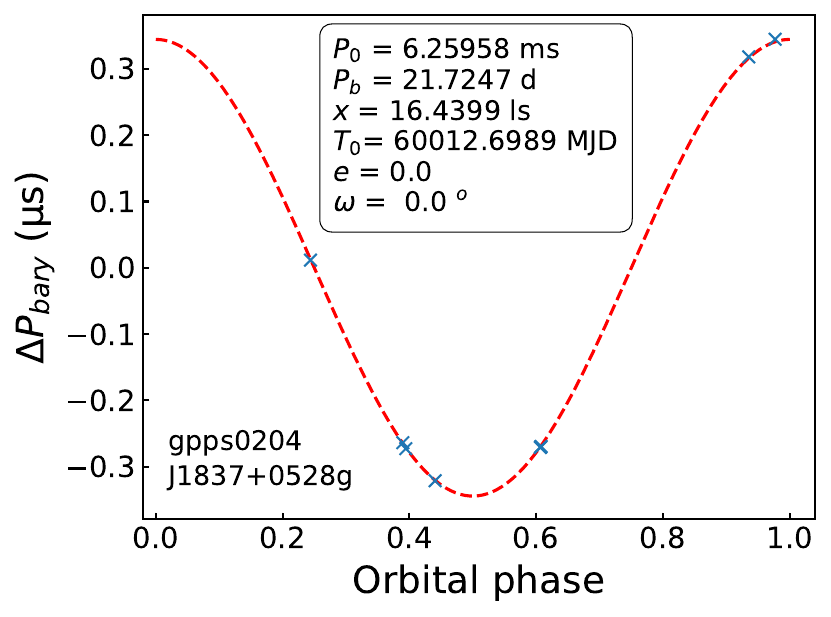}
  \includegraphics[height = 0.18\textheight] {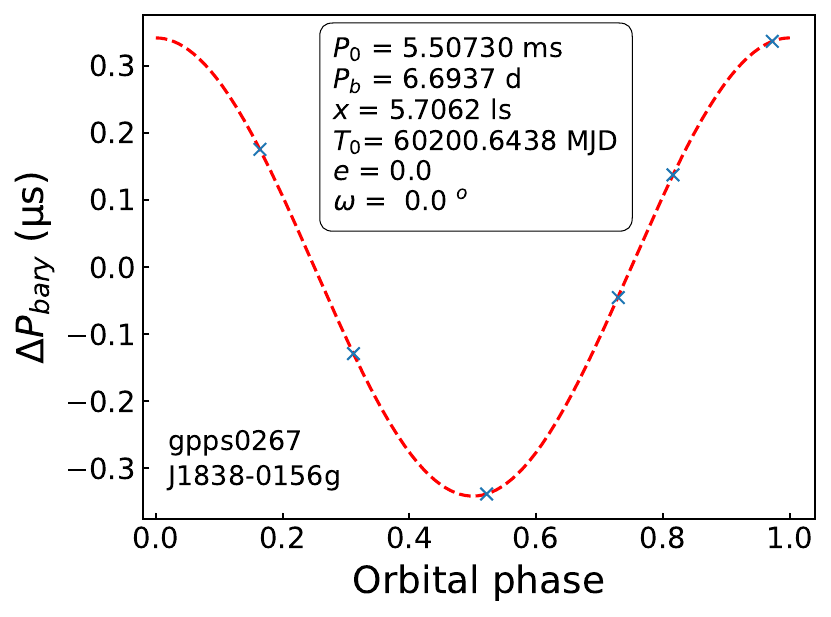}\\
\caption{Variations of barycentric periods for 78 binary pulsars across the orbit phase. For each pulsar, the observed barycentric periods are marked by ``x" after the average period $P_0$ is subtracted. The error-bars are marked but too small to see for most data. Dashed line is the best-fit by using the preliminary Keplerian model with orbital parameters ($P_b$, $x$, $T_0$, $e$ and $\omega$) listed inside the panel. The orbital phase is referred to the periastron of the orbit.}
  \label{fig:Pbary_appendix}
\end{figure*}
\addtocounter{figure}{-1}
\begin{figure*}
  \centering
  \includegraphics[height = 0.18\textheight] {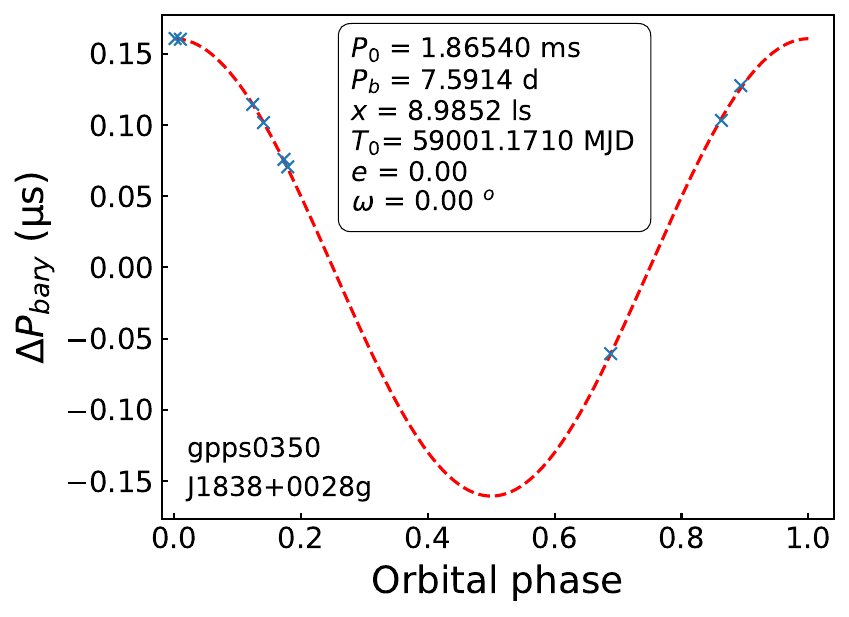}
  \includegraphics[height = 0.18\textheight] {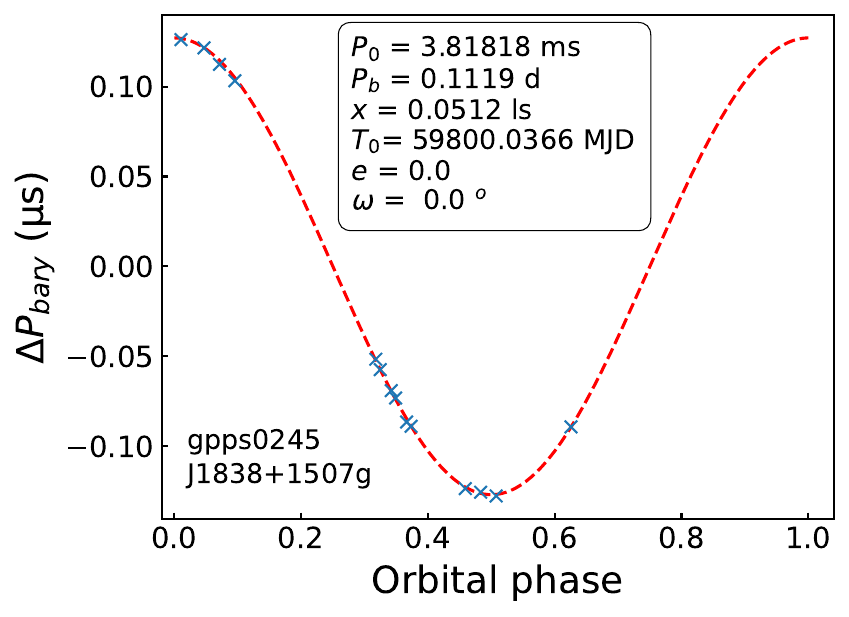}
  \includegraphics[height = 0.18\textheight] {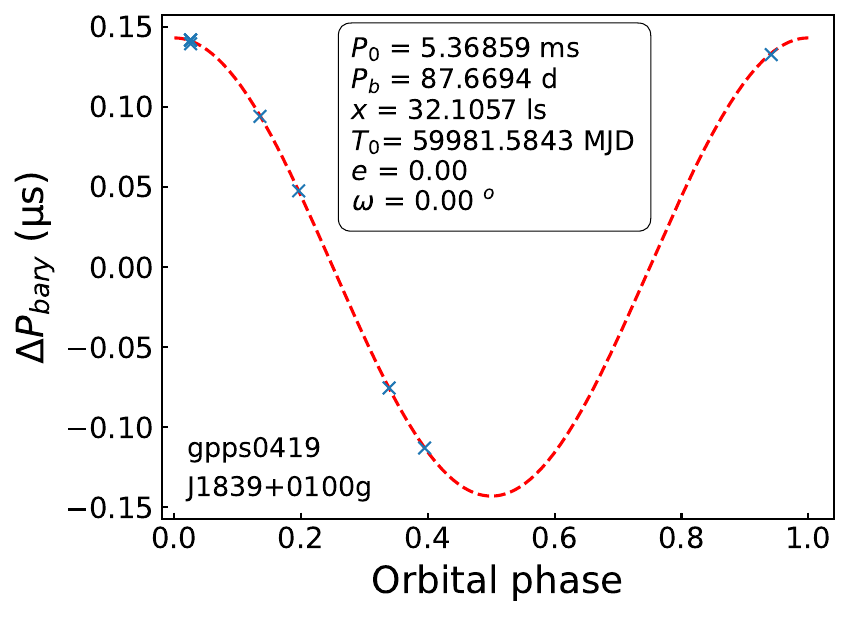}\\
  \includegraphics[height = 0.18\textheight] {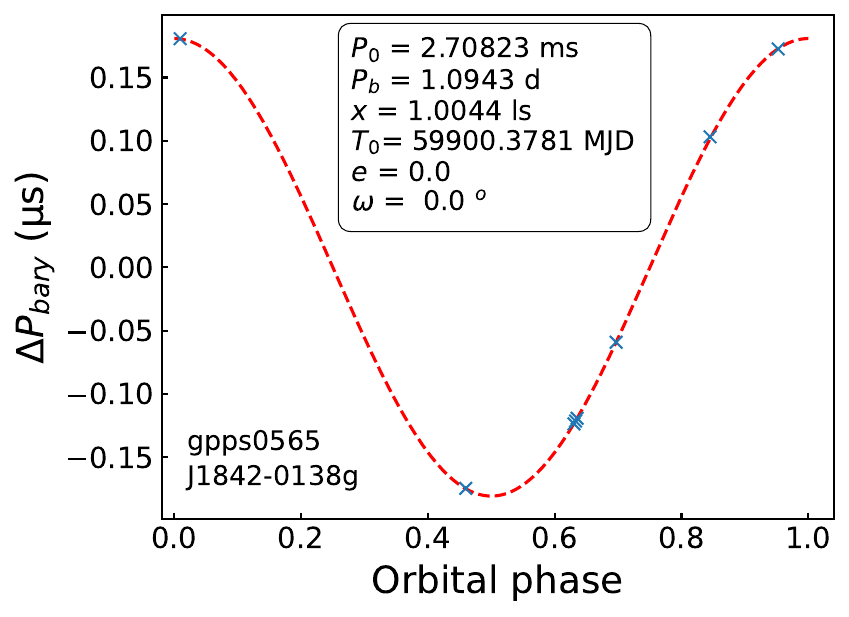} 
  \includegraphics[height = 0.18\textheight] {pngs/J1842+0407g_gpps0417_Pbary.pdf}
  \includegraphics[height = 0.18\textheight] {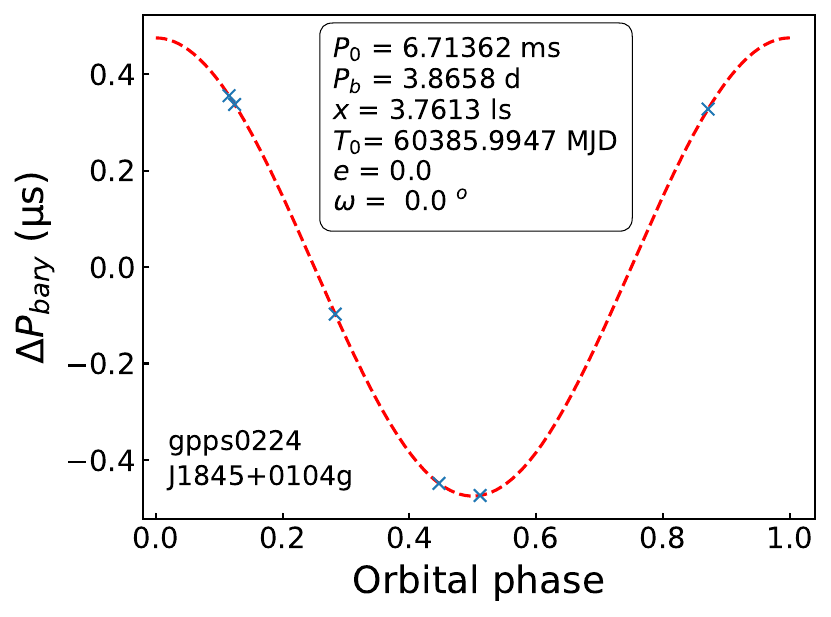}\\ 
  \includegraphics[height = 0.18\textheight] {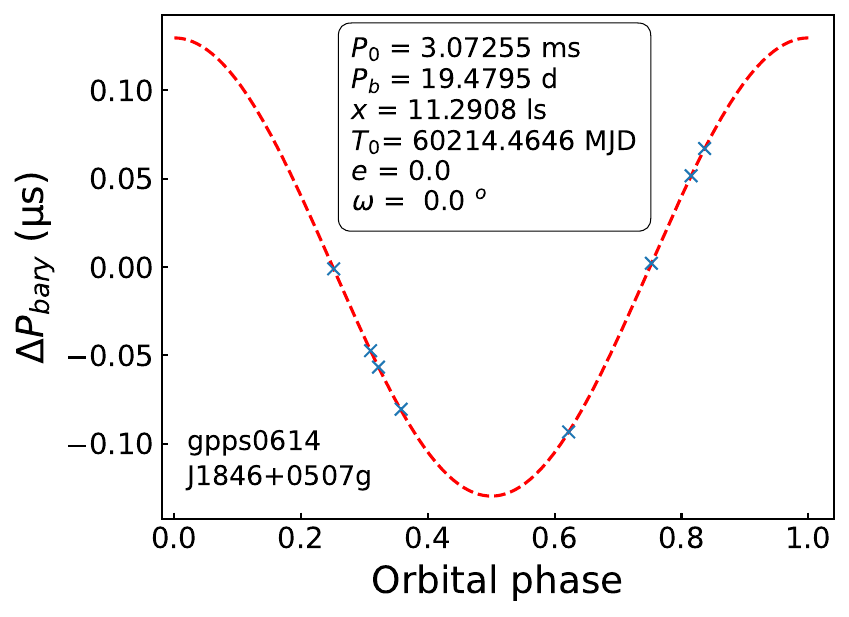}
  \includegraphics[height = 0.18\textheight] {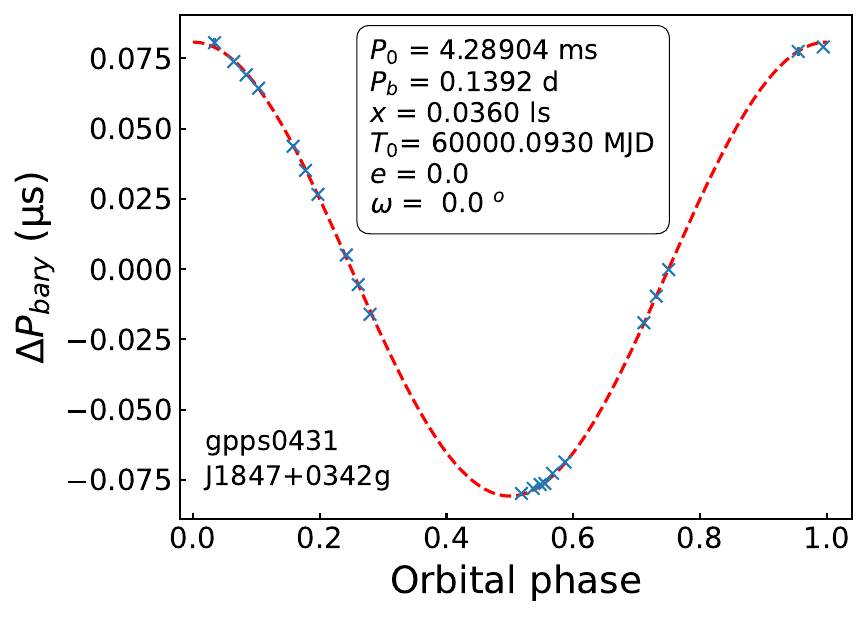}
  \includegraphics[height = 0.18\textheight] {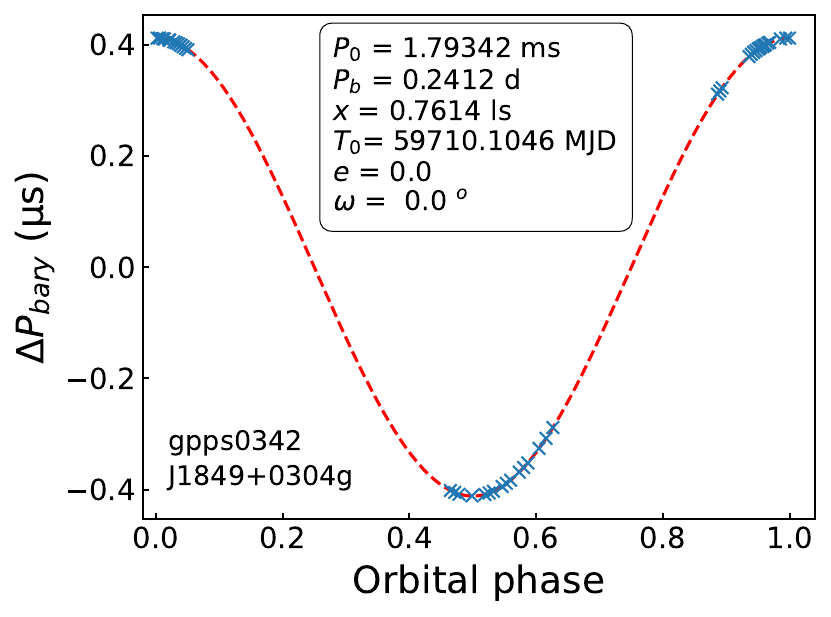}\\
  \includegraphics[height = 0.18\textheight] {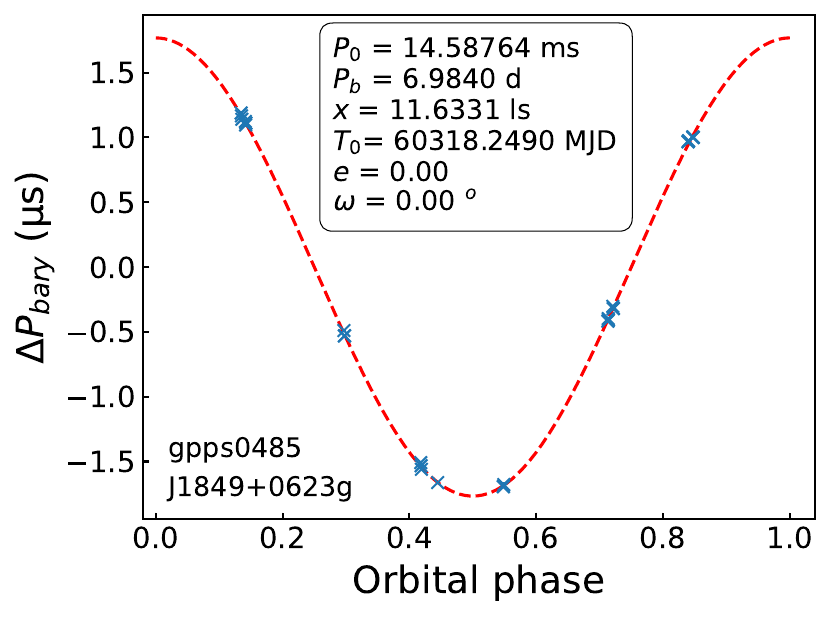}
  \includegraphics[height = 0.18\textheight] {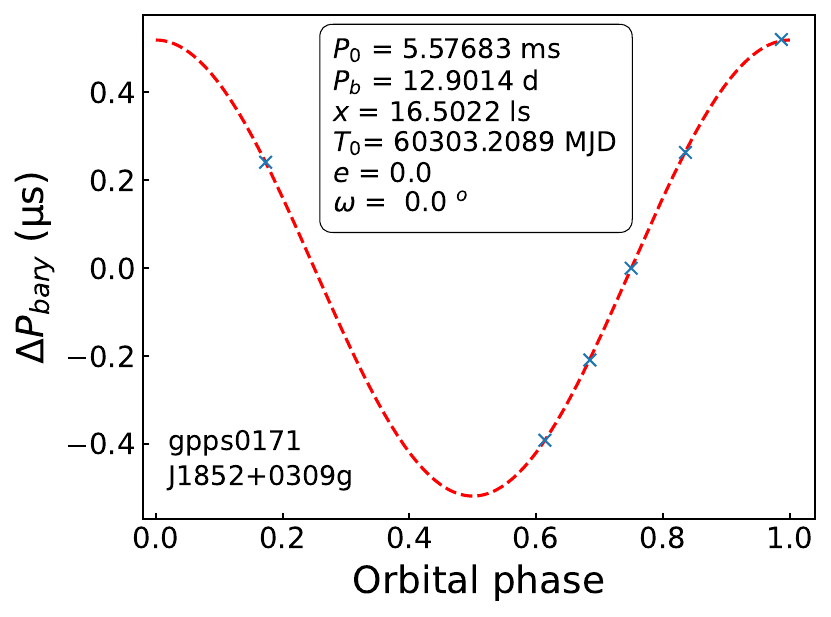}
  \includegraphics[height = 0.18\textheight] {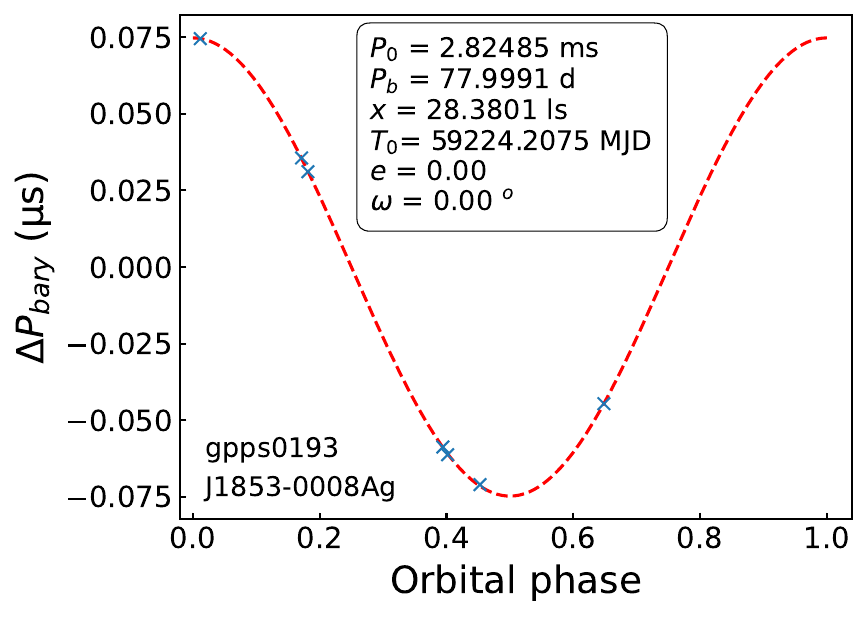}\\
  \includegraphics[height = 0.18\textheight] {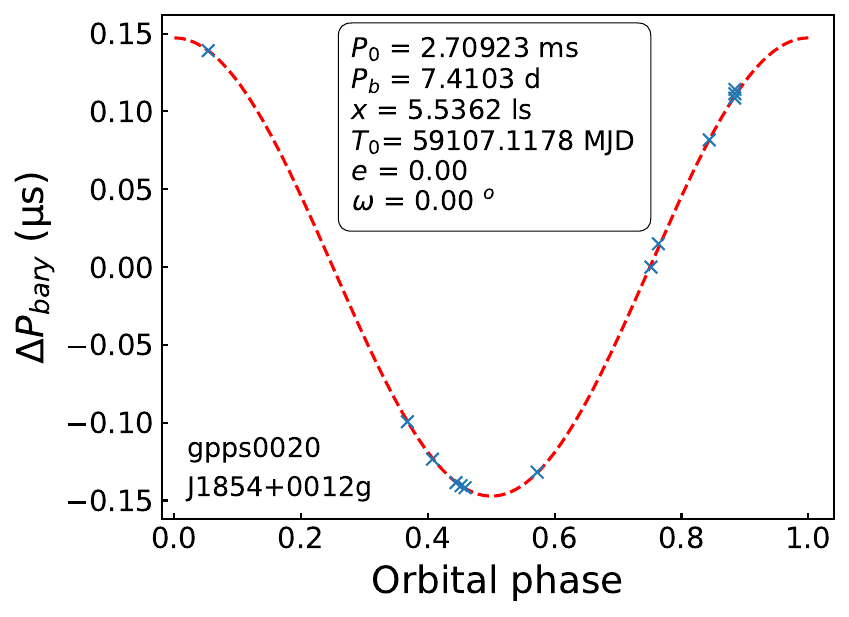}
  \includegraphics[height = 0.18\textheight] {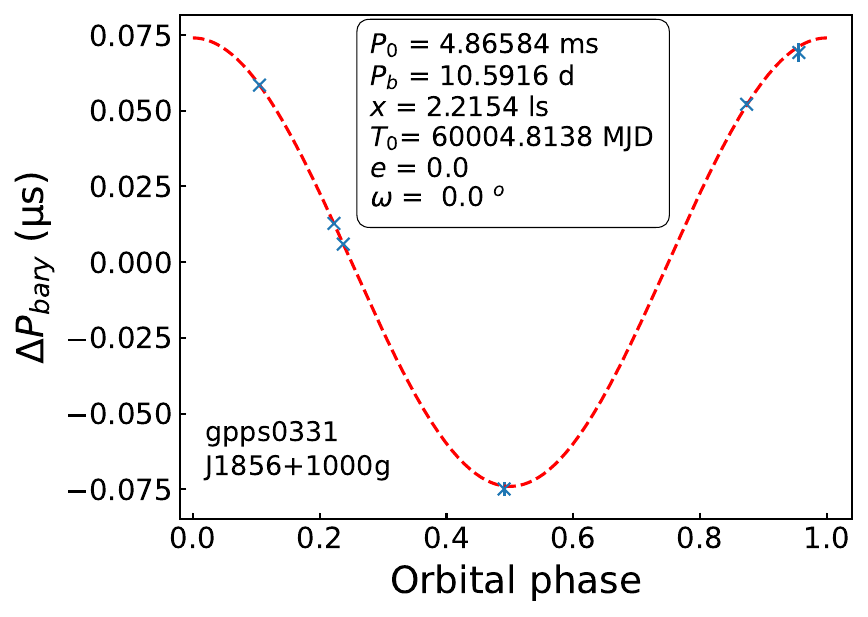}
  \includegraphics[height = 0.18\textheight] {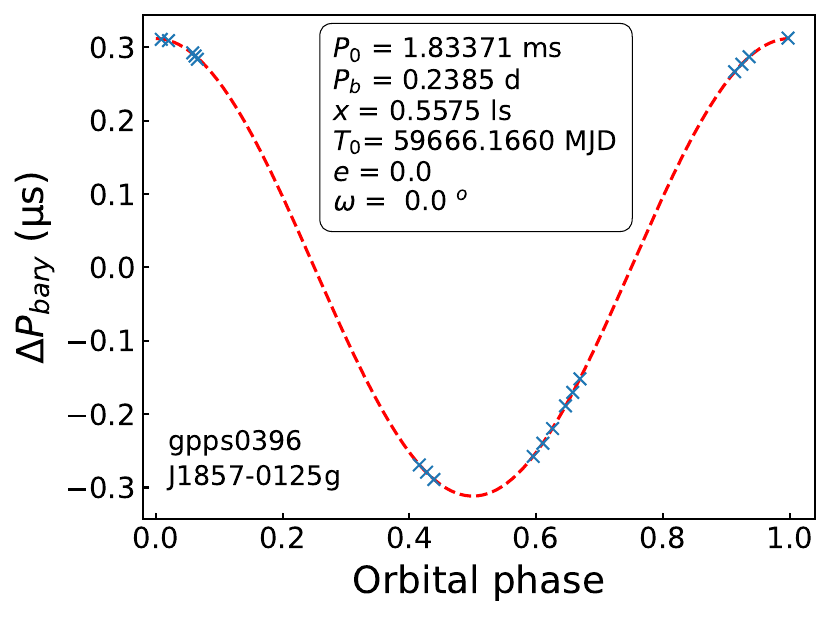} 
 \caption{--continued--}
\end{figure*}
\addtocounter{figure}{-1}
\begin{figure*}
  \centering
  \includegraphics[height = 0.18\textheight] {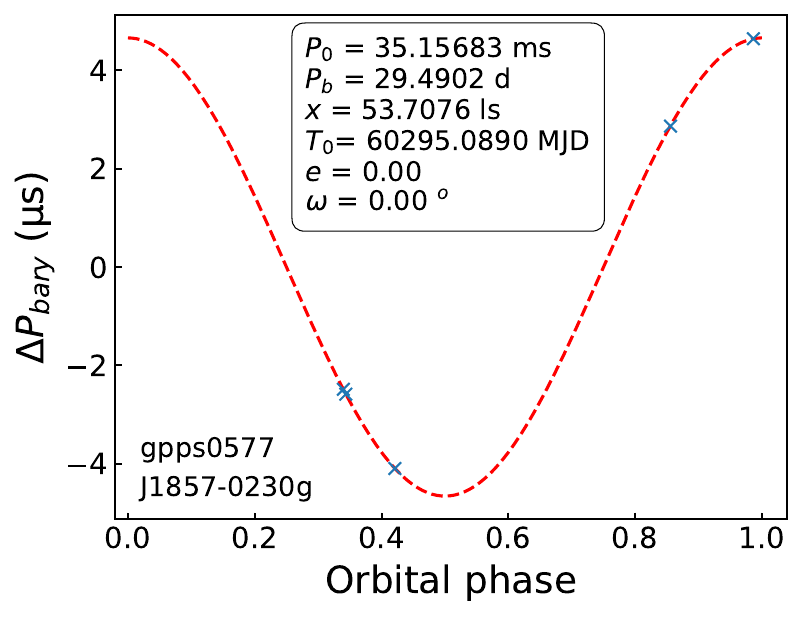}
  \includegraphics[height = 0.18\textheight] {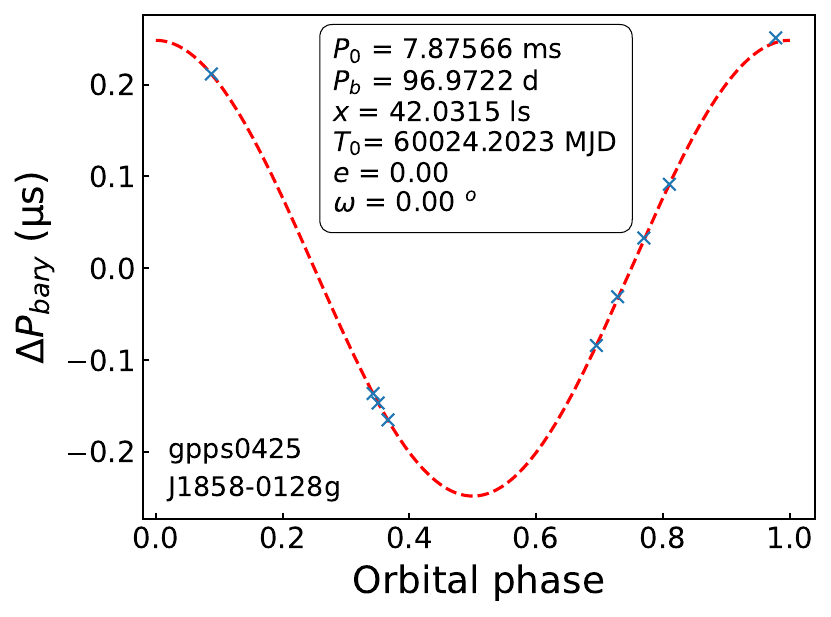} 
  \includegraphics[height = 0.18\textheight] {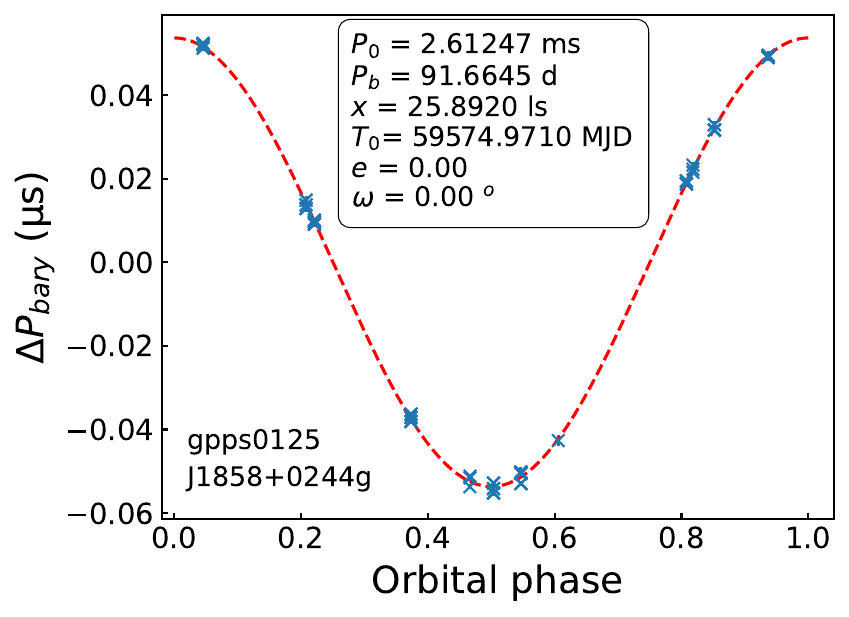}\\ 
  \includegraphics[height = 0.18\textheight] {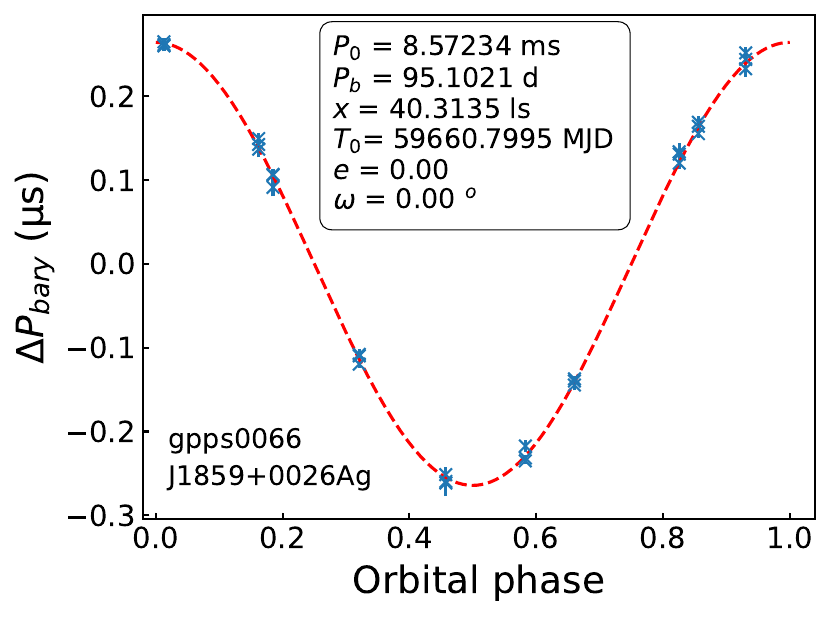}
  \includegraphics[height = 0.18\textheight] {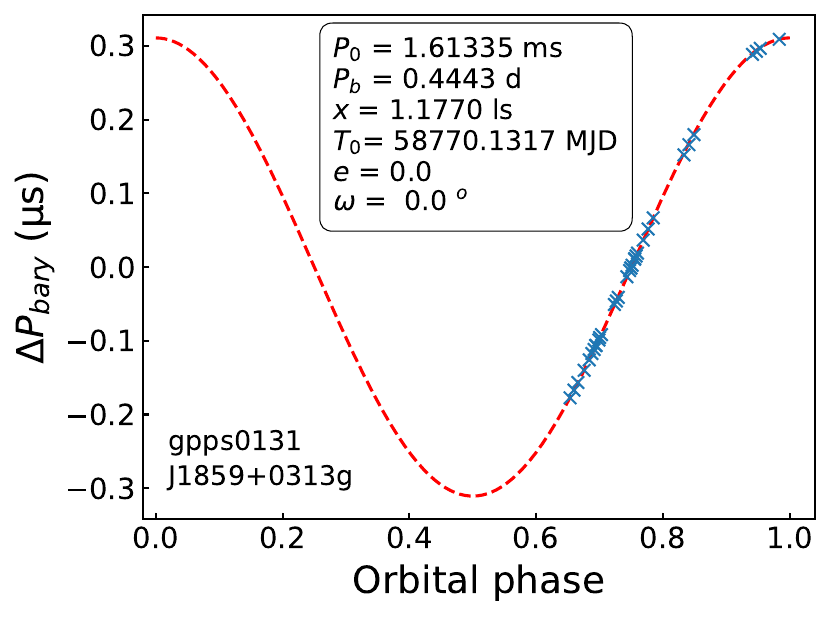}
  \includegraphics[height = 0.18\textheight] {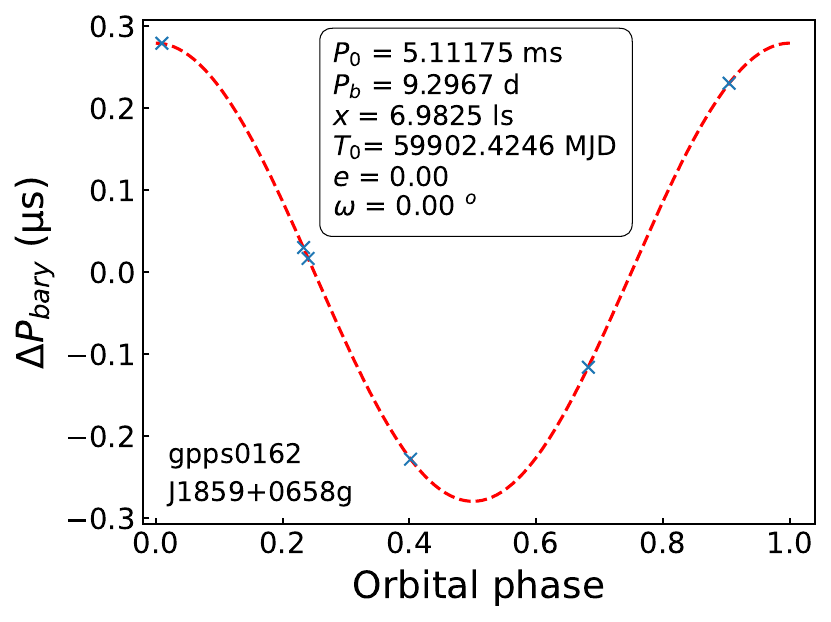}\\ 
  \includegraphics[height = 0.18\textheight] {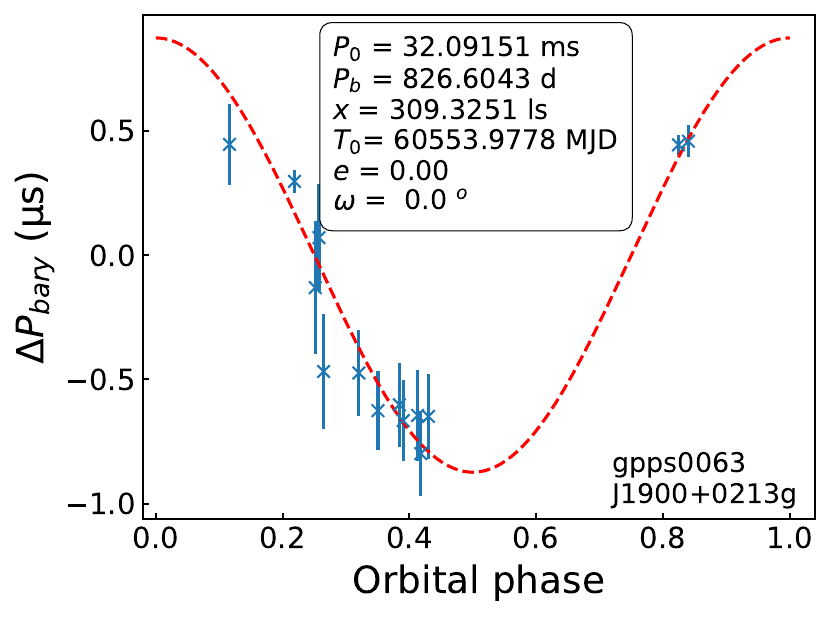}
  \includegraphics[height = 0.18\textheight] {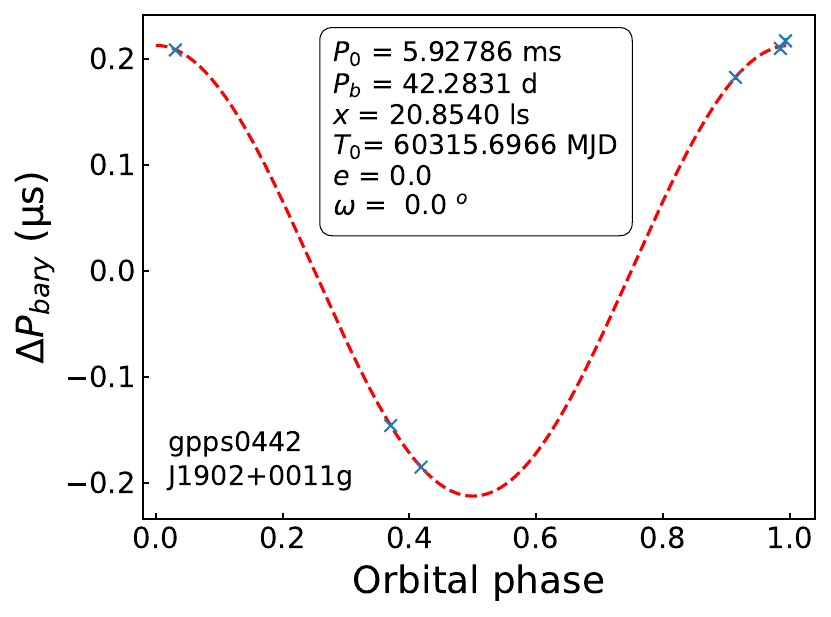} 
  \includegraphics[height = 0.18\textheight] {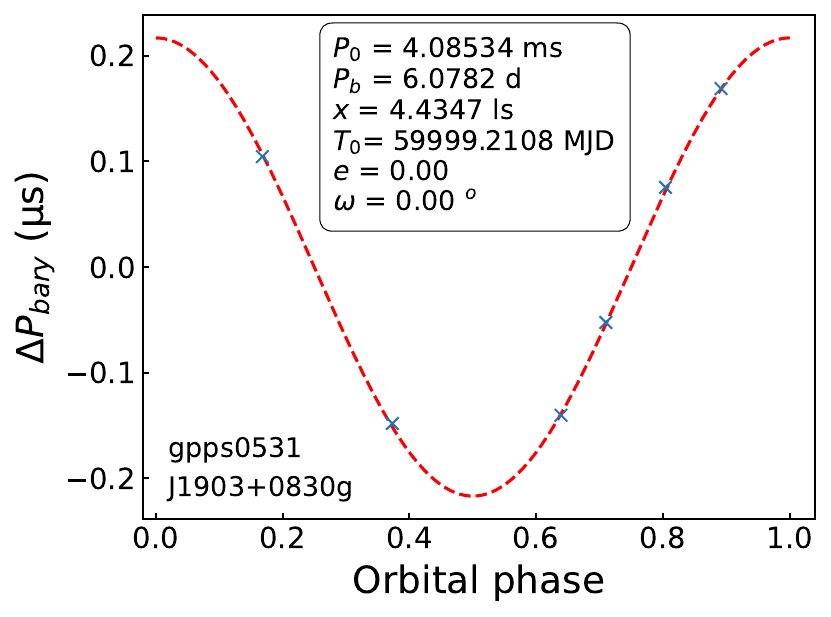}\\
  \includegraphics[height = 0.18\textheight] {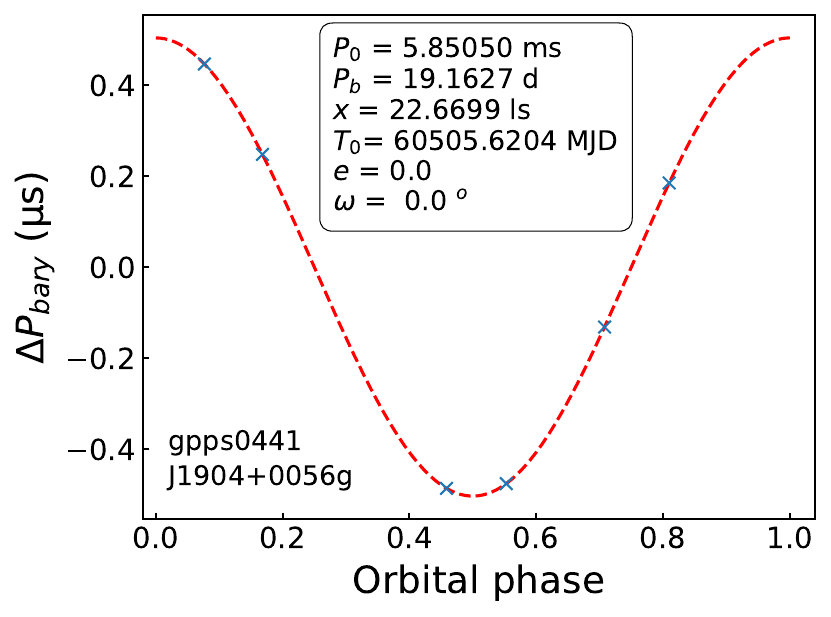}
  \includegraphics[height = 0.18\textheight] {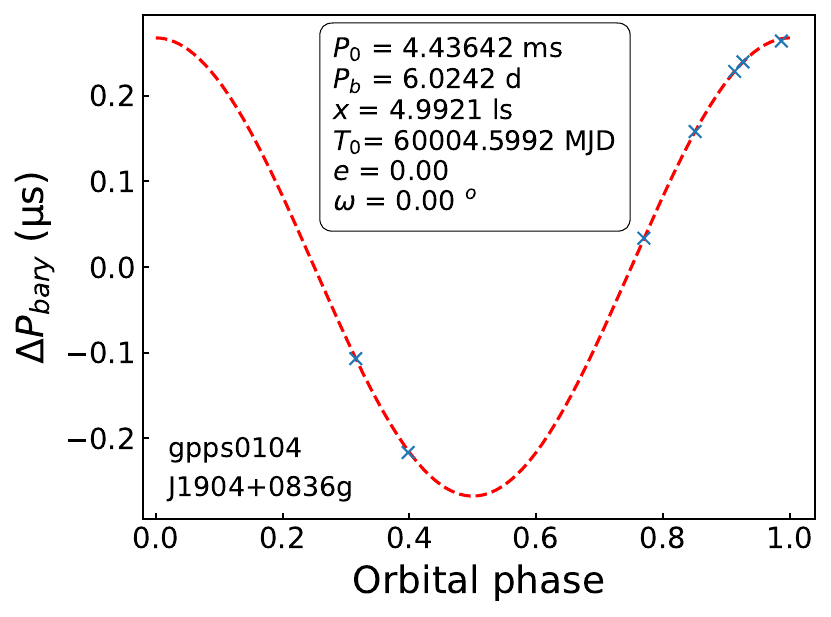}
  \includegraphics[height = 0.18\textheight] {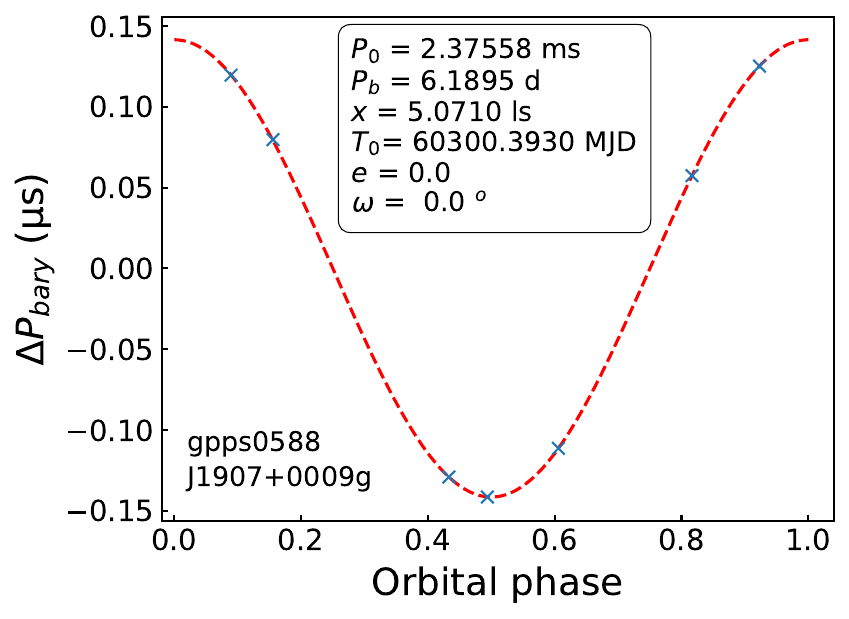}\\  
  \includegraphics[height = 0.18\textheight] {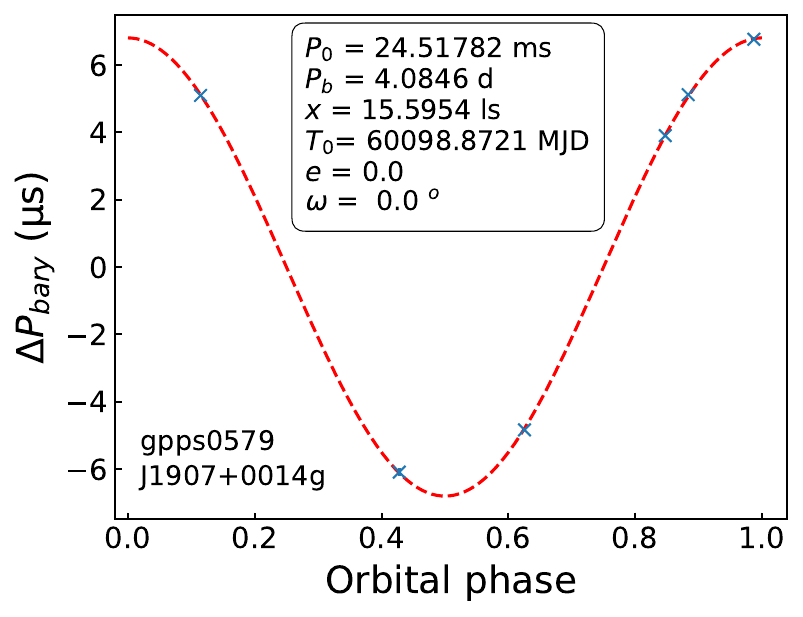}
  \includegraphics[height = 0.18\textheight] {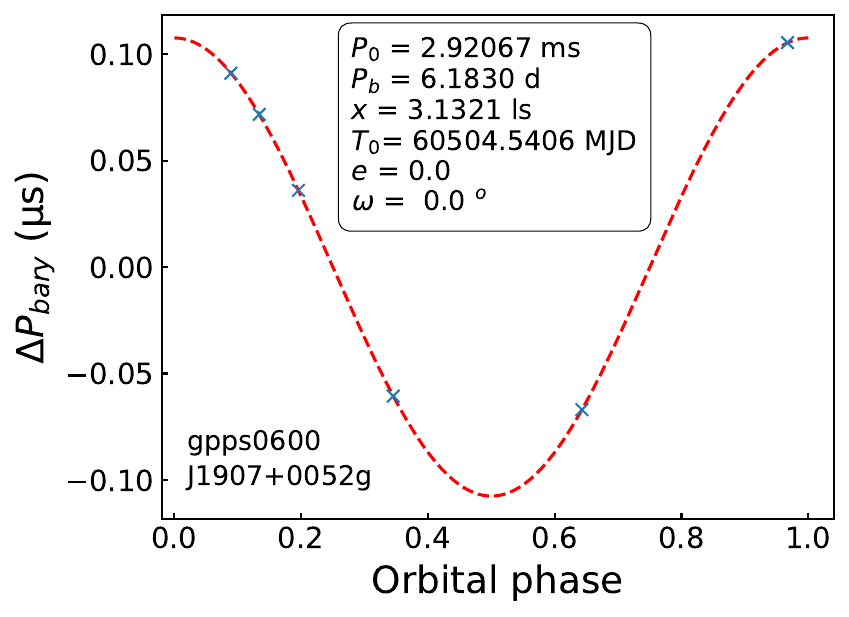}  
  \includegraphics[height = 0.18\textheight] {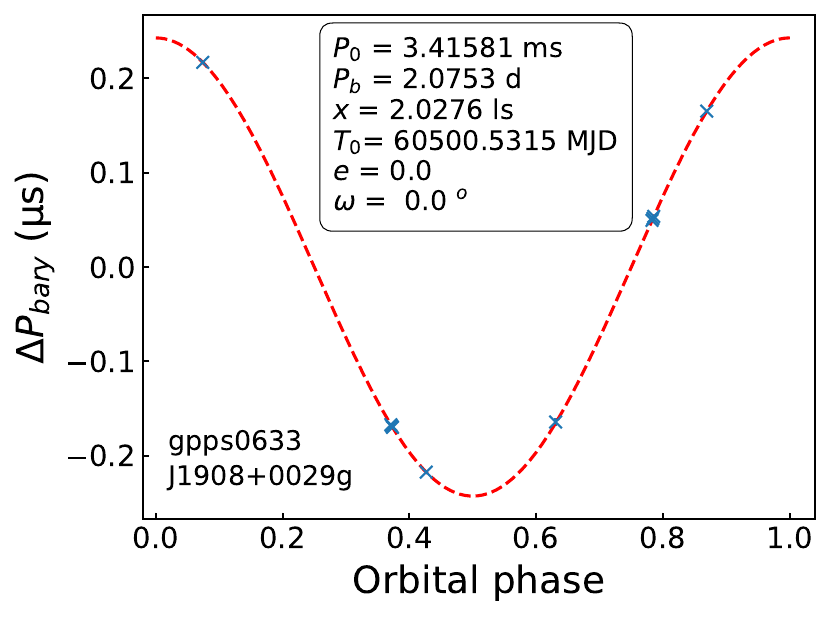}\\
  \caption{--continued--}
\end{figure*}
\addtocounter{figure}{-1}
\begin{figure*}
  \centering
  \includegraphics[height = 0.18\textheight] {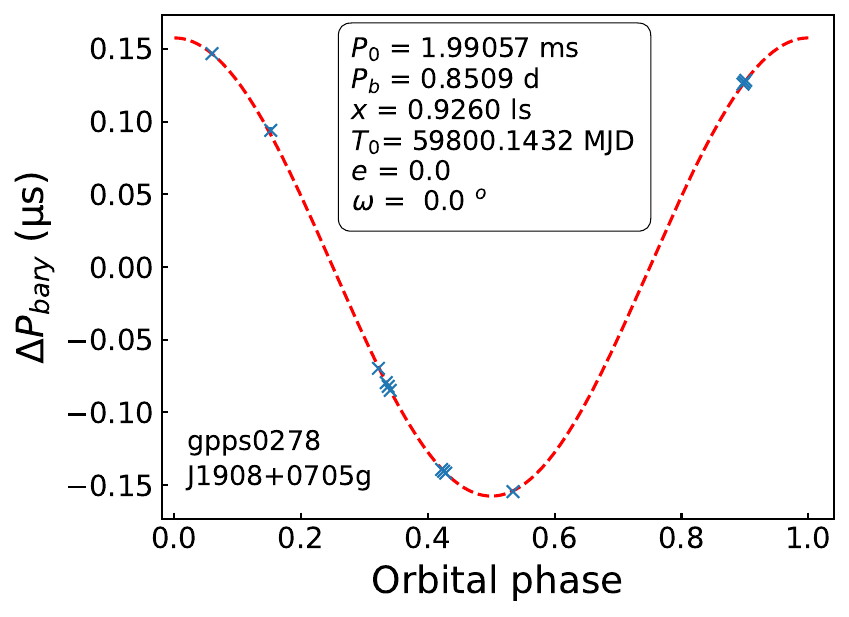}
  \includegraphics[height = 0.18\textheight] {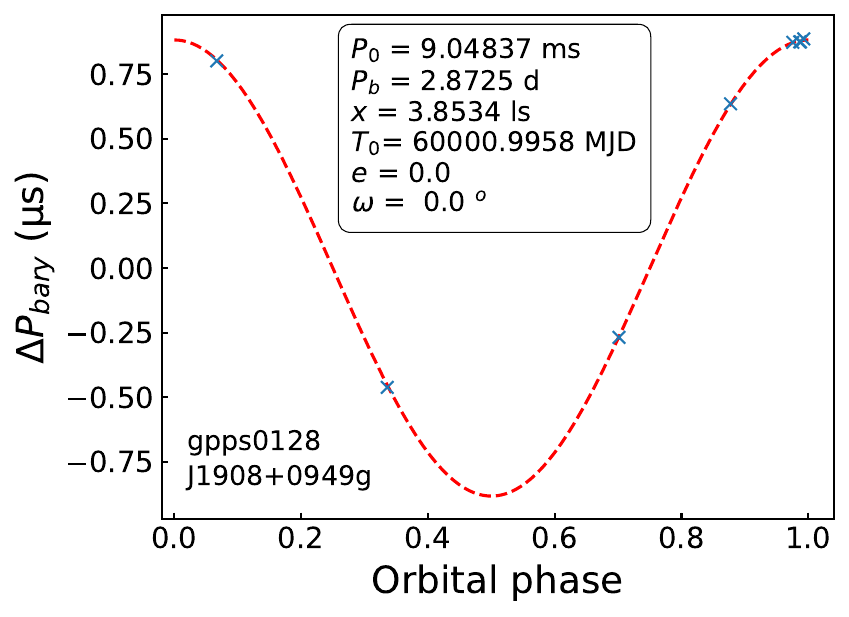}
  \includegraphics[height = 0.18\textheight] {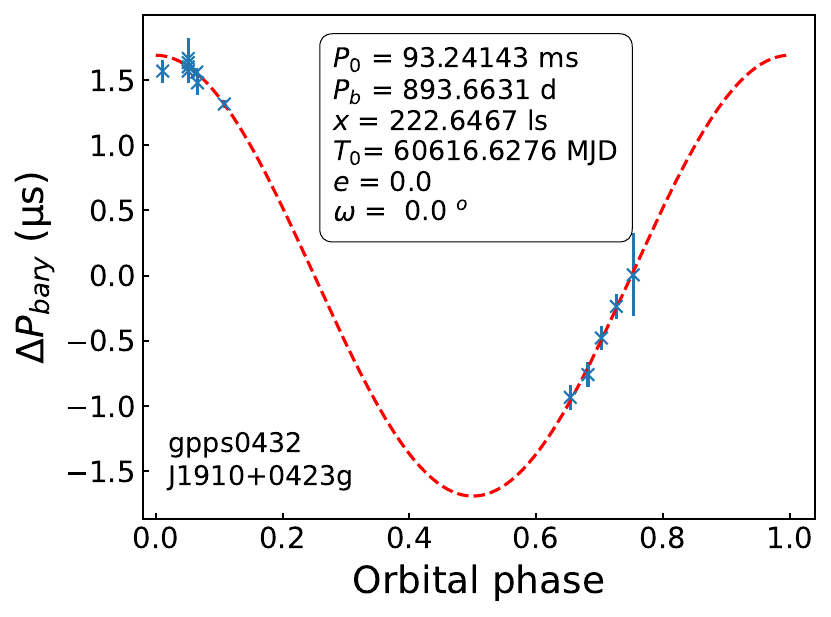}\\ 
  \includegraphics[height = 0.18\textheight] {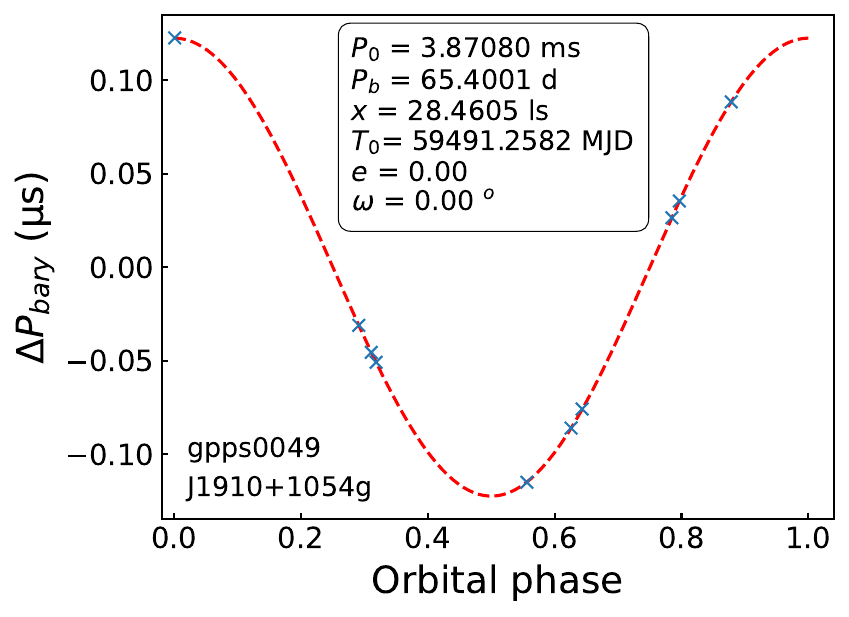} 
  \includegraphics[height = 0.18\textheight] {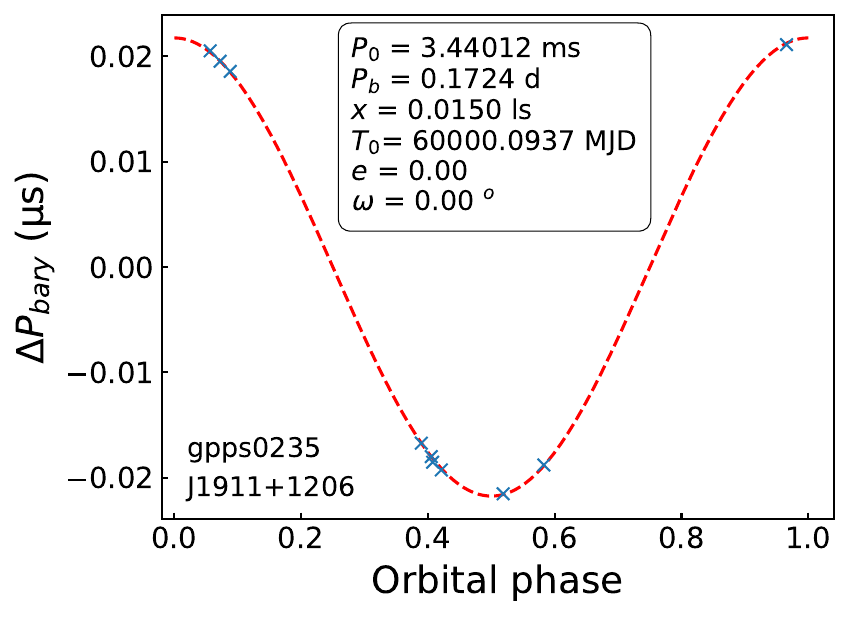}
  \includegraphics[height = 0.18\textheight] {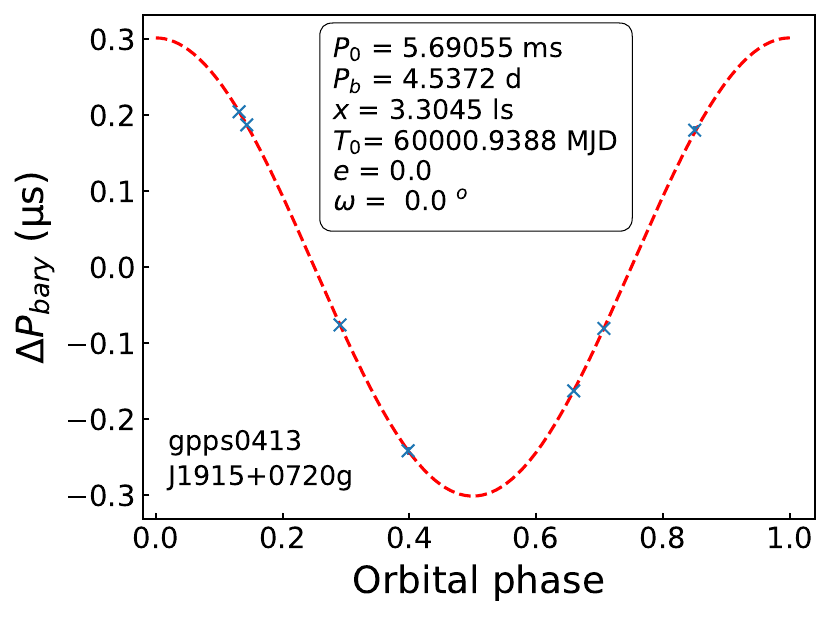}\\ 
  \includegraphics[height = 0.18\textheight] {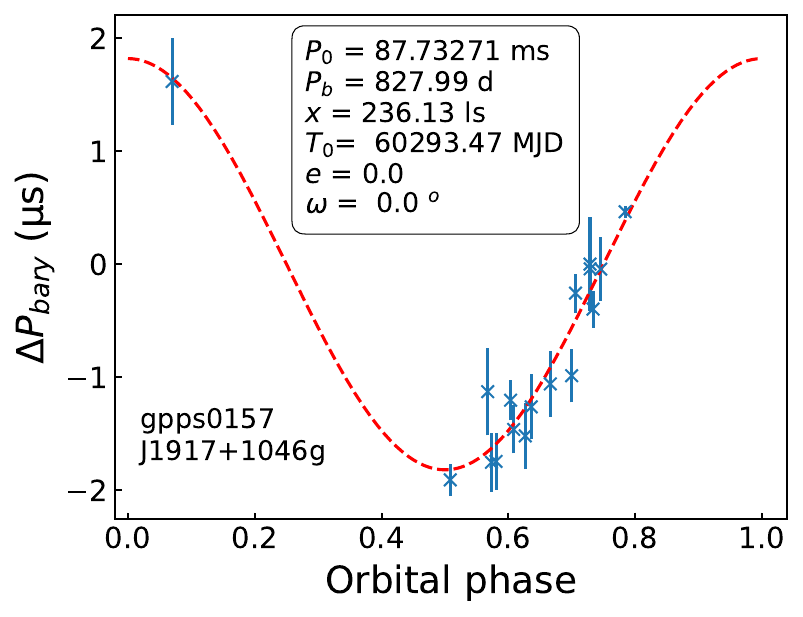}
  \includegraphics[height = 0.18\textheight] {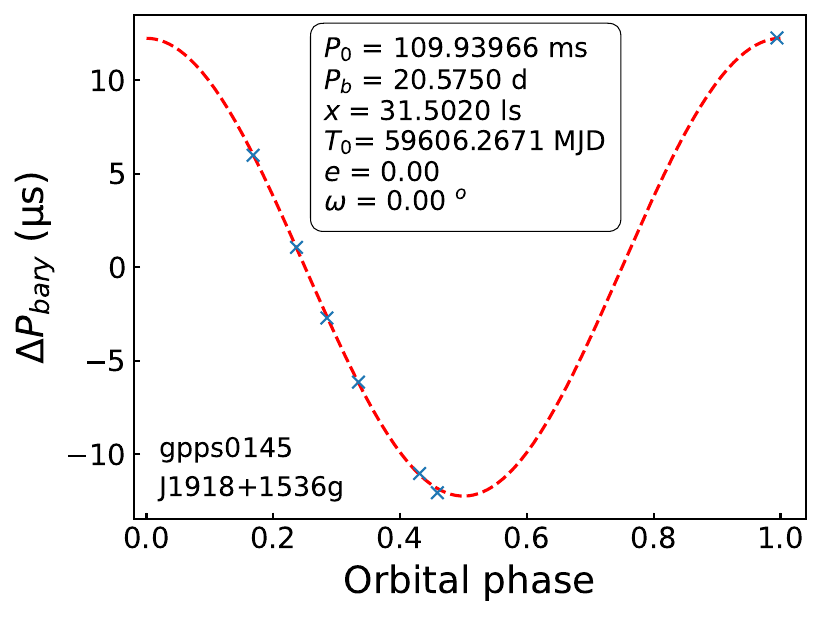}
  \includegraphics[height = 0.18\textheight] {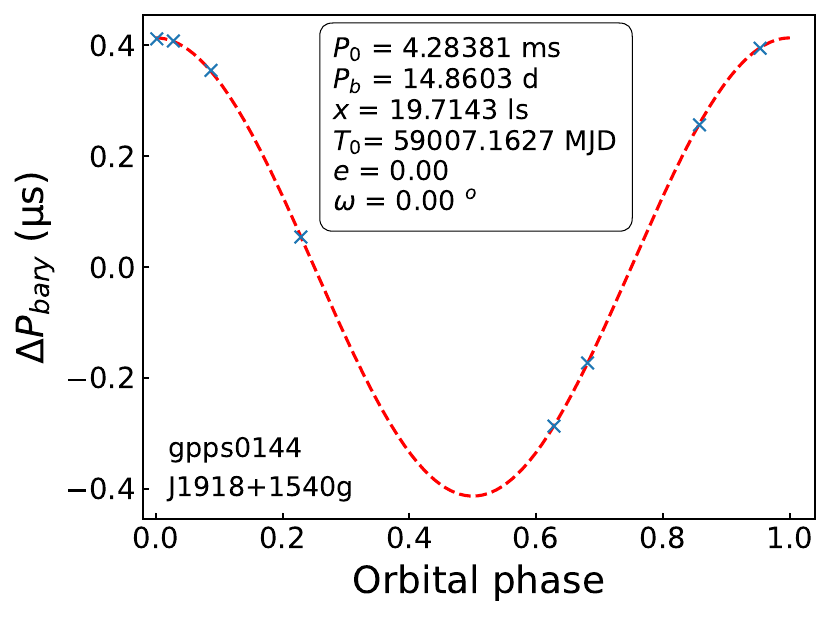}\\
  \includegraphics[height = 0.18\textheight] {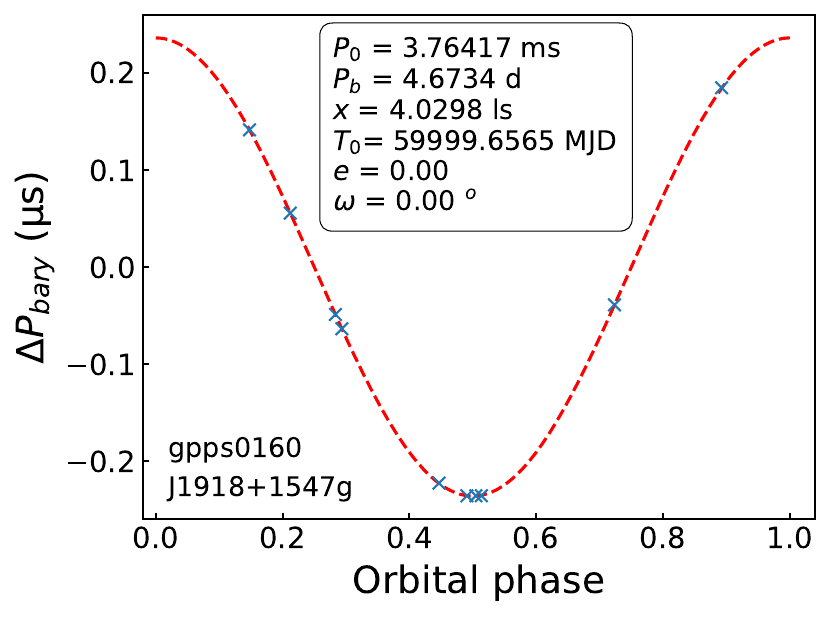}
  \includegraphics[height = 0.18\textheight] {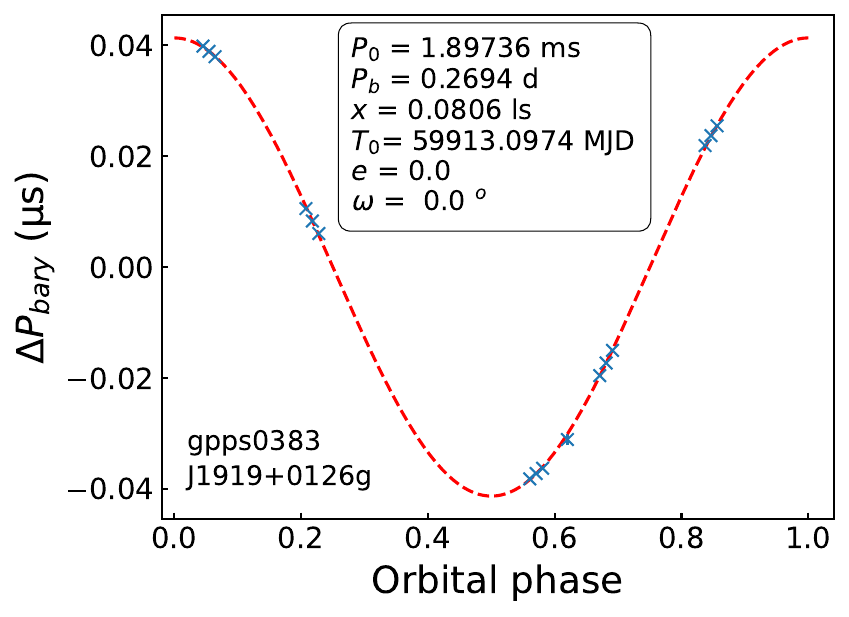}
  \includegraphics[height = 0.18\textheight] {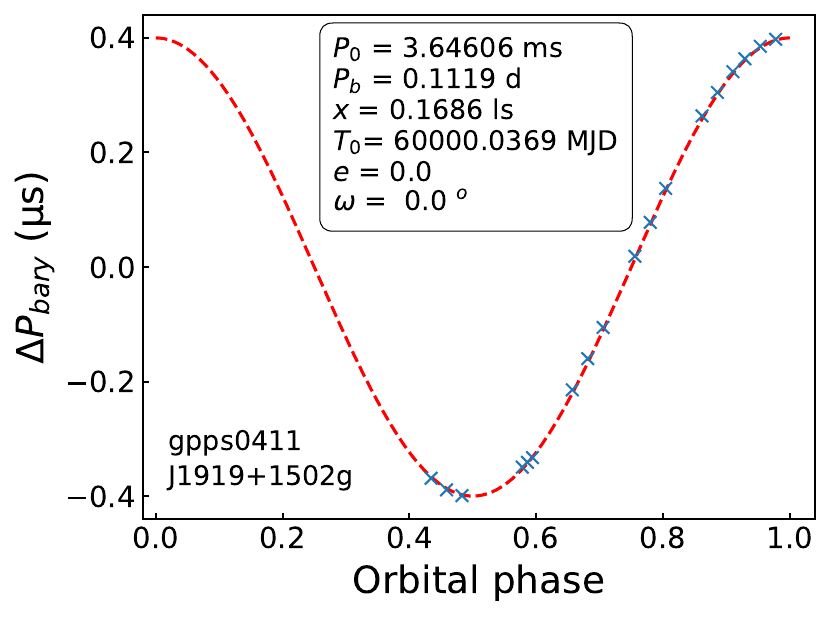}\\ 
  \includegraphics[height = 0.18\textheight] {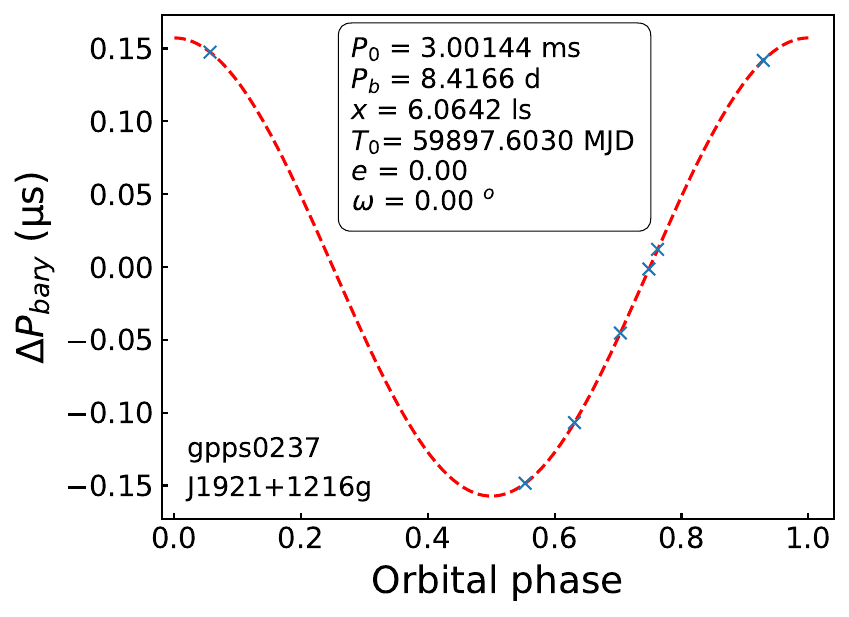} 
  \includegraphics[height = 0.18\textheight] {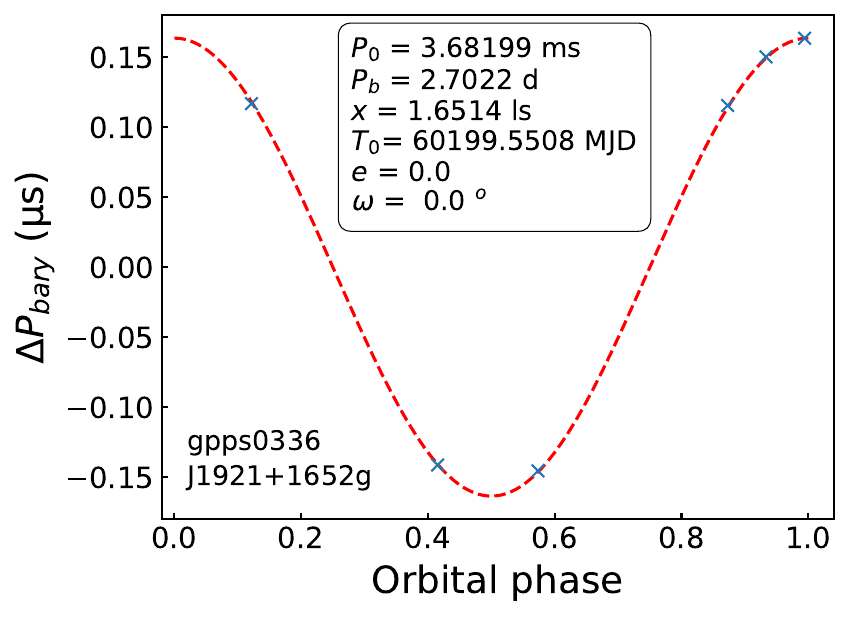}
  \includegraphics[height = 0.18\textheight] {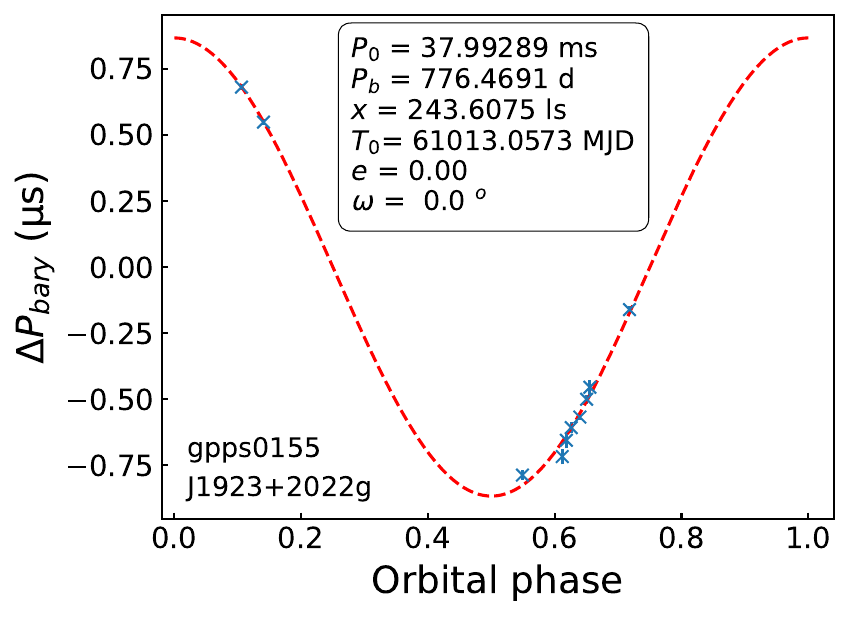}\\

  \caption{--continued--}
\end{figure*}
\addtocounter{figure}{-1}
\begin{figure*}
  \centering
  \includegraphics[height = 0.18\textheight] {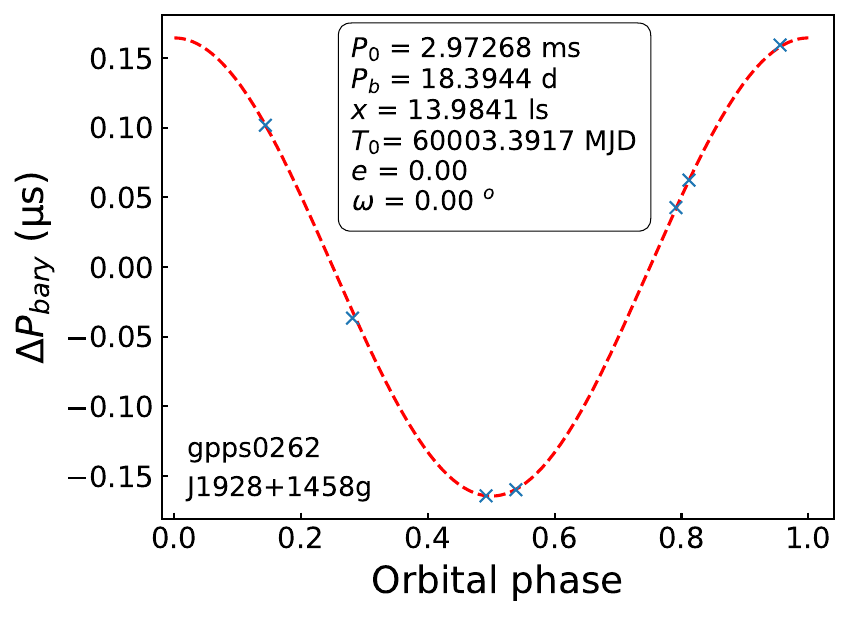}
  \includegraphics[height = 0.18\textheight] {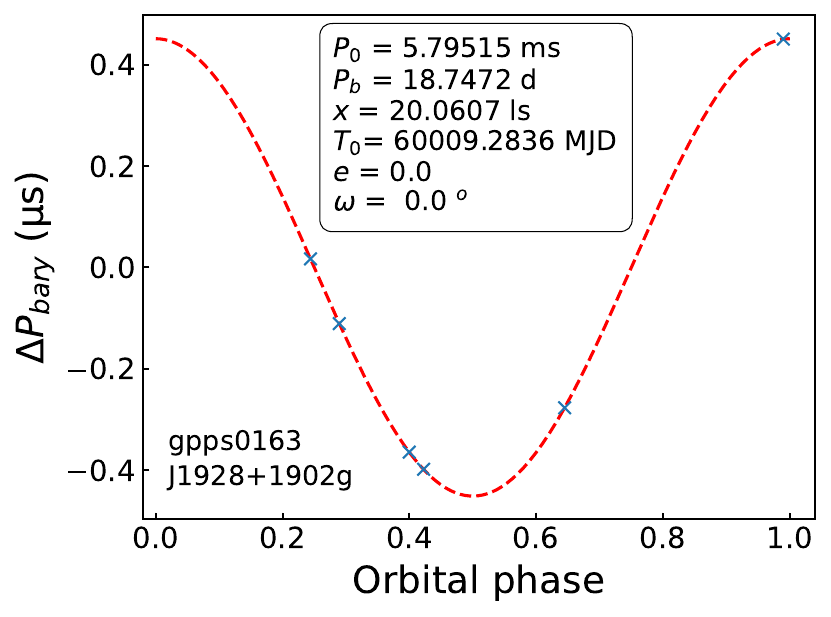}
  \includegraphics[height = 0.18\textheight] {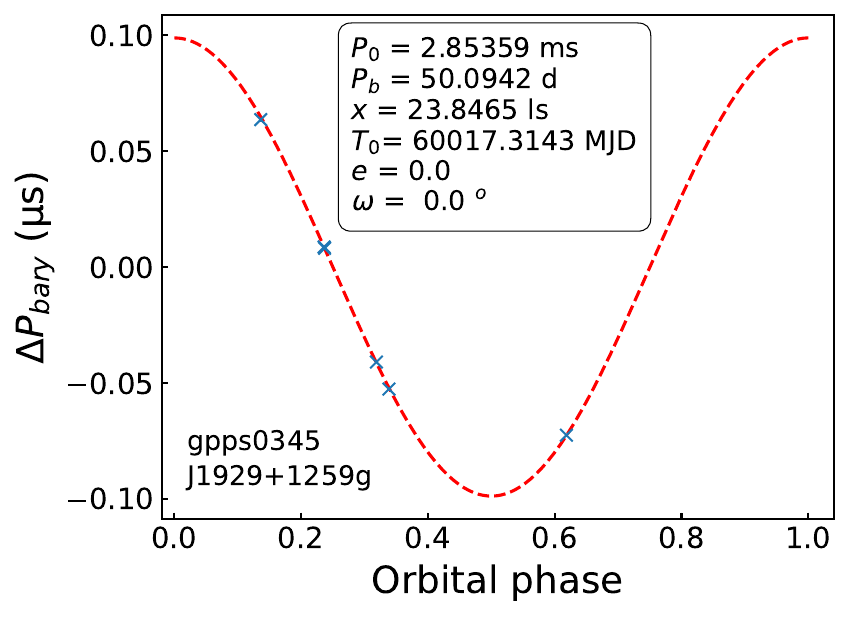}\\ 
  \includegraphics[height = 0.18\textheight] {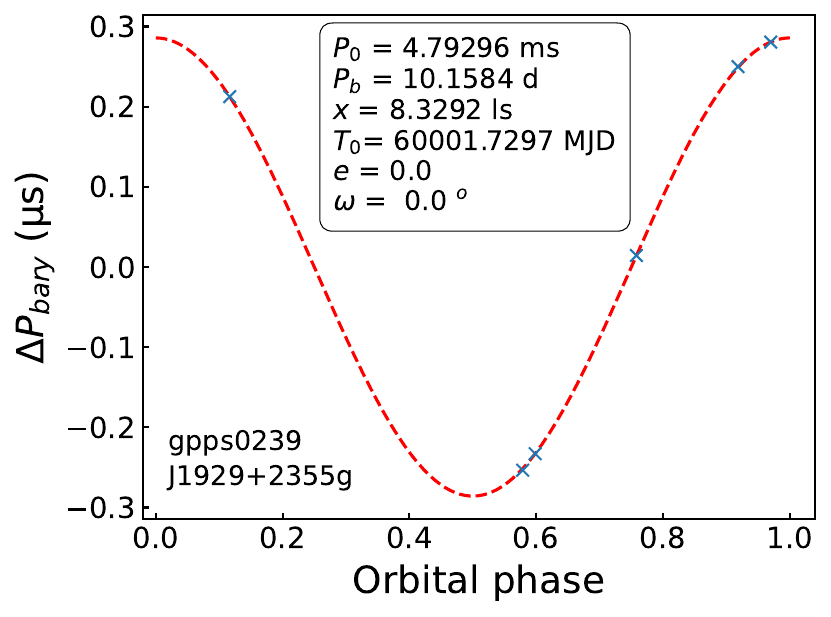}
  \includegraphics[height = 0.18\textheight] {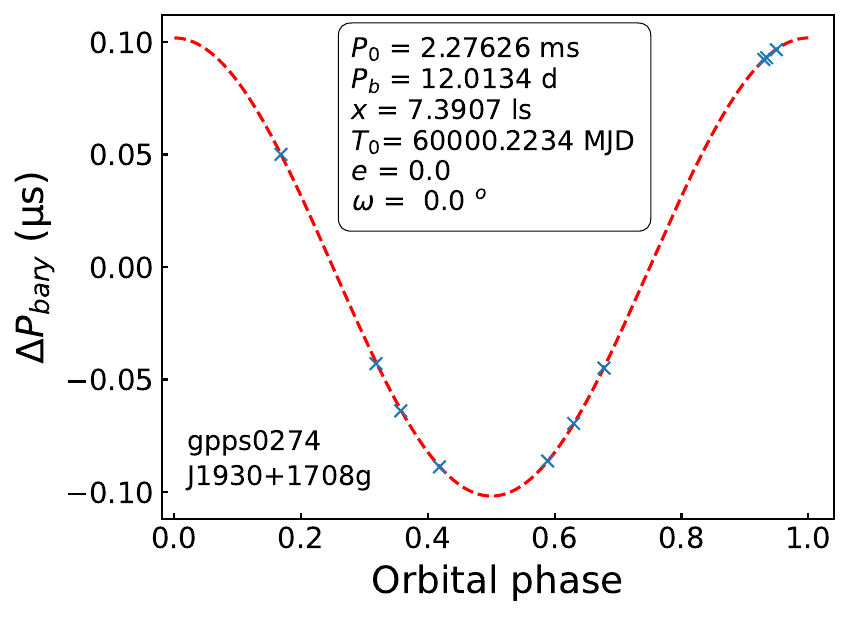}
  \includegraphics[height = 0.18\textheight] {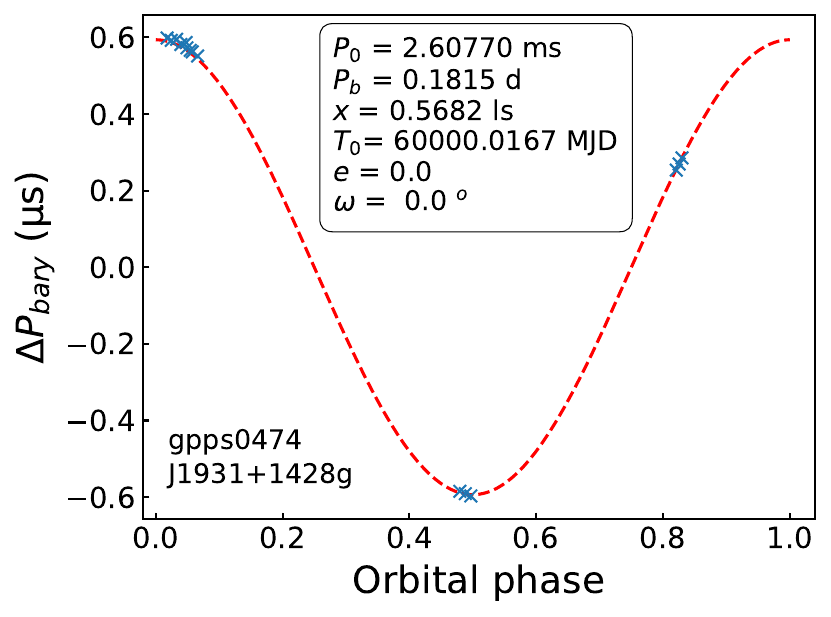}\\
  \includegraphics[height = 0.18\textheight] {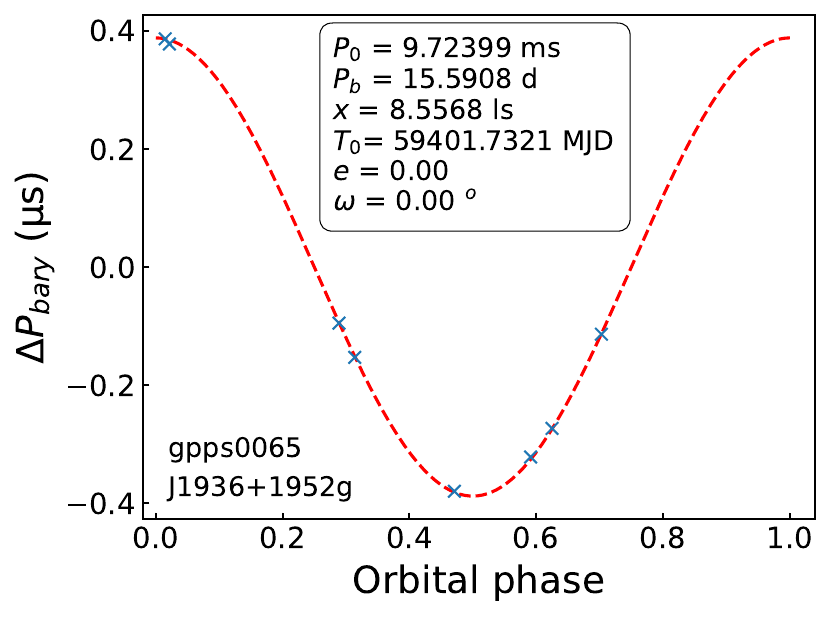}
  \includegraphics[height = 0.18\textheight] {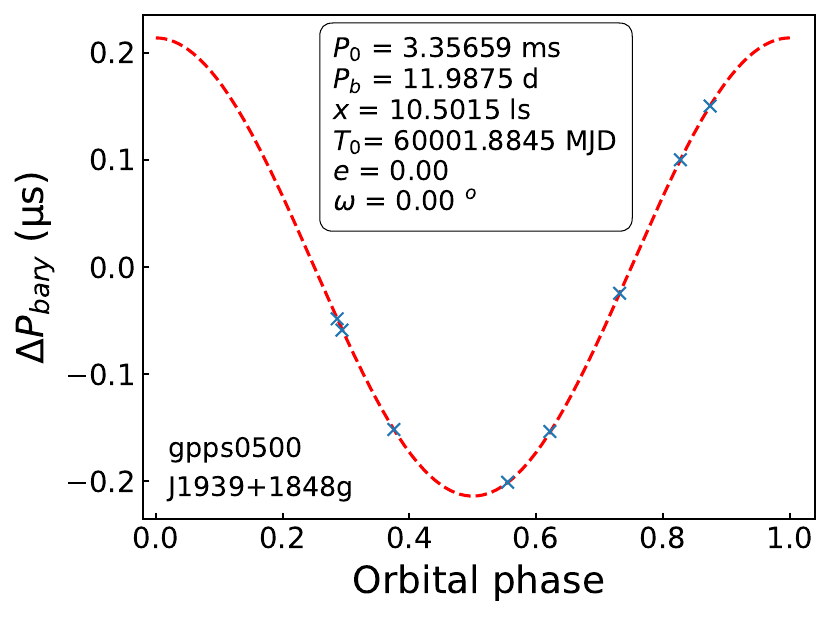}
  \includegraphics[height = 0.18\textheight] {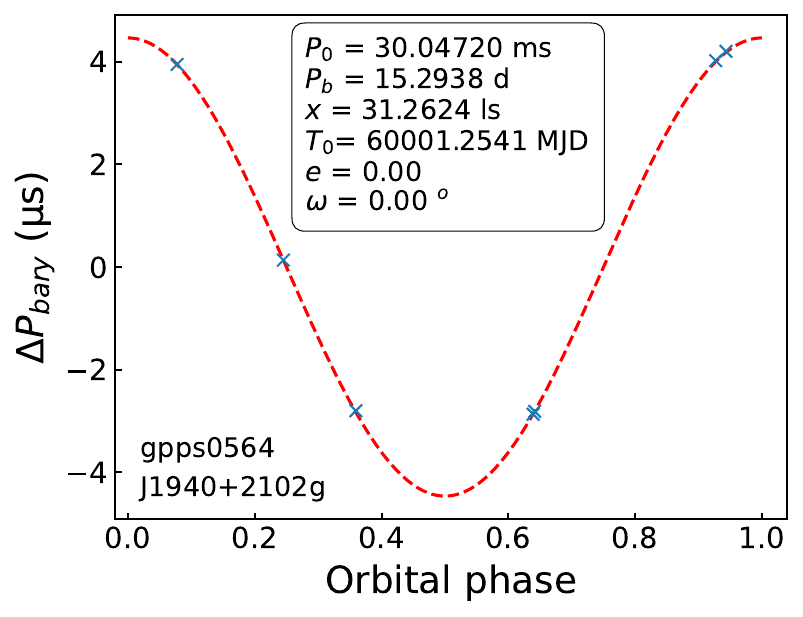}\\
  \includegraphics[height = 0.18\textheight] {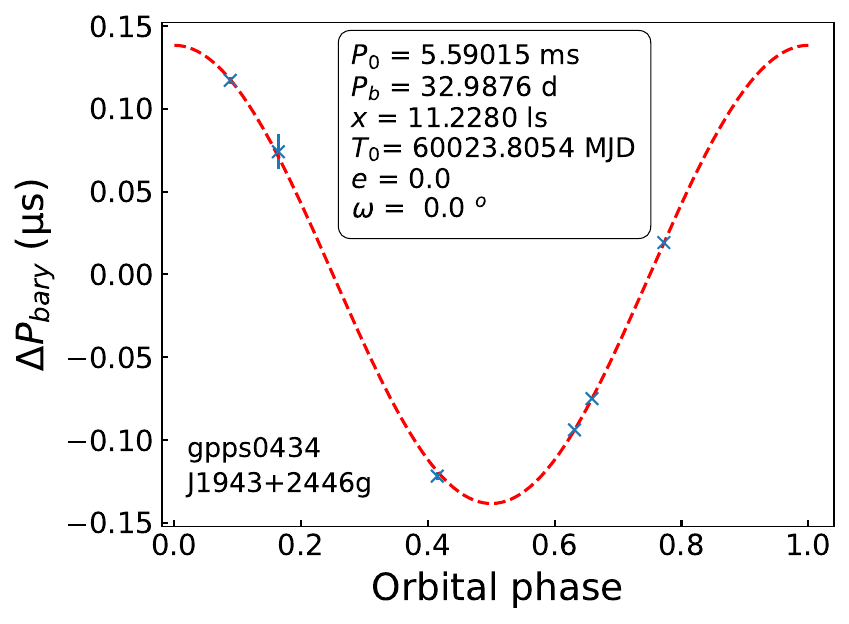}
  \includegraphics[height = 0.18\textheight] {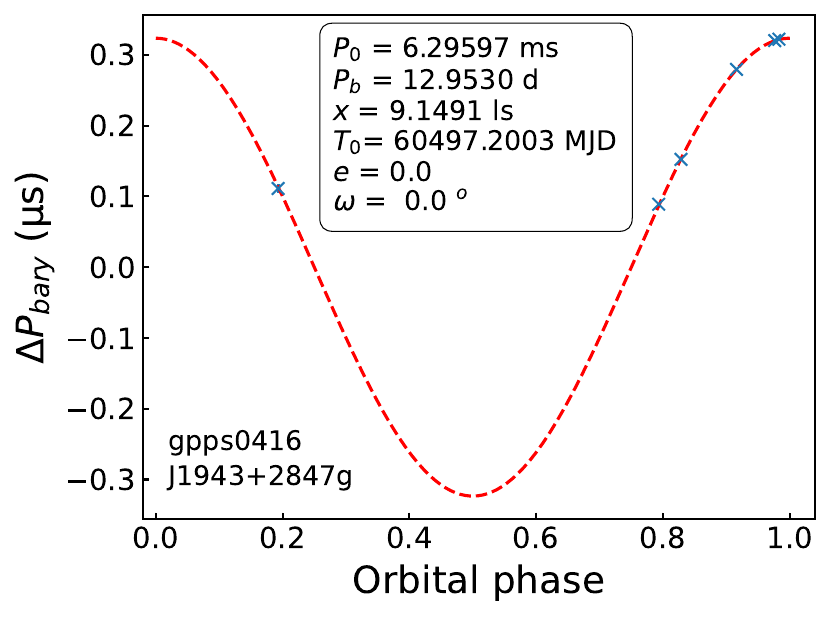} 
  \includegraphics[height = 0.18\textheight] {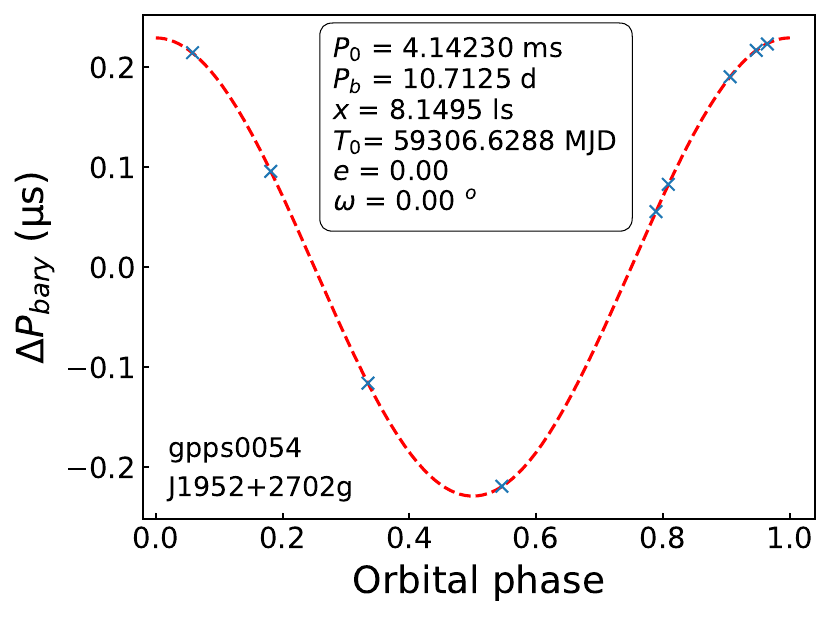}\\
  \includegraphics[height = 0.18\textheight] {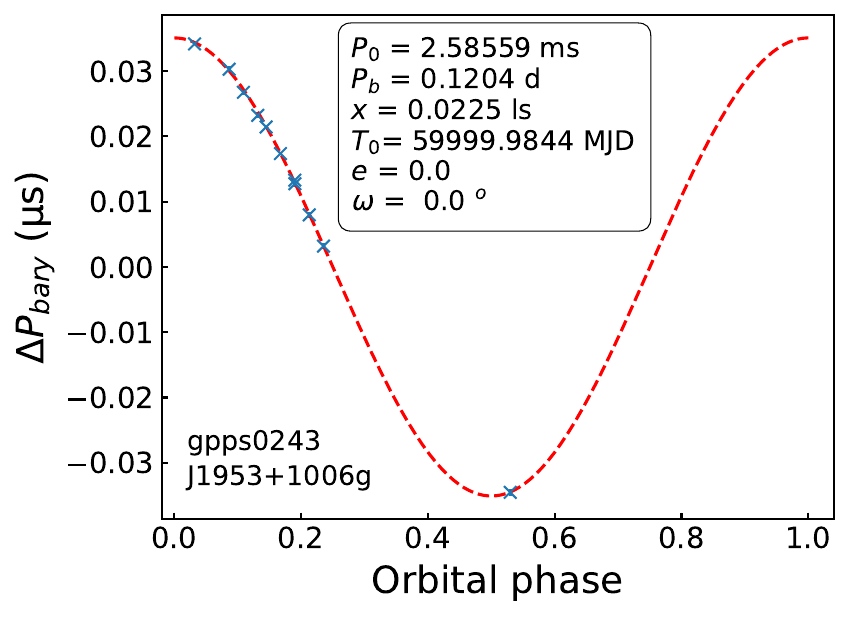}
  \includegraphics[height = 0.18\textheight] {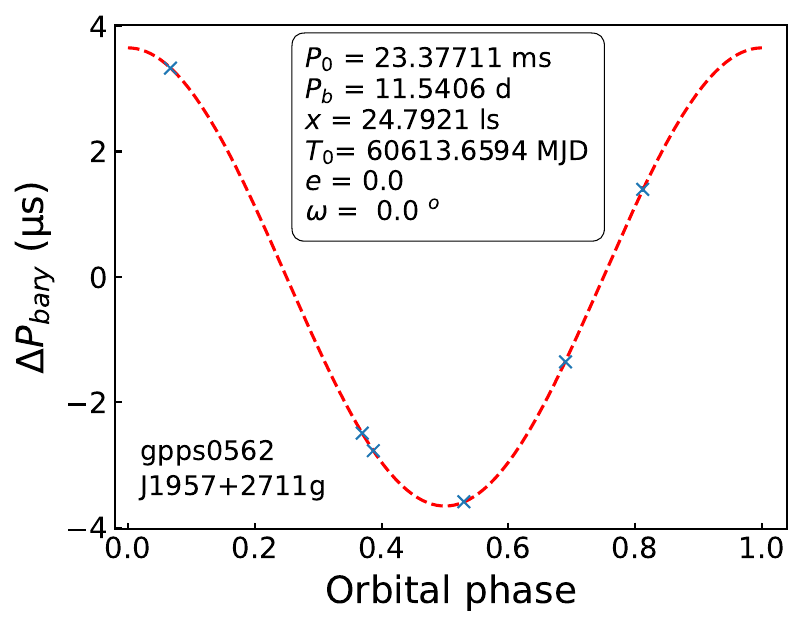}
  \includegraphics[height = 0.18\textheight] {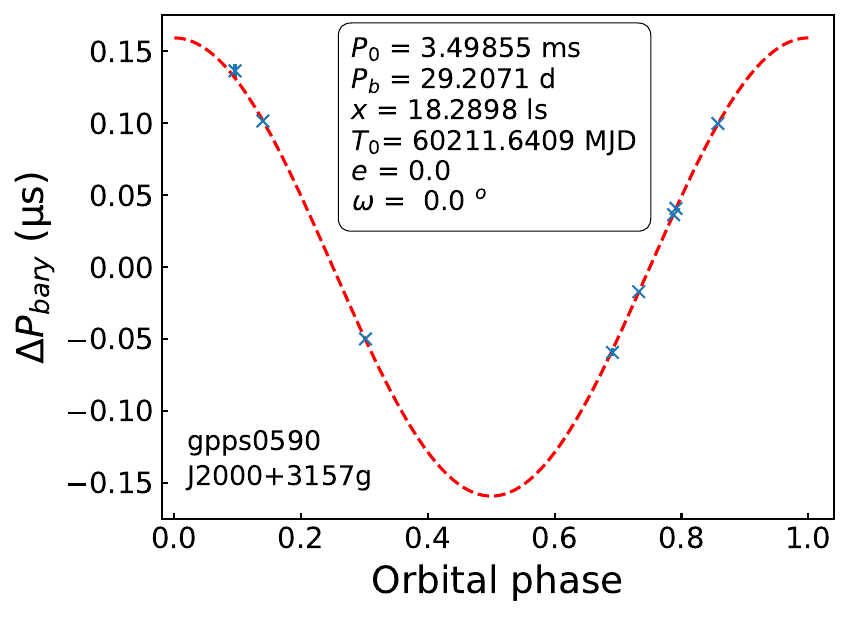}\\
  \caption{--continued--}
\end{figure*}
\addtocounter{figure}{-1}
\begin{figure*}
  \centering 

  \includegraphics[height = 0.18\textheight] {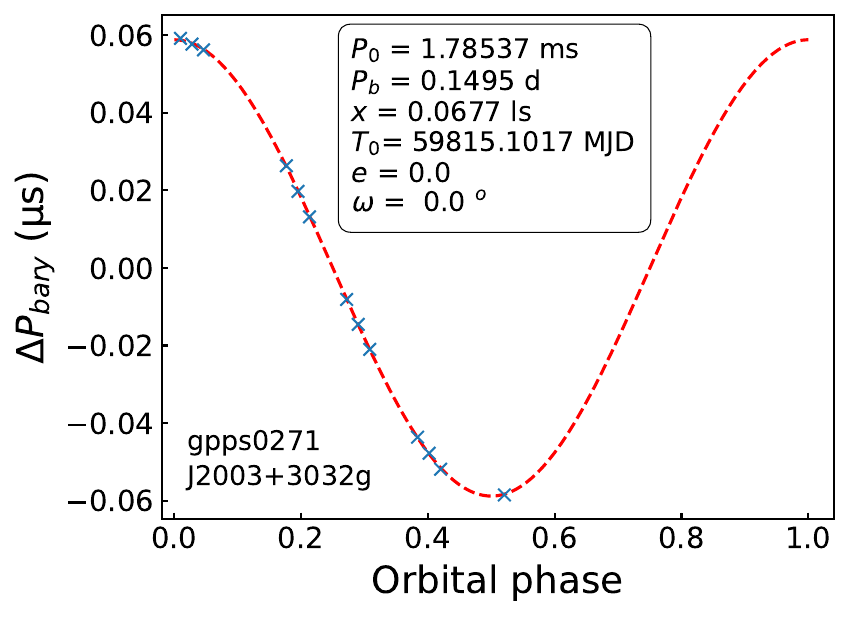}
  \includegraphics[height = 0.18\textheight] {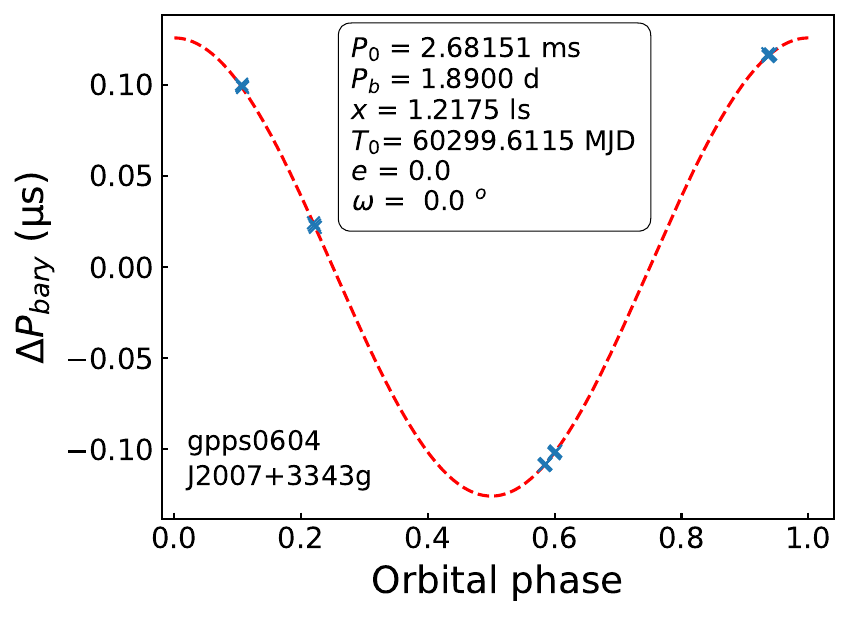} 
  \includegraphics[height = 0.18\textheight] {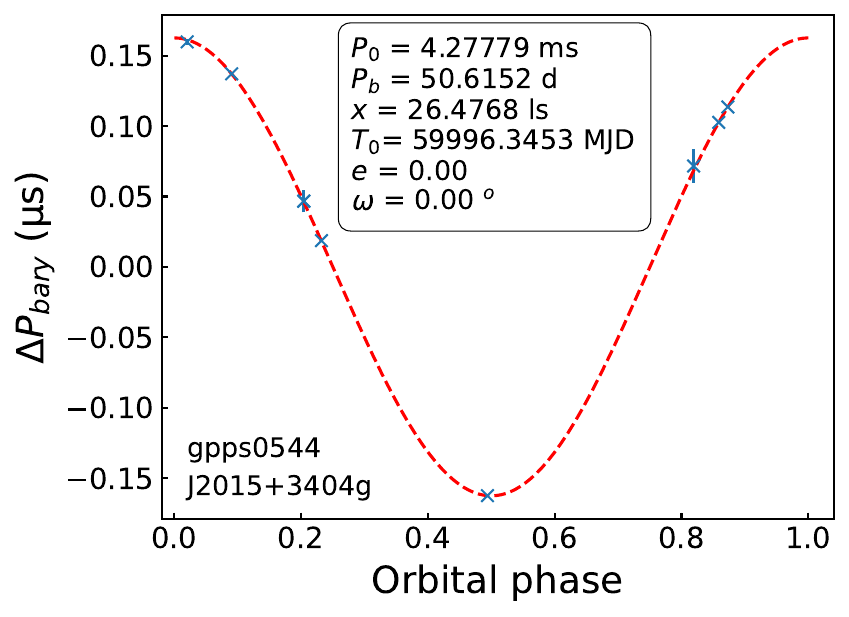}\\
  \caption{--end.}
\end{figure*}

\end{document}